\documentclass[a4paper,fleqn]{cas-sc}

\usepackage[T1]{fontenc}
\usepackage[polutonikogreek,italian,english]{babel}
\usepackage[page,toc,titletoc,title]{appendix}
\usepackage[numbers,sort&compress]{natbib}
\usepackage{graphicx}
\usepackage{graphics}
\usepackage{braket}
\usepackage{bbold}
\usepackage{amssymb}
\usepackage{amsmath}
\usepackage{nicefrac}
\usepackage{dcolumn}
\usepackage{bm}
\usepackage{slashed}
\usepackage{datetime}
\usepackage{mciteplus}
\usepackage{multirow}
\usepackage{bbold}
\usepackage{color}
\usepackage{xcolor}
\usepackage{float}
\usepackage[utf8]{inputenc}
\usepackage[normalem]{ulem}
\usepackage{acro}

\DeclareAcronym{AI}{
  short=AI,
  long=Artificial Intelligence,
}
\DeclareAcronym{BSM}{
  short=BSM,
  long=Beyond-the-Standard-Model,
}
\DeclareAcronym{QCD}{
  short=QCD,
  long=Quantum ChromoDynamics,
}
\DeclareAcronym{SM}{
  short=SM,
  long=Standard Model,
}
\DeclareAcronym{CP}{
  short=CP,
  long=Charge-Parity,
}
\DeclareAcronym{HF-NRevo}{
  short=HF-NRevo,
  long=Heavy-flavor NonRelativistic evolution,
}
\DeclareAcronym{SLAC}{
  short=SLAC,
  long=Stanford Linear Accelerator Center,
}
\DeclareAcronym{BNL}{
  short=BNL,
  long=Brookhaven National Laboratory,
}
\DeclareAcronym{FCNCs}{
  short=FCNCs,
  long=Flavor-Changing Neutral Currents,
}
\DeclareAcronym{GIM}{
  short=GIM,
  long=Glashow--Iliopoulos--Maiani,
}
\DeclareAcronym{CEM}{
  short=CEM,
  long= Color Evaporation Model,
}
\DeclareAcronym{CSM}{
  short=CSM,
  long= Color Singlet Mechanism,
}
\DeclareAcronym{CO}{
  short=CO,
  long=Color Octet,
}
\DeclareAcronym{NRQCD}{
  short=NRQCD,
  long=NonRelativistic QCD,
}
\DeclareAcronym{SDC}{
  short=SDC,
  long=Short-Distance Coefficient,
}
\DeclareAcronym{SDCs}{
  short=SDCs,
  long=Short-Distance Coefficients,
}
\DeclareAcronym{LDME}{
  short=LDME,
  long=Long-Distance Matrix Element,
}
\DeclareAcronym{LDMEs}{
  short=LDMEs,
  long=Long-Distance Matrix Elements,
}
\DeclareAcronym{LO}{
  short=LO,
  long=Leading Order,
}
\DeclareAcronym{NLO}{
  short=NLO,
  long=Next-to-Leading Order,
}
\DeclareAcronym{NNLO}{
  short=NNLO,
  long=Next-to-NLO,
}
\DeclareAcronym{MHOUs}{
  short=MHOUs,
  long=Missing Higher-Order Uncertainties,
}
\DeclareAcronym{DIS}{
  short=DIS,
  long=Deep Inelastic Scattering,
}
\DeclareAcronym{DGLAP}{
  short=DGLAP,
  long=Dokshitzer--Gribov--Lipatov--Altarelli--Parisi,
}
\DeclareAcronym{PDFs}{
  short=PDFs,
  long=Parton Distribution Functions,
}
\DeclareAcronym{FFs}{
  short=FFs,
  long=Fragmentation Functions,
}
\DeclareAcronym{MPIs}{
  short=MPIs,
  long=Multi-Parton Interactions,
}
\DeclareAcronym{DPS}{
  short=DPS,
  long=Double-Parton Scattering,
}
\DeclareAcronym{SCET}{
  short=SCET,
  long=Soft and Collinear Effective Theory,
}
\DeclareAcronym{TM}{
  short=TM,
  long=Transverse-Momentum,
}
\DeclareAcronym{TMD}{
  short=TMD,
  long=Transverse-Momentum-Dependent,
}
\DeclareAcronym{FFNS}{
  short=FFNS,
  long=Fixed-Flavor Number Scheme,
}
\DeclareAcronym{VFNS}{
  short=VFNS,
  long=Variable-Flavor Number Scheme,
}
\DeclareAcronym{ZM-VFNS}{
  short=ZM-VFNS,
  long=Zero-Mass-Variable-Flavor Number Scheme,
}
\DeclareAcronym{GM-VFNS}{
  short=GM-VFNS,
  long=General-Mass Variable-Flavor Number Scheme,
}
\DeclareAcronym{ABF}{
  short=ABF,
  long=Altarelli--Ball--Forte,
}
\DeclareAcronym{BFKL}{
  short=BFKL,
  long=Balitsky--Fadin--Kuraev--Lipatov,
}
\DeclareAcronym{LL}{
  short=LL,
  long=Leading Logarithmic,
}
\DeclareAcronym{NLL}{
  short=NLL,
  long=Next-to-Leading Logarithmic,
}
\DeclareAcronym{NNLL}{
  short=NNLL,
  long=Next-to-NLL,
}
\DeclareAcronym{LVM}{
  short=LVM,
  long=Light Vector Meson,
}
\DeclareAcronym{UGD}{
  short=UGD,
  long=Unintegrated Gluon Distribution,
}
\DeclareAcronym{LHC}{
  short=LHC,
  long=Large Hadron Collider,
}
\DeclareAcronym{HL-LHC}{
  short=HL-LHC,
  long=High-Luminosity Large Hadron Collider,
}
\DeclareAcronym{FCC}{
  short=FCC,
  long=Future Circular Collider,
}
\DeclareAcronym{EIC}{
  short=EIC,
  long=Electron-Ion Collider,
}
\DeclareAcronym{HFAG}{
  short=HFAG,
  long=Heavy Flavor Averaging Group,
}
\DeclareAcronym{SCA}{
  short=SCA,
  long=Small-Cone Algorithm
}
\DeclareAcronym{BLM}{
  short=BLM,
  long=Brodsky--Lepage--Mackenzie,
}
\DeclareAcronym{SNAJ}{
  short=SNAJ,
  long=Suzuki--Nejad--Amiri--Ji,
}
\DeclareAcronym{MSb}{
  short={$\boldsymbol{\overline{\rm MS}}$},
  long=Modified Minimal Subtraction,
}
\DeclareAcronym{MOM}{
  short=MOM,
  long=MOMentum,
}
\DeclareAcronym{HELL}{
  short={HELL},
  long=High Energy Large Logarithms,
}

\newcommand{\deffont}[1]{\begin{otherlanguage*}{polutonikogreek}#1\end{otherlanguage*}}

\def\tsc#1{\csdef{#1}{\textsc{\lowercase{#1}}\xspace}}
\tsc{WGM}
\tsc{QE}

\newcommand{\drv}{{\rm d}}

\newcommand{\LQCD}{\Lambda_{\rm QCD}}
\newcommand{\MSb}{\overline{\rm MS}}

\newcommand{\NLO}{{\rm NLO}}

\newcommand{\LL}{{\rm LL/LO}}
\newcommand{\NLL}{{\rm NLL/NLO}}
\newcommand{\NLLp}{{\rm NLL/NLO^+}}
\newcommand{\NLLpp}{{\rm NLL/NLO^{(+)}}}
\newcommand{\HENLOp}{{\rm HE}\mbox{-}{\rm NLO^+}}

\newcommand{\CnLL}{{\cal C}_n^\LL}

\newcommand{\CnNLLp}{{\cal C}_n^\NLLp}

\newcommand{\CnHENLOp}{{\cal C}_n^{{\rm HE}\text{-}{\rm NLO}^+}}
\newcommand{\DY}{\Delta Y}

\newcommand{\vqTTa}{\langle {\vec q}_T^{\;2} \rangle}

\newcommand{\F}{{\cal F}}

\newcommand{\Jpsi}{J/\psi}

\newcommand{\BCs}{B_c(^1S_0)}
\newcommand{\Bss}{B_c(^3S_1)}

\newcommand{\XQq}{X_{Qq\bar{Q}\bar{q}}}

\newcommand{\Xcs}{X_{cs\bar{c}\bar{s}}}

\newcommand{\Xbs}{X_{bs\bar{b}\bar{s}}}

\newcommand{\TQQ}{T_{4Q}}
\newcommand{\TQc}{T_{4c}}
\newcommand{\TQcZpp}{T_{4c}(0^{++})}

\newcommand{\TQbZpp}{T_{4b}(0^{++})}

\newcommand{\PQQ}{P_{5Q}}
\newcommand{\PQc}{P_{5c}}
\newcommand{\PQb}{P_{5b}}
\newcommand{\bPQQ}{\bar{P}_{5Q}}

\newcommand{{\HFNRevo}}{\textsc{HF-NRevo}}

\newcommand{{\Jethad}}{\textsc{Jethad}}
\newcommand{{\symJethad}}{\textsc{symJethad}}
\newcommand{{\psymJethad}}{\textsc{(sym)Jethad}}
\newcommand{{\Hell}}{\textsc{Hell}}
\newcommand{{\RadISH}}{\textsc{RadISH}}
\newcommand{{\Pegasus}}{\textsc{QCD-PEGASUS}}
\newcommand{{\HOPPET}}{\textsc{HOPPET}}
\newcommand{{\QCDNUM}}{\textsc{QCDNUM}}
\newcommand{{\APFEL}}{\textsc{APFEL}}
\newcommand{{\APFELpp}}{\textsc{APFEL++}}
\newcommand{{\APFELppp}}{\textsc{APFEL(++)}}
\newcommand{{\EKO}}{\textsc{EKO}}
\newcommand{{\FeynCalc}}{\textsc{FeynCalc}}

\newcommand{{\HCFF}}{{\tt HCFF1.0}}
\newcommand{{\NRFF}}{{\tt NRFF1.0}}

\begin{document}
\let\WriteBookmarks\relax
\def\floatpagepagefraction{1}
\def\textpagefraction{.001}

\shorttitle{Fully Heavy Pentaquarks with {\Jethad}: A high-Energy Viewpoint}    

\shortauthors{Celiberto, Francesco Giovanni}  

\title []{\Huge Fully Heavy Pentaquarks with {\Jethad}: \\ A High-Energy Viewpoint}  

\author[1]{Francesco Giovanni Celiberto}[orcid=0000-0003-3299-2203]

\cormark[1]


\ead{francesco.celiberto@uah.es}


\affiliation[1]{organization={Universidad de Alcal\'a (UAH), Departamento de F\'isica y Matem\'aticas},
            addressline={Campus Universitario}, 
            city={Alcal\'a de Henares},
            postcode={E-28805}, 
            state={Madrid},
            country={Spain}}




\begin{abstract}
We examine the leading-power fragmentation of fully heavy pentaquarks in high-energy hadronic collisions.  
To this end, we complete the release of the hadron-structure-oriented {\tt PQ5Q1.0} fragmentation functions, by discussing the $\PQc$ set and delivering the $\PQb$ one.  
These functions incorporate an improved computation of the initial-scale input for the constituent heavy-quark fragmentation channel, making them particularly suitable for describing both the direct formation of a compact multicharm state and the hadronization from a diquark-antiquark-diquark configuration.  
For phenomenological applications, we employ the data-validated \textsc{(sym)Jethad} framework to compute and analyze NLL/NLO$^+$ semi-inclusive production rates of pentaquark-plus-jet systems at the upcoming HL-LHC and the future FCC.  
This study marks a further step toward connecting hadronic structure, precision QCD, and the emerging physics of exotic matter.
\end{abstract}



\begin{keywords}
 {\tt PQ5Q1.0} FF release \sep
 Hadronic structure \sep
 Precision QCD \sep
 Exotic matter \sep
 Pentacharms \sep
 Heavy flavor \sep
 Fragmentation \sep
 Resummation \sep
 HL-LHC \sep
 FCC
\end{keywords}

\maketitle

\newcounter{appcnt}


\tableofcontents
\clearpage

\setlength{\parskip}{3pt}%

\section{Introductory remarks}
\label{sec:introduction}

The production of heavy-flavored particles in high-energy hadronic collisions plays a key role in unraveling the fundamental dynamics of strong interactions.
Heavy-quark signatures serve as key probes for exploring the core principles of particle physics, offering potential signatures of New Physics through interactions between heavy quarks and \ac{BSM} particles.
At the same time, the large masses of heavy quarks allow for precise perturbative \ac{QCD} calculations, enabling detailed studies of strong interactions at next-generation colliding machines, such as the \ac{HL-LHC}~\cite{Apollinari:2015wtw,Apollinari:2015bam,Apollinari:2017lan,Chapon:2020heu}, the \ac{EIC}~\cite{AbdulKhalek:2021gbh,Khalek:2022bzd,Hentschinski:2022xnd,Amoroso:2022eow,Abir:2023fpo,Allaire:2023fgp}, and the \ac{FCC}~\cite{FCC:2018byv,FCC:2018evy,FCC:2018vvp,FCC:2018bvk,FCC:2025lpp,FCC:2025uan,FCC:2025jtd}.

QCD, the theory governing the strong force, is a fundamental pillar of the \ac{SM} of particle physics.
Formulated as a non-Abelian gauge theory based on the $SU(N_c)$ group with $N_c = 3$, QCD describes quarks-fermionic fields in the fundamental triplet representation-and gluons-massless spin-1 bosons mediating strong interactions, represented in the adjoint octet representation~\cite{Gell-Mann:1962yej,Gell-Mann:1964ewy,Zweig:1964jf,Fritzsch:1973pi}.
Beyond its established role in the SM, QCD also provides fertile ground for potential BSM extensions, including axions for resolving the strong CP problem~\cite{Peccei:1977hh,Peccei:1977ur,Peccei:2006as,Duffy:2009ig}, non-Abelian dark gauge forces~\cite{Forestell:2017wov,Huang:2020crf}, quarkyonic matter~\cite{McLerran:2007qj,Hidaka:2008yy,McLerran:2018hbz}, and higher-dimensional QCD operators~\cite{Buchmuller:1985jz,Witten:1979kh,Dudek:2010wm,Afonin:2019unu}.
These possibilities open new avenues for exploring the frontiers of particle physics and searching for physics beyond the SM.

A crucial domain within QCD is the study of hadrons containing two or more heavy quarks.
Among these, mesons formed by a heavy quark-antiquark pair ($Q\bar{Q}$) are known as quarkonia.
The origins of quarkonium studies date back to the so-called ``Quarkonium November Revolution'' of 1974, when the $\Jpsi$ meson was independently discovered at SLAC~\cite{SLAC-SP-017:1974ind} and BNL~\cite{E598:1974sol}, later confirmed by the Frascati ADONE experiment~\cite{Bacci:1974za}.
This breakthrough significantly advanced the understanding of the strong interaction and the quark substructure of hadrons.

While quarkonia are categorized as conventional hadrons, the color-neutrality of QCD allows for the formation of more intricate bound states, leading to the emergence of exotic hadrons.
These states possess quantum numbers that cannot be accommodated within the traditional quark-antiquark or three-quark configurations.
Exotic hadrons can be further divided into two primary classes: those containing explicit gluonic degrees of freedom, such as hybrids and glueballs~\cite{Close:1991pf,Close:1997qda,Close:1998zz,Minkowski:1998mf,Close:2000yg,Mathieu:2008me,Hsiao:2013dta,D0:2020tig,Csorgo:2019ewn}, and those formed by multiple quarks, such as tetraquarks and pentaquarks~\cite{Gell-Mann:1964ewy,Jaffe:1976ig,Jaffe:1976ih,Ader:1981db}.
The first experimental observation of an exotic hadron, the $X(3872)$, occurred in 2003 at Belle~\cite{Belle:2003nnu}, marking the start of the ``Second Quarkonium Revolution''.
More recently, in 2021, the $X(2900)$ was observed by LHCb~\cite{LHCb:2020bls}, providing the first evidence of an exotic state with open-charm flavor.

Despite the progress in characterizing their mass spectra and decay channels, the production mechanisms of exotic hadrons remain an open question.
Theoretical descriptions of these states include meson molecules~\cite{Tornqvist:1993ng,Braaten:2003he,Braaten:2010mg,Braaten:2020iye,Guo:2013sya,Guo:2013xga,Cleven:2015era,Fleming:2021wmk,Dai:2023mxm,Fleming:2007rp,Fleming:2008yn,Fleming:2011xa,Mehen:2015efa,Mutuk:2022ckn,Wang:2020dgr,Wang:2013daa,Xin:2021wcr,Wang:2020cme,Wang:2014gwa}, compact diquark models~\cite{Maiani:2004vq,tHooft:2008rus,Maiani:2013nmn,Maiani:2014aja,Maiani:2017kyi,Mutuk:2021hmi,Mutuk:2021epz,Mutuk:2022zgn,Mutuk:2022nkw,Wang:2023sii,Wang:2019tlw,Wang:2013vex,Wang:2013llv,Wang:2013exa}, and hadroquarkonium configurations~\cite{Dubynskiy:2008mq,Dubynskiy:2008di,Li:2013ssa,Voloshin:2013dpa,Guo:2017jvc,Ferretti:2018ojb,Ferretti:2018tco,Ferretti:2020ewe}, each offering different perspectives on their internal dynamics. 

A decisive breakthrough in the exploration of fully heavy exotics was achieved with the LHC observation of a family of fully-charmed tetraquark candidates---labeled as $X(6600)$, $X(6900)$, and $X(7100)$---in the double-$\Jpsi$ and $\Jpsi$-plus-$\psi(2S)$ invariant-mass channels~\cite{CMS:2023owd}. These resonances, lying in the region $4m_c < M < 7\,{\rm GeV}$, are currently the only confirmed fully-heavy exotic states. Their experimental interpretation has recently been strengthened by a dedicated spin-parity analysis~\cite{CMS:2025fpt} favoring a $J^{PC} = 2^{++}$ assignment for the leading structure. Such findings disfavor loosely bound molecular explanations and support compact tetraquark configurations, possibly dominated by spin-1 diquark components. These results provide strong motivation for extending phenomenological studies to other multiquark systems with heavy content.

The quest for exotic hadrons has remained a central theme in experimental particle physics over the past two decades.
A decade after the landmark discovery of the $X(3872)$, further evidence supporting the existence of tetraquarks came with the simultaneous observation of the charged charmonium-like state $Z_c(3900)$ by BESIII~\cite{BESIII:2013ris} and Belle~\cite{Belle:2013yex}.
This resonance, emerging as a decay product of the anomalous $Y(4260)$, strengthened the case for multiquark configurations beyond the conventional meson-baryon classification.
More recently, in 2021, the LHCb Collaboration reported the observation of the $X(2900)$ resonance~\cite{LHCb:2020bls}, marking the first clear instance of an exotic hadron with open-charm flavor.

Pentaquark candidates have also attracted considerable experimental interest.
However, their detection poses additional challenges.
The presence of an antiquark in the minimal Fock configuration opens the possibility of flavor cancellation: if the antiquark shares the same flavor as one of the four quarks in the system, the resulting state may mimic the quantum numbers of an ordinary three-quark baryon, making it experimentally elusive.
To mitigate this ambiguity, early experimental searches concentrated on configurations where such cancellations were not possible.

The first claimed pentaquark signal dates back to 2003, when the LEPS and DIANA Collaborations reported evidence for the $\Theta^+$ baryon~\cite{LEPS:2003wug,DIANA:2003uet}, consistent with a theoretical prediction made in 1997 for a pentaquark state near 1530~MeV~\cite{Diakonov:1997mm}.
However, subsequent searches produced conflicting results~\cite{CLAS:2005koo,Hicks:2004ge,ParticleDataGroup:2008zun,Praszalowicz:2024mji}, leading to widespread skepticism and a reassessment of early claims.

A decisive breakthrough came in 2015, when the LHCb Collaboration discovered two pentaquark candidates, $P_c(4380)^+$ and $P_c(4450)^+$, in the invariant-mass distribution of $J/\psi + p$ pairs from $\Lambda_b \to J/\psi\, p\, K^-$ decays~\cite{Aaij:2015tga}. These structures, interpreted as hidden-charm pentaquark states, marked the first robust identification of five-quark configurations with a clear experimental signature and narrow width.

This discovery was later consolidated by further observations.
In 2019, LHCb announced the detection of three additional pentaquark states, $P_c(4312)^+$, $P_c(4440)^+$, and $P_c(4457)^+$, with improved mass resolution and larger statistics~\cite{Aaij:2019vzc}.
In 2021, the same experiment reported the first evidence for a strange charmonium-like pentaquark, $P_{cs}(4459)$, seen in the $J/\psi \,+\, \Lambda$ spectrum from $[\Xi_b^- \to J/\psi \,+\, \Lambda \,+\, K^-]$ decays~\cite{LHCb:2020jpq}.
A year later, in 2022, the $P_{\psi_s}^{\Lambda}(4338)$ resonance was observed in flavor-untagged $[B^- \to J/\psi \,+\, p \,+\, \bar{p}]$ decays~\cite{LHCb:2022ogu}.
Additional experimental studies on the $P_c$ sector can be found in Refs.~\cite{LHCb:2016lve,LHCb:2021chn}.

Despite significant advances in identifying mass spectra and decay modes of exotic baryons, the underlying mechanisms responsible for their formation remain elusive.
Only a limited number of theoretical frameworks have been proposed to describe their production, most of which rely on model-dependent assumptions.
Among these are approaches based on color evaporation~\cite{Maciula:2020wri} and hadron-quark duality~\cite{Karliner:2016zzc,Becchi:2020mjz}.

Complementary insights have been provided by Lattice QCD calculations and effective field theories, which have deepened our understanding of multiquark interactions and the conditions for resonance formation~\cite{Francis:2018jyb,Leskovec:2019ioa,Liu:2019tjn,Bicudo:2022cqi,Alexandrou:2024iwi,Prelovsek:2023sta}.
More recently, studies have emphasized the critical role of heavy-quark dynamics in pentaquark systems, particularly the interactions between heavy mesons and baryons~\cite{Xiao:2019aya}.
In parallel, the interplay between long-range pion exchange and short-distance quark-gluon dynamics has been investigated as a possible binding mechanism, yielding predictions for new states with diverse spin-parity quantum numbers~\cite{Ali:2017jda,Yamaguchi:2017zmn}.

From a theoretical perspective, fully heavy tetraquarks and pentaquarks represent some of the most tractable exotic states to study. 
Owing to the large heavy-quark mass $m_Q$, these systems are well above the perturbative threshold and can be modeled as nonrelativistic bound states: $|QQ\bar{Q}\bar{Q}\rangle$ for tetraquarks and $|Q\bar{Q}QQQ\rangle$ for pentaquarks. 
Their lowest Fock states lack valence light quarks and dynamical gluons, closely resembling quarkonia, whose dominant component is $|Q\bar{Q}\rangle$.

This analogy implies that methods developed for quarkonia can be extended to the fully heavy sector.
While charmonia are often seen as QCD ``hydrogen atoms''~\cite{Pineda:2011dg}, fully charmed tetraquarks could be interpreted as QCD ``helium-2 atoms'' or ``hydrogen molecules'', and their pentaquark counterparts as “helium-3 atoms”.

Large cross sections for $X(3872)$ production at high transverse momentum observed at the LHC~\cite{CMS:2013fpt,ATLAS:2016kwu,LHCb:2021ten} offer key insight into production mechanisms, highlighting the relevance of leading-power, single-parton \emph{fragmentation} in exotic hadron formation.

To systematically address this, we recently introduced two novel families of \ac{ZM-VFNS} \ac{FFs}: {\tt TQHL1.1} for doubly heavy tetraquarks ($\XQq$) and {\tt TQ4Q1.1} for fully heavy ones ($\TQQ$)~\cite{Celiberto:2024beg,Celiberto:2024_TQHL11,Celiberto:2024_TQ4Q11}. 
These functions merge nonrelativistic initial-scale inputs~\cite{Suzuki:1977km,Nejad:2021mmp,Feng:2020riv,Bai:2024ezn} with a \ac{NLO} DGLAP-based evolution framework~\cite{Celiberto:2024mex,Celiberto:2024bxu,Celiberto:2024rxa,Celiberto:2025xvy} that properly includes heavy-quark thresholds (see Refs.~\cite{Ma:2025ryo,Nakhaei:2025zty} for recent applications).

More broadly, heavy-hadron fragmentation at leading power bridges hadronic structure and perturbative QCD. 
The presence of heavy quarks in the lowest Fock state demands a dual approach: nonperturbative modeling at the input and high-precision perturbative treatment at short distances. 
A reliable description requires the synergy of both components, thus combining structural insight with precision evolution.

In Ref.~\cite{Celiberto:2025ipt}, we investigated the leading-power fragmentation of fully charmed pentaquark states ($S$-wave $|c\bar{c}ccc\rangle$ pentacharms, or simply $\PQc$) at new-generation hadron colliders.
There, we introduced a new set of collinear FFs, denoted as {\tt PQ5Q1.0}, based on an enhanced treatment of the initial-scale input for the (anti)charm fragmentation channel.
These functions were designed to describe the short-distance emission of either compact multicharm states or dicharm-charm-dicharm configurations.

Establishing a flexible framework that accommodates multiple initial-scale inputs within leading-power fragmentation is essential to capture the full dynamics of pentacharm production.
Due to the finite mass and spatial motion of quarks, even a tightly bound state may evolve into a meson-cluster configuration.
At the quantum level, such transitions arise from fluctuations that couple the compact state to its internal meson substructures, gradually deforming it into a more extended, diquark- or molecular-like object~\cite{Sazdjian:2022kaf}.

This phenomenon represents a dynamical mechanism whose precise nature requires solving the four- or five-body bound-state problem in the presence of confining forces.  
A general and complete solution to this problem remains elusive.
Therefore, our {\tt PQ5Q1.0} \emph{multimodal} FFs can serve as a useful guidance for future analyses aimed at discriminating between the (relative weight and connections among) distinct exotic formation dynamics.
In this context, we use the term ``multimodal'' to indicate that the initial fragmentation input is modeled through multiple, physically motivated structural components, reflecting different possible internal configurations of the exotic hadron.

In the present review, we extend and complete the study initiated in Ref.~\cite{Celiberto:2025ipt}, finalizing the release of the hadron-structure-oriented {\tt PQ5Q1.0} functions by fully presenting the $\PQc$ set and delivering, for the first time, their bottom-flavored counterpart, $\PQb$.

Our phenomenological analysis relies on the $\NLLp$ hybrid-factorization scheme, which embeds the resummation of leading \ac{LL}, \ac{NLL}, and selected higher-order (NLL$^+$) high-energy logarithms into collinear factorization at NLO.\footnote{The $\NLLp$ notation, first introduced in studies of Mueller--Navelet jets~\cite{Celiberto:2022gji} and later extended to exotic hadron production~\cite{Celiberto:2024mab,Celiberto:2024beg}, updates traditional terminology to reflect both logarithmic and fixed-order accuracy, in line with modern QCD-resummation standards.}
Among various hybrid approaches developed for forward production, we adopt one where high-energy resummation is consistently integrated with fixed-order collinear elements.
This structure, already applied in prior studies~\cite{Bolognino:2021mrc,Celiberto:2022dyf,Celiberto:2025euy}, ensures a coherent treatment of rapidity-differential observables while preserving compatibility with collinear inputs.
Here, we specifically focus on the production of fully heavy pentaquarks with bottom flavor, $\PQb$, exploring their fragmentation and inclusive features within this hybrid framework.

The structure of this review is organized as follows.
Section~\ref{sec:HF_fragmentation} outlines the technical foundations of our strategy for modeling pentaquark collinear fragmentation.
In Section~\ref{sec:HE_resummation}, we introduce the $\NLLpp$ hybrid scheme, which combines collinear factorization with high-energy resummation.
Our phenomenological results and final conclusions are presented in Sections~\ref{sec:phenomenology} and~\ref{sec:conclusions}, respectively.

\section{Heavy-flavor fragmentation: From heavy-light hadrons to tetraquarks}
\label{sec:HF_fragmentation}

In this section, we outline the strategy adopted to construct the hadron-structure-oriented, \emph{multimodal} {\tt PQ5Q1.0} FF family.
These functions describe the collinear ZM-VFNS fragmentation of $S$-wave $\PQc$ and $\PQb$ pentaquarks, starting from an initial-scale input for the charm channel that can be modeled either through a direct fragmentation picture or via a scalar-diquark configuration.

For completeness, we begin in Section~\ref{ssec:FFs-intro} with a brief overview of the main features of heavy-flavor fragmentation, spanning heavy-light hadrons, quarkonia, and exotic states.
We then turn to the pentacharm case: the direct and diquark-inspired inputs are detailed in Section~\ref{ssec:FFs-direct-diquark}.
Finally, in Section~\ref{ssec:FFs-PQ5Q10}, we present and discuss the energy evolution of the {\tt PQ5Q1.0} FFs.

All symbolic manipulations required for the construction of the {\tt PQ5Q1.0} set were carried out using {\symJethad}, a dedicated \textsc{Mathematica}~\cite{Mathematica_V14-2} plugin of the data-validated {\Jethad} framework~\cite{Celiberto:2020wpk,Celiberto:2022rfj,Celiberto:2023fzz,Celiberto:2024mrq,Celiberto:2024swu,Celiberto:2025csa}, designed for the analytical treatment of hadronic-structure and precision-QCD expressions.

\subsection{Essentials of heavy-flavor fragmentation}
\label{ssec:FFs-intro}

Unlike light hadrons, the fragmentation of heavy-flavored hadrons involves a layered mechanism due to the perturbative nature of heavy-quark masses in the lowest Fock state.
This sets them apart from light-hadron FFs, which are entirely nonperturbative, as the initial-scale input for heavy-hadron FFs must account for a perturbative component.

For singly heavy systems like $D$, $B$, or $\Lambda_{c,b}$, the input structure can be understood as a sequential process~\cite{Cacciari:1996wr,Cacciari:1993mq,Jaffe:1993ie,Kniehl:2005mk,Helenius:2018uul,Helenius:2023wkn,Generet:2023vte}.
In the first step, a high-$p_T$ parton fragments into a heavy quark $Q$.
This part is governed by perturbative dynamics, as the strong coupling evaluated at $m_Q$ remains moderate.
The so-called \ac{SDC} describe the $[i \to Q]$ transition, calculated over a short time scale before hadronization sets in.
The NLO formulation of this stage was first established in~\cite{Mele:1990yq,Mele:1990cw}, with next-to-NLO refinements presented in~\cite{Rijken:1996vr,Mitov:2006wy,Blumlein:2006rr,Melnikov:2004bm,Mitov:2004du,Biello:2024zti} (see also Refs.~\cite{Fickinger:2016rfd,Maltoni:2022bpy,Czakon:2021ohs,Czakon:2022pyz,Generet:2023vte,Aglietti:2007bp,Aglietti:2022rcm,Gaggero:2022hmv,Ghira:2023bxr,Bonino:2023icn,Cacciari:2024kaa}).

In the second stage, the heavy quark undergoes hadronization into a color-neutral state.
This part is inherently nonperturbative and modeled using either phenomenological parametrizations~\cite{Kartvelishvili:1977pi,Bowler:1981sb,Peterson:1982ak,Andersson:1983jt,Collins:1984ms,Colangelo:1992kh} or EFT-inspired approaches~\cite{Georgi:1990um,Eichten:1989zv,Grinstein:1992ss,Neubert:1993mb,Jaffe:1993ie}.

To obtain a complete ZM-VFNS FF set, evolution effects must be included.
Starting from initial inputs with no scaling violations, the FFs are evolved numerically by solving the DGLAP equations at the target perturbative accuracy.

This two-step structure extends naturally to quarkonium fragmentation, where the lowest Fock state involves a $Q\bar{Q}$ pair.
Its treatment requires additional care, as both production and binding mechanisms must be accounted for.
NRQCD provides the appropriate formalism~\cite{Caswell:1985ui,Thacker:1990bm,Bodwin:1994jh,Cho:1995vh,Cho:1995ce,Leibovich:1996pa,Bodwin:2005hm}, offering a systematic expansion in $\alpha_s$ and $v_{\cal Q}$ (the relative velocity of the heavy constituents).
The $Q\bar{Q}$ pair is produced at short distances (encoded in SDCs), while the formation of the physical bound state is captured by the so-called \ac{LDMEs}.
A full state is built as a sum over Fock components with increasing complexity~\cite{Grinstein:1998xb,Kramer:2001hh,QuarkoniumWorkingGroup:2004kpm,Lansberg:2005aw,Lansberg:2019adr}.

NRQCD offers a powerful framework to study quarkonium production across a wide kinematic range in transverse momentum ($p_T$).
At low $p_T$, the prevailing mechanism is the short-distance creation of the $(Q\bar{Q})$ pair within the hard subprocess, followed by its nonperturbative evolution into a physical quarkonium state.
As $p_T$ increases, this picture changes: the fragmentation of a single parton into the final-state hadron (accompanied by inclusive radiation) becomes progressively more important and eventually takes over as the dominant production mode.

Short-distance production aligns with the \ac{FFNS} approach (see, \emph{e.g.}, Ref.~\cite{Alekhin:2009ni}), and can be interpreted as a two-parton fragmentation mechanism, encoding genuine power-suppressed contributions~\cite{Fleming:2012wy,Kang:2014tta,Echevarria:2019ynx,Boer:2023zit,Celiberto:2024mex,Celiberto:2024bxu,Celiberto:2024rxa,Celiberto:2025xvy}.
By contrast, the single-parton fragmentation process is embedded within the ZM-VFNS framework and evolves through the DGLAP formalism.

Initial \ac{LO} evaluations of the fragmentation input for gluon and constituent heavy-quark channels into $S$-wave vector quarkonia (in color-singlet states) date back to the early 1990s~\cite{Braaten:1993rw,Braaten:1993mp}, while NLO refinements have only emerged in more recent years~\cite{Zheng:2019gnb,Zheng:2021sdo,Feng:2021qjm,Zheng:2026clv}.
Building upon these, the {\tt ZCW19$^+$} ZM-VFNS fragmentation set was constructed in Refs.~\cite{Celiberto:2022dyf,Celiberto:2023fzz}, incorporating DGLAP evolution for vector quarkonia.
The formalism was then extended to $\BCs$ and $\Bss$ states via the {\tt ZCFW22} set~\cite{Celiberto:2022keu,Celiberto:2024omj}.

Phenomenological analyses using {\tt ZCFW22} revealed that the $p_T$ and rapidity spectra of charmed $B$ mesons reproduced the LHCb findings~\cite{LHCb:2014iah,LHCb:2016qpe,Celiberto:2024omj}, which reported that the production rate of $\BCs$ mesons is below 0.1\% of that for singly bottomed $B$ mesons~\cite{Celiberto:2024omj}.
This result served as an important benchmark for validating the ZM-VFNS approach at high transverse momentum.

Shifting to the exotic sector, recent works suggest that NRQCD factorization can shed light on the nature of double-$\Jpsi$ final states~\cite{LHCb:2020bwg,ATLAS:2023bft,CMS:2023owd}, which may be interpreted as manifestations of fully charmed compact tetraquarks~\cite{Zhang:2020hoh,Zhu:2020xni}.

In this framework, the production of a $\TQc$ state originates from a short-distance mechanism in which two charm and two anticharm quarks are generated at a scale set by the inverse charm mass.
Much like in the case of singly heavy hadrons and quarkonia, asymptotic freedom allows one to model heavy-tetraquark fragmentation as a two-step convolution process: a perturbatively controlled short-distance contribution followed by a long-distance, nonperturbative transition.

The first NRQCD-based calculation of the initial input for the $[g \to \TQc]$ $S$-wave fragmentation channel in the color-singlet configuration was reported in Ref.~\cite{Feng:2020riv}.
Subsequently, our work on the {\tt TQ4Q1.0} ZM-VFNS FFs combined this gluon-induced input with a $[c \to \TQc]$ component obtained by adapting a calculation based on the Suzuki model~\cite{Suzuki:1977km,Suzuki:1985up,Amiri:1986zv,Nejad:2021mmp}, previously employed to describe the fragmentation into doubly heavy $\XQq$ states~\cite{Nejad:2021mmp}.

Building on this strategy, the first ZM-VFNS FF set for heavy-light tetraquarks---the {\tt TQHL1.0} family--- was introduced in Ref.~\cite{Celiberto:2023rzw} (see also the review in Ref.~\cite{Celiberto:2024mrq}).
Later, the public release of {\tt TQ4Q1.1} and {\tt TQHL1.1} extended this program: NRQCD-based modeling of the $[Q \to \TQQ]$ input was included~\cite{Bai:2024ezn}, the fragmentation into $\XQq$ states was refined, and the analysis was generalized to include axial-vector and bottomoniumlike tetraquarks~\cite{Celiberto:2024beg,Celiberto:2025dfe,Celiberto:2025ziy,Celiberto:2025vra}.

\subsection{From direct to diquark: exploring competing channels}
\label{ssec:FFs-direct-diquark}

\begin{figure*}[!t]
\centering
\includegraphics[width=0.475\textwidth]{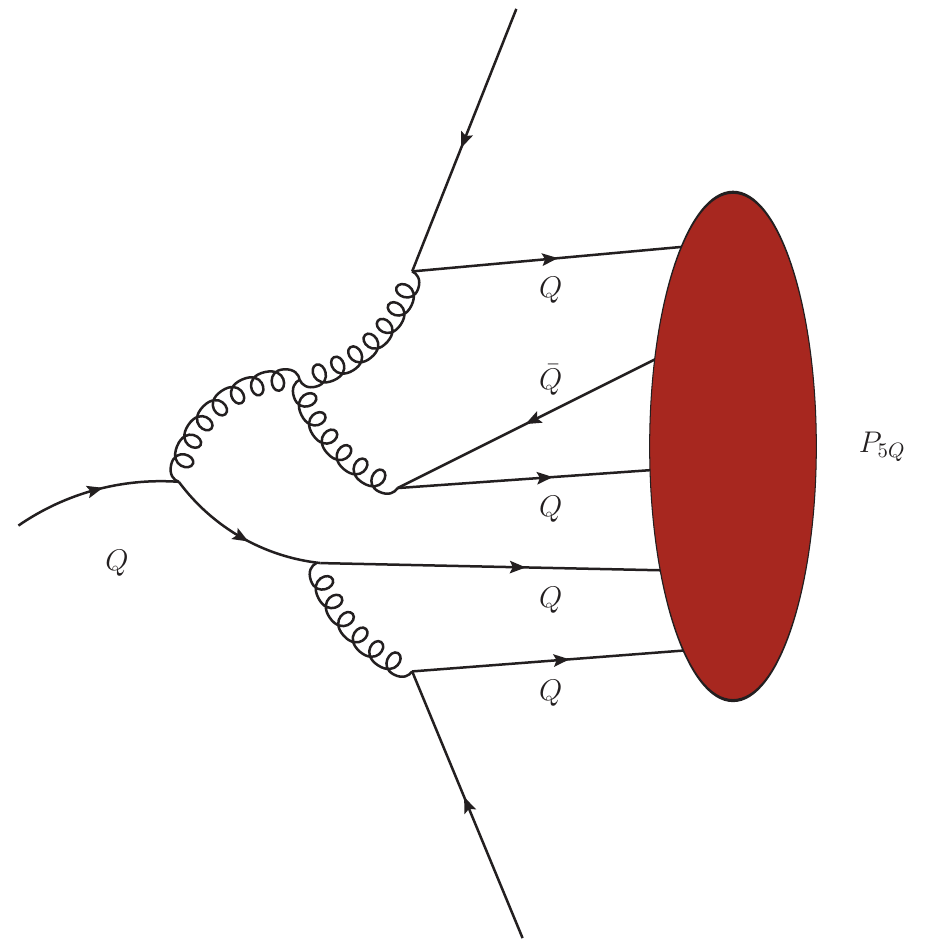}
\hspace{0.40cm}
\includegraphics[width=0.475\textwidth]{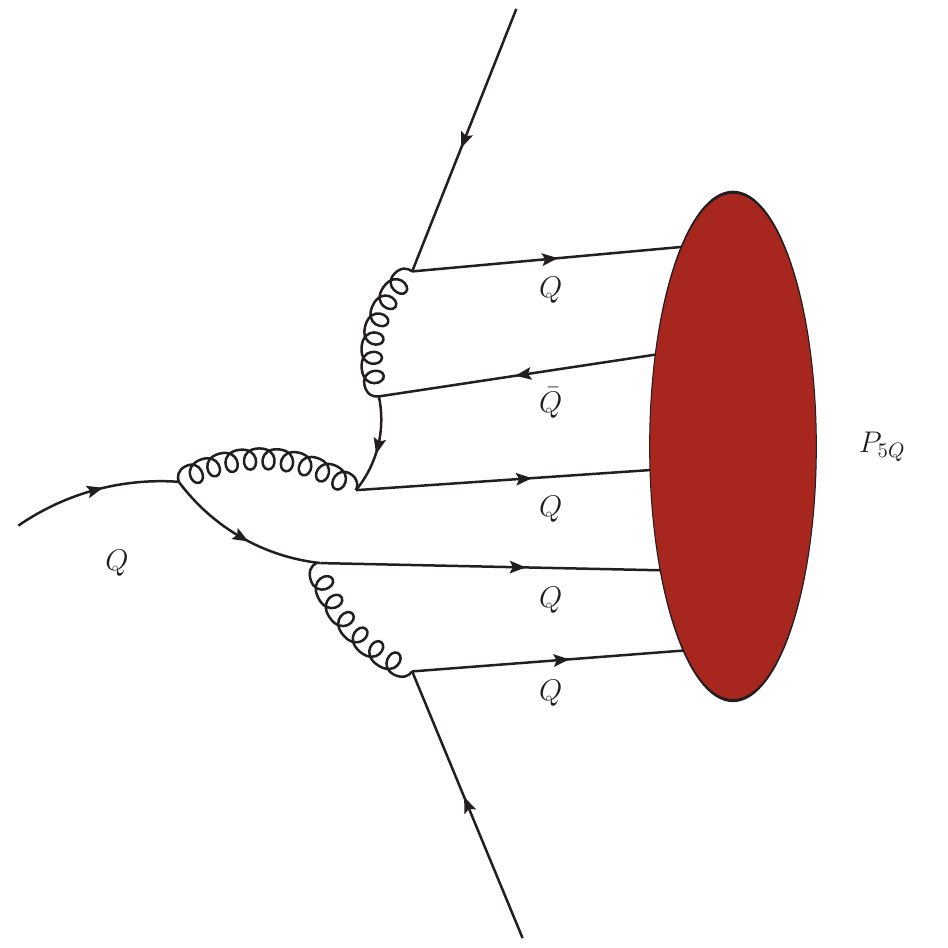}

\vspace{0.20cm}

\includegraphics[width=0.475\textwidth]{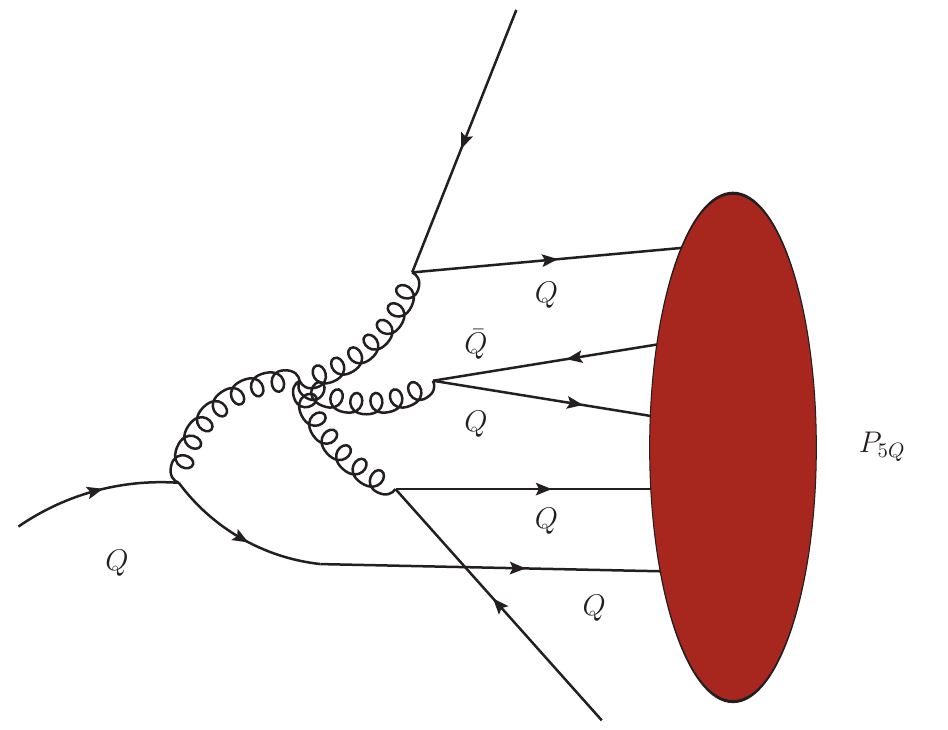}

\caption{Representative leading-order diagrams for the initial-scale collinear fragmentation of a constituent heavy antiquark into a color-singlet $S$-wave $\PQQ$ state within the \emph{direct multiquark} scenario.
The nonperturbative hadronization component of the corresponding FFs is depicted by blue ovals.
Diagrams produced using {\tt JaxoDraw 2.0}~\cite{Binosi:2008ig}.}
\label{fig:PQQ_FF_direct}
\end{figure*}

We begin by presenting the case of \emph{direct multiquark} fragmentation.
Our modeling strategy for the initial-scale input of constituent heavy-quark fragmentation into a color-singlet $S$-wave pentaquark state, within the direct scenario (see Fig.\ref{fig:PQQ_FF_direct}), builds upon the formalism developed in Ref.~\cite{Farashaeian:2024son}.
That work employed a Suzuki-inspired framework~\cite{Suzuki:1977km,Suzuki:1985up,Amiri:1986zv}, enriched with spin-structure considerations and transverse-momentum dependence.
The collinear limit was obtained by neglecting the internal motion of the constituent quarks inside the bound state~\cite{Lepage:1980fj,Brodsky:1985cr,Amiri:1986zv}.

This approach mirrors the NRQCD factorization structure, where the perturbative production of a $(Q\bar{Q})$ pair is followed by its nonperturbative transition into a quarkonium via LDMEs.
Analogously, in our framework, a multiquark system is generated through above-threshold splittings of an outgoing heavy quark $Q$, as illustrated in Fig.~\ref{fig:PQQ_FF_direct}.
The resulting amplitude is then convoluted with a bound-state wave function that encodes the nonperturbative hadronization dynamics, following the Suzuki prescription.

The representative channels depicted in Fig.~\ref{fig:PQQ_FF_direct} correspond to the $[Q \to (Q \bar{Q} Q Q Q) + \bar{Q} \bar{Q}]$ subprocess---that is, the SDC for direct $[Q \to \PQQ]$ fragmentation.
By exchanging all heavy quarks and antiquarks, one obtains the corresponding SDC for $[\bar{Q} \to \bPQQ]$ fragmentation.

In this study, we assume symmetry in the production mechanisms of $\PQQ$ and $\bPQQ$ states, and thus in their fragmentation probabilities.
We restrict our phenomenological analysis to inclusive observables that average over both particle and antiparticle emissions.
Under this assumption, FFs initiated by $Q$ and $\bar{Q}$ are treated on equal footing (for comparison with light-hadron cases, see, \emph{e.g.}, Ref.~\cite{Bertone:2018ecm}).

By working with {\symJethad}~\cite{Celiberto:2020wpk,Celiberto:2022rfj,Celiberto:2023fzz,Celiberto:2024mrq,Celiberto:2024swu,Celiberto:2025csa}, interfaced with {\FeynCalc}~\cite{Mertig:1990an,Shtabovenko:2016sxi,Shtabovenko:2020gxv}, we obtained the explicit form of the $[Q,\bar{Q} \to \PQQ]$ initial-scale {\tt PQ5Q1.0} FF in the direct scenario (not given in Ref.~\cite{Farashaeian:2024son}).
It reads
\begin{equation}
\begin{split}
 \label{PQQ_FF_initial-scale_Q_direct}
 D^{\PQQ}_{Q,\,{\rm [direct]}}(z,\mu_{F,0}) \,=\,
 {\cal N}_{P,\,{\rm [direct]}}^{(Q)} \,
 (1-z)^4 z^4
 \,{\cal R}_{P/Q}^2
 \,\frac{{\cal S}_{P,\,{\rm [direct]}}^{(Q)}(z; {\cal R}_{q_T/Q}) }{{\cal T}_{P,\,{\rm [direct]}}^{(Q)}(z; {\cal R}_{q_T/Q}, {\cal R}_{P/Q})}
 \;,
\end{split}
\end{equation}
where we have defined the dimensionless mass and transverse-momentum ratios as ${\cal R}_{P/Q} = M_{\PQQ}/m_Q$ and ${\cal R}_{q_T/Q} = \sqrt{\vqTTa}/m_Q$, with $m_Q \equiv m_c = 1.5$~GeV and $m_Q \equiv m_b = 4.5$~GeV being the charm-quark mass and the bottom-quark one, respectively.
As a reasonable assumption suitable for exploratory studies, we set the pentacharm mass to $M_{\PQQ} = 5m_Q$.
The overall factor in Eq.~\eqref{PQQ_FF_initial-scale_Q_direct} is
\begin{equation}
 \label{PQQ_FF_initial-scale_Q_N_direct}
 {\cal N}_{P,\,{\rm [direct]}}^{(Q)} \, = \,
 \left\{ 320 \sqrt{5} \pi^2 \, f_{\cal B} \, C_F \big[ \alpha_s\big(\mu_{R,0}^{\rm [direct]}\big) \big]^3 \right\}^2
 \,,
\end{equation}
where $f_{\cal B} = 0.25$~GeV is the hadron decay constant~\cite{ParticleDataGroup:2020ssz} and $C_F = (N_c^2-1)/(2N_c)$ denotes the Casimir factor for gluon emission from a quark.
Then, the numerator of Eq.~\eqref{PQQ_FF_initial-scale_Q_direct} reads 
\begin{equation}
\label{PQQ_FF_initial-scale_Q_num_direct}
\begin{split}
 {\cal S}_{P,\,{\rm [direct]}}^{(Q)}(z; {\cal R}_{q_T/Q}) 
\,&=\,
 \sum\limits_{k=0}^9 \, z^{2k} \, \gamma_{P,\,{\rm [direct]}}^{(Q)}(z; k)
 \left({\cal R}_{q_T/Q}\right)^{2k}
\;,
\end{split}
\end{equation}
with the $\gamma_{P,\,{\rm [direct]}}^{(Q)}(z; k)$ coefficients listed in Appendix~A of Ref.~\cite{Celiberto:2025ipt}.

Our treatment of the $D^{\PQQ}_{Q,\,{\rm [diquark]}}(z,\mu_{F,0})$ function enhances Ref.~\cite{Farashaeian:2024son}.
First, in line with the expectations from LO kinematics (see Fig.~\ref{fig:PQQ_FF_direct}), we do not impose a fixed normalization on the overall constant in Eqs.~\eqref{PQQ_FF_initial-scale_Q_direct} and~\eqref{PQQ_FF_initial-scale_Q_N_direct}.
Instead, we compute it explicitly at the initial scale $\mu_{R,0}^{\rm [diquark]} = 7 m_Q$.

Then, the original formulation by Suzuki accounts for spin correlations and serves as an effective proxy for \ac{TMD} FFs.
To extract the collinear limit, rather than integrating over the transverse-momentum squared of the outgoing heavy quark, one replaces it with an average value, denoted as $\vqTTa$.
This average then acts as a free parameter, to be fixed through phenomenological arguments.
As noted in Ref.~\cite{GomshiNobary:1994eq}, increasing $\vqTTa$ shifts the peak of the resulting FF toward lower $z$ and simultaneously suppresses its normalization.

The study of Ref.~\cite{Farashaeian:2024son} adopted the choice $\vqTTa = 1 \mbox{ GeV}^2$ as a maximal estimate, corresponding to an upper bound on the squared transverse momentum.
In the present work, we refine this prescription by introducing a more balanced choice of $\vqTTa$, aligned with the exploratory character of our analysis.

This refinement follows the strategy previously introduced in the study of $\TQQ$ fragmentation~\cite{Celiberto:2024mab}, where a phenomenological calibration was proposed.
In that context, data-driven considerations on fragmentation in hadronic collisions suggested that heavy-quark FFs, both for light hadrons~\cite{Celiberto:2016hae,Celiberto:2017ptm,Bolognino:2018oth,Celiberto:2020wpk} and heavy ones~\cite{Celiberto:2021dzy,Celiberto:2021fdp,Celiberto:2022dyf,Celiberto:2022keu,Celiberto:2025euy}, are typically probed at longitudinal momentum fractions above $\langle z \rangle > 0.4$.

Moreover, it was assumed that the constituent-quark FFs are comparable in size to their gluon counterparts.
This assumption is supported by the case of scalar $S$-wave quarkonium, such as $\eta_c$, whose gluon and charm FFs at LO are similar in size for $z > 0.4$~\cite{Braaten:1993rw,Braaten:1993mp}.

For $\TQQ$ fragmentation, numerical scans in Ref.~\cite{Celiberto:2024mab} showed that setting $\vqTTa_{\TQQ} = 70 \mbox{ GeV}^2$ ensures both $\langle z \rangle \gtrsim 0.4$ and consistency in size between the quark and gluon channels.

At present, no dedicated calculations are available for gluon-initiated FFs into pentacharm states.
As a result, the value of $\vqTTa_{\PQQ}$ must be chosen based solely on the features of the constituent-quark channel.
Following the same procedure used for $\vqTTa_{\TQQ}$, we performed a scan over $\vqTTa_{\PQQ}$ and selected the value $\vqTTa_{\PQQ} = 90 \mbox{ GeV}^2$ as a phenomenologically motivated benchmark.

\begin{figure*}[!t]
\centering
\includegraphics[width=0.475\textwidth]{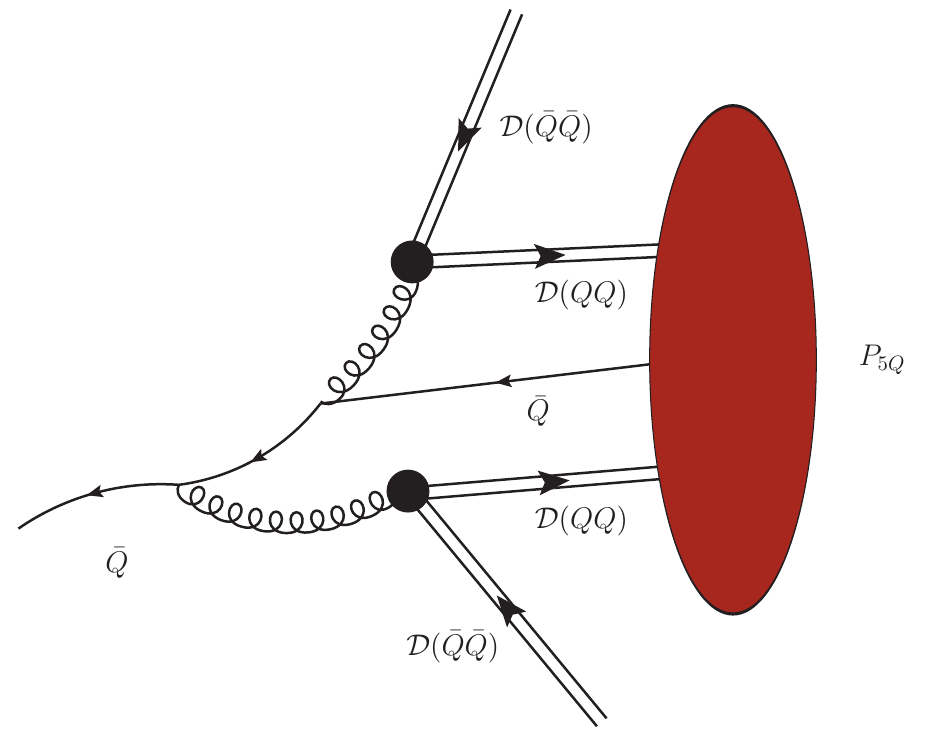}
\hspace{0.40cm}
\includegraphics[width=0.475\textwidth]{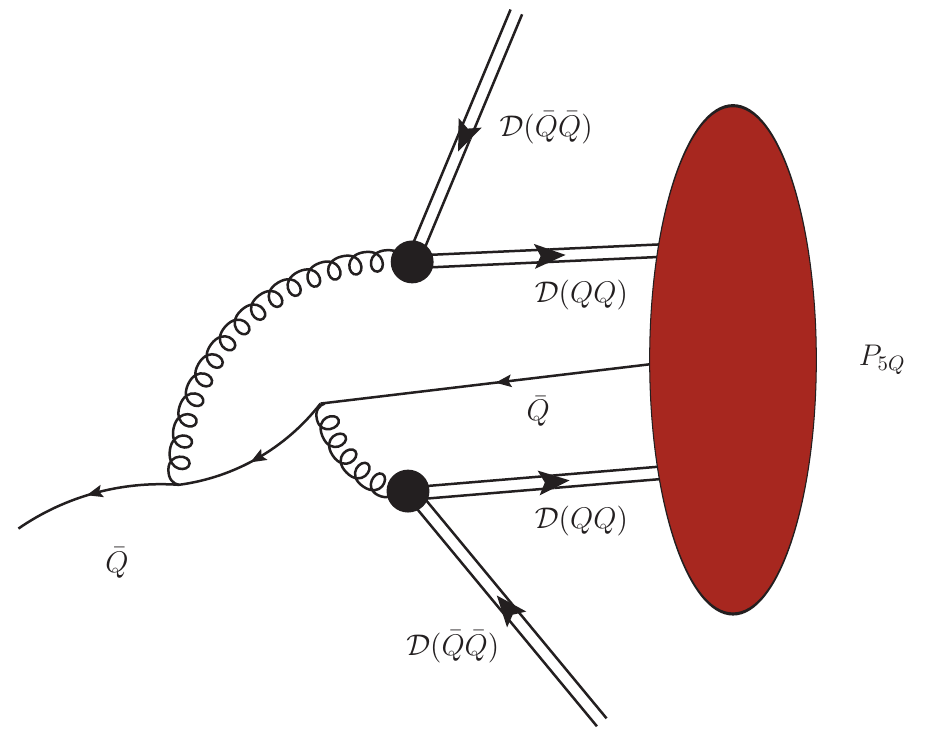}

\vspace{0.20cm}

\includegraphics[width=0.475\textwidth]{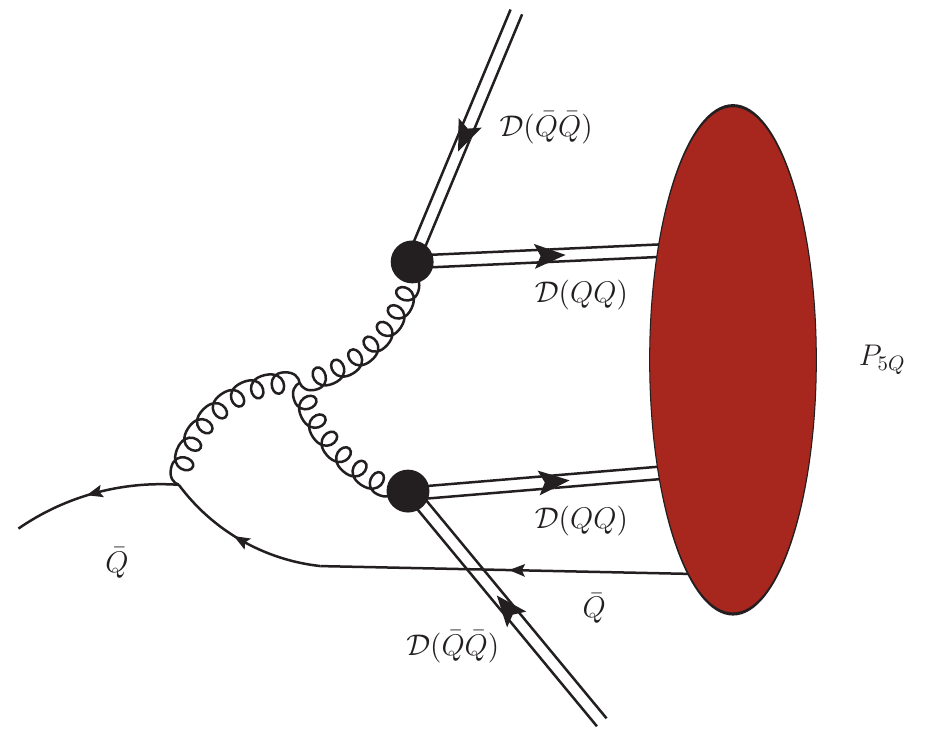}

\caption{Representative leading-order diagrams for the initial-scale collinear fragmentation of a constituent heavy antiquark into a color-singlet $S$-wave $\PQQ$ pentaquark in the scalar-diquark scenario.
Double lines denote ${\cal D}(QQ)$ or ${\cal D}(\bar{Q}\bar{Q})$ heavy-diquark configurations, while black dots represent effective gluon-diquark-antidiquark interaction vertices.
The nonperturbative hadronization stage of the associated FFs is indicated by blue ovals.
Diagrams produced using {\tt JaxoDraw 2.0}~\cite{Binosi:2008ig}.}
\label{fig:PQQ_FF_diquark}
\end{figure*}

We now turn our attention to the \emph{diquark-based} scenario.
The interacting quark-diquark picture assumes that two quarks can form a colored, quasi-bound state---the diquark---as first proposed in Ref.~\cite{Gell-Mann:1964ewy}.
This framework has seen wide application in hadron spectroscopy and has been extended to describe the production and decay of baryons, especially those containing heavy flavors~\cite{Maiani:2004vq,Jaffe:2003sg,Guo:2013xga,DeSanctis:2016zph}.
Diquarks are typically classified as scalar (spin-0) or axial-vector (spin-1) states.
Their internal structure is incorporated via nonperturbative form factors, which encode their compositeness.
Scalar diquarks require a single form factor, while axial-vector configurations typically involve multiple ones.

Early applications of the diquark model to describe fragmentation into octet baryons and singly heavy baryons were developed in Refs.~\cite{Nzar:1995wb,Ma:2001ri,Yang:2002gh} and~\cite{Falk:1993gb,Adamov:1997yk,MoosaviNejad:2017rvi,Delpasand:2019xpk}, respectively.
Studies of the mass spectra of doubly and fully heavy tetraquarks using a relativistic quasi-potential diquark-antidiquark framework were presented in Refs.~\cite{Faustov:2020qfm,Faustov:2021hjs,Faustov:2022mvs}.
A detailed discussion of pentaquarks within a diquark-based approach can be found in Ref.~\cite{Maiani:2015vwa}.

In this work, we adopt the methodology developed in Ref.~\cite{Farashaeian:2024cpd} to model the initial-scale input for constituent heavy-quark fragmentation into $S$-wave, color-singlet pentaquarks within the \emph{diquark-based} picture (see Fig.~\ref{fig:PQQ_FF_diquark}).
Following the same Suzuki-inspired framework used for the direct-channel analysis~\cite{Suzuki:1977km,Suzuki:1985up,Amiri:1986zv}, we now assume that the dominant Fock-state configuration of the pentaquark is $|{\cal D}(QQ),\bar{Q},{\cal D}(QQ)\rangle$, where ${\cal D}(QQ)$ denotes a colored heavy diquark composed of two heavy quarks.
At LO, the splitting $[\bar{Q} \to |{\cal D}(QQ),\bar{Q},{\cal D}(QQ)\rangle + {\cal D}(\bar{Q}\bar{Q}),{\cal D}(\bar{Q}\bar{Q})]$ is described by three main classes of diagrams, as illustrated in Fig.~\ref{fig:PQQ_FF_diquark}.

We note that the representative channels depicted in Fig.~\ref{fig:PQQ_FF_diquark} correspond to the diquark-based $[\bar{Q} + \PQQ]$ SDC.
By reversing all heavy-quark flavors---interchanging [$Q \leftrightarrow \bar{Q}$] throughout---one obtains the complementary $[Q \to |{\cal D}(\bar{Q}\bar{Q}),Q,{\cal D}(\bar{Q}\bar{Q})\rangle + {\cal D}(QQ),{\cal D}(QQ)]$ transition, describing the $[Q \to \bPQQ]$ SDC within the diquark picture.
As in the direct scenario discussed above, we adopt a symmetric treatment of particle and antiparticle formation mechanisms.
Accordingly, we assume equal production rates for pentaquarks and antipentaquarks, and thus work with fully symmetric $Q$ and $\bar{Q}$ fragmentation channels.

Describing the lowest Fock component of the pentaquark as a diquark-quark-diquark system leads to significantly more compact analytical expressions compared to the direct multiquark setup.
This simplification is particularly beneficial in symbolic computations, which were carried out using {\symJethad}~\cite{Celiberto:2020wpk,Celiberto:2022rfj,Celiberto:2023fzz,Celiberto:2024mrq,Celiberto:2024swu,Celiberto:2025csa} in combination with {\FeynCalc}~\cite{Mertig:1990an,Shtabovenko:2016sxi,Shtabovenko:2020gxv}.
Following this strategy, we reproduced the explicit analytical form of the $[Q,\bar{Q} \to \PQQ]$ initial-scale FF in the diquark-based {\tt PQ5Q1.0} family.
We get
\begin{equation}
\begin{split}
 \label{PQQ_FF_initial-scale_Q_diquark}
 D^{\PQQ}_{Q,\,{\rm [diquark]}}(z,\mu_{F,0}) \,=\,
 {\cal N}_{P,\,{\rm [diquark]}}^{(Q)} \,
 \left[\frac{z^2(1 - z)}{z + 2}\right]^2
 \,
 \frac{{\cal S}_{P,\,{\rm [diquark]}}^{(Q)}(z; {\cal R}_{q_T/Q}) }{{\cal T}_{P,\,{\rm [diquark]}}^{(Q)}(z; {\cal R}_{q_T/Q})}
 \;.
\end{split}
\end{equation}
with ${\cal R}_{q_T/Q} = \sqrt{\vqTTa}/m_Q$\,. 
The overall factor in Eq.~\eqref{PQQ_FF_initial-scale_Q_diquark} is
\begin{equation}
 \label{PQQ_FF_initial-scale_Q_N_diquark}
 {\cal N}_{P,\,{\rm [diquark]}}^{(Q)} \, = \,
 \left[\frac{{\cal V}_{P,\,\rm [diquark]}^{(g{\cal D}\bar{{\cal D}})}}{m_c}\right]^4
 \!
 \left\{ \frac{625 \pi^2}{6 \sqrt{2}} \, f_{\cal B} \, C_F \big[ \alpha_s\big(\mu_{R,0}^{\rm [diquark]}\big) \big]^2 \right\}^2
 \,,
\end{equation}
with ${\cal V}_{P,\,\rm [diquark]}^{(g{\cal D}\bar{{\cal D}})}$ being a form factor entering the description of the gluon-diquark-antidiquark effective vertex (the black bullet in Fig.~\ref{fig:PQQ_FF_diquark}). 
Then, the numerator and the denominator respectively read 
\begin{equation}
\label{PQQ_FF_initial-scale_Q_num_diquark}
\begin{split}
 {\cal S}_{P,\,{\rm [diquark]}}^{(Q)}(z; {\cal R}_{q_T/Q}) 
\,&=\,
 \left\{ 64z^4 + 356z^3 - 99z^2 - 2128z + 1540 + \frac{672}{z} \right. \\[0.20cm]
\,&+\, 
 225 \left[ \frac{128z(1 - z)^6}{[{\cal R}_{q_T/Q}^2 \, z^2 + (5 - z)^2]^2} - \frac{45(z + 2)^2(8z^2 + 60z - 33 - 60/z)}{[{\cal R}_{q_T/Q}^2 \, z^2 + z^2 - 10z + 100]^2} \right] \\[0.20cm]
\,&+\,
 30 \left[ \frac{16(1 - z)^4 (z^2 - 20z - 35)}{{\cal R}_{q_T/Q}^2 \, z^2 + (5 - z)^2} \right] \\[0.20cm]
\,&+\,
 \left. \frac{(z + 2)(100z^4 + 256z^3 - 588z^2 + 1859z - 922 - 1380/z)}{{\cal R}_{q_T/Q}^2 \, z^2 + z^2 - 10z + 100} \right\}
\end{split}
\end{equation}
and
\begin{equation}
\label{PQQ_FF_initial-scale_Q_den_diquark}
\begin{split}
 {\cal T}_{P,\,{\rm [diquark]}}^{(Q)}(z; {\cal R}_{q_T/Q}) 
  \,&=\, 
 [{\cal R}_{q_T/Q}^2 \, z^2 + (5-z)^2]^2
  \;.
\end{split}
\end{equation}

As discussed earlier, our analysis adopts the calculation in Ref.~\cite{Farashaeian:2024cpd} as a benchmark reference for modeling the initial-scale quark-to-pentaquark FF in the diquark scenario.
That function was derived under the simplifying assumption that only scalar diquark states contribute, with pseudovector configurations being neglected.

Our treatment of the $D^{\PQQ}_{Q,\,{\rm [diquark]}}(z,\mu_{F,0})$ function refines and extends the strategy of Ref.~\cite{Farashaeian:2024cpd} in several key aspects.
First, as in the direct-channel analysis of the previous section, we do not impose a fixed normalization on the overall constant in Eqs.~\eqref{PQQ_FF_initial-scale_Q_diquark} and~\eqref{PQQ_FF_initial-scale_Q_N_diquark}.
Instead, we compute it explicitly at the initial scale $\mu_{R,0}^{\rm [diquark]} = 9 m_Q$, in accordance with LO kinematics (see Fig.~\ref{fig:PQQ_FF_diquark}).

Second, rather than treating $\vqTTa$ as a free parameter, we fix its value by applying the same phenomenologically motivated procedure used in the direct case.
Third, to isolate the dynamical features of the genuine scalar-diquark setup with respect to the multiquark picture, we set the gluon-diquark-antidiquark form factor to ${\cal V}_{P,\,\rm [diquark]}^{(g{\cal D}\bar{{\cal D}})} = 1 \mbox{ GeV}$, a notable reduction from the ${\cal O}(5 \mbox{ GeV})$ value adopted in Ref.~\cite{Farashaeian:2024cpd}.

\begin{figure*}[!t]
\centering

\includegraphics[scale=0.625,clip]{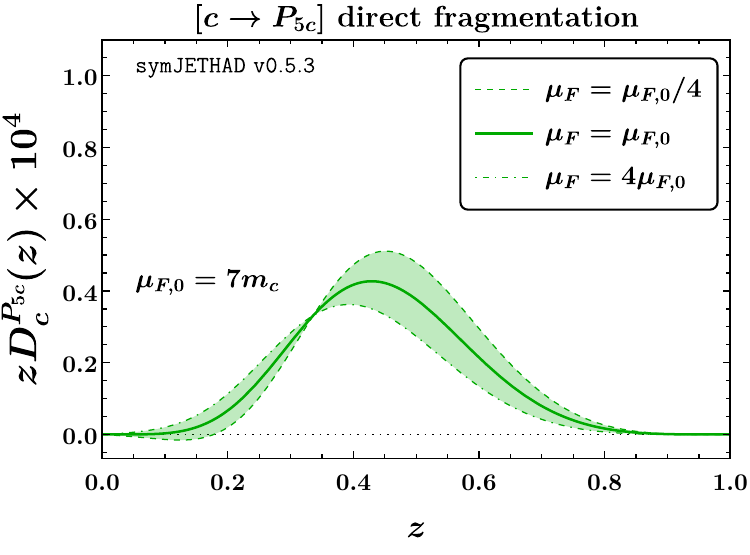}
\hspace{0.25cm}
\includegraphics[scale=0.625,clip]{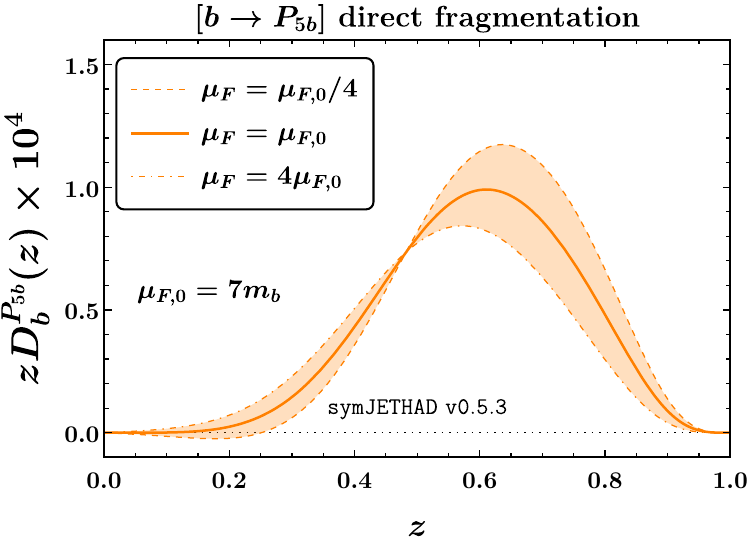}

\vspace{0.45cm}

\includegraphics[scale=0.625,clip]{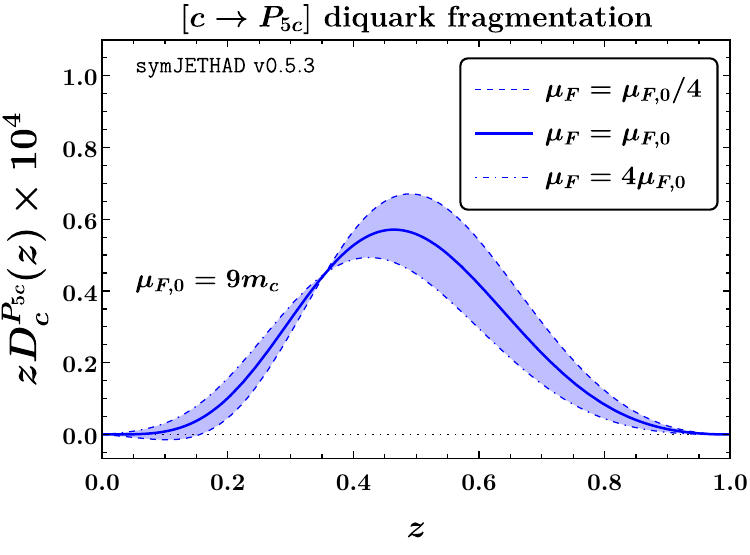}
\hspace{0.25cm}
\includegraphics[scale=0.625,clip]{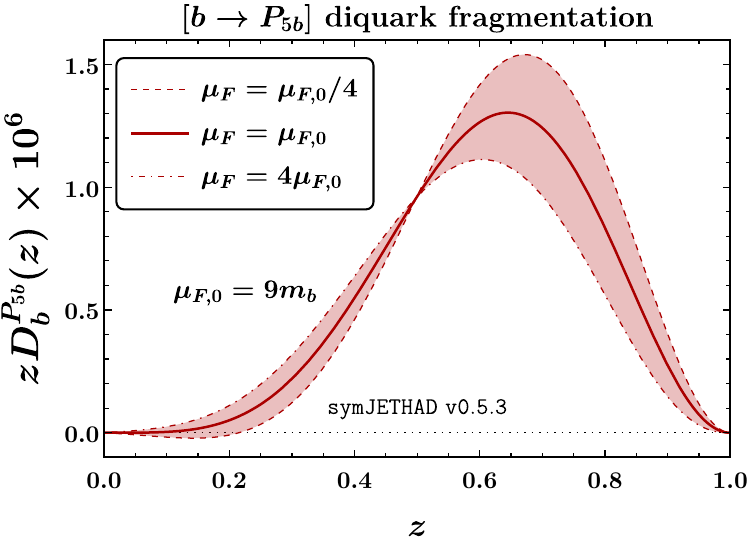}

\caption{Constituent heavy-quark to $\PQc$ (left) and $\PQb$ (right) initial-scale fragmentation channels in the direct (upper) and scalar-diquark (lower) picture. 
For the sake of illustration, an expanded diagonal DGLAP evolution is performed in the range $\mu_{F,0}/4$ to $4\mu_{F,0}$.}
\label{fig:PQQ_FF_initial-scale_Q}
\end{figure*}

Figure~\ref{fig:PQQ_FF_initial-scale_Q} shows a comparative overview of initial-scale constituent-quark FFs for scalar pentaquarks $\PQc$ (left panels) and $\PQb$ (right panels), modeled within the direct (upper) and scalar-diquark (lower) frameworks. 
All distributions are multiplied by $z$, and shown after diagonal DGLAP evolution from the reference scale $\mu_{F,0}$, varied in the range $\mu_{F,0}/4$ to $4\mu_{F,0}$ to estimate the impact of factorization-scale choices.

In the direct scenario (upper panels), the initial scale is set to $\mu_{F,0}^{\rm [direct]} = 7 m_Q$, where $Q = c$ (left) or $b$ (right), in accordance with the treatment discussed in Section~\ref{ssec:FFs-direct-diquark}. 
In both cases, the FFs exhibit a pronounced peak in the moderate-$z$ region, and rapidly vanish toward the endpoints $[z \to 0]$ and $[z \to 1]$. 

The peaks of the bottom-quark FFs are noticeably shifted toward higher $z$ values compared to their charm counterparts.
This behavior is consistent with theoretical expectations: heavier quarks tend to fragment with a harder spectrum, transferring a larger fraction of their initial momentum to the produced hadron~\cite{Mele:1990cw,Cacciari:2005uk,Kniehl:2008zza}.
Such trend is a direct consequence of the reduced phase space for collinear and soft radiation in the fragmentation of heavier partons.

In the diquark scenario (lower panels), the initial scale is increased to $\mu_{F,0}^{\rm [diquark]} = 9 m_Q$, reflecting the higher invariant mass of the pentaquark Fock state involving two heavy diquarks. 
While the shape of the FFs remains similar to the direct case, the overall magnitude is enhanced---particularly in the charm sector---with a peak value $\sim 10 \div 15\%$ higher than the direct counterpart. 
This result suggests that, under otherwise equal assumptions, the diquark production mechanism may yield slightly higher cross sections at moderate to high energies.

We note that bottom FFs (right panels) are significantly more peaked and broader than their charm analogs, consistent with the expected hierarchy in heavy-quark fragmentation. 
These findings will be quantitatively explored in the phenomenological study presented in Section~\ref{sec:phenomenology}.

\subsection{The {\tt PQ5Q1.0} determinations from {\HFNRevo}}
\label{ssec:FFs-PQ5Q10}

\begin{figure*}[!t]
\centering

   \includegraphics[scale=0.400,clip]{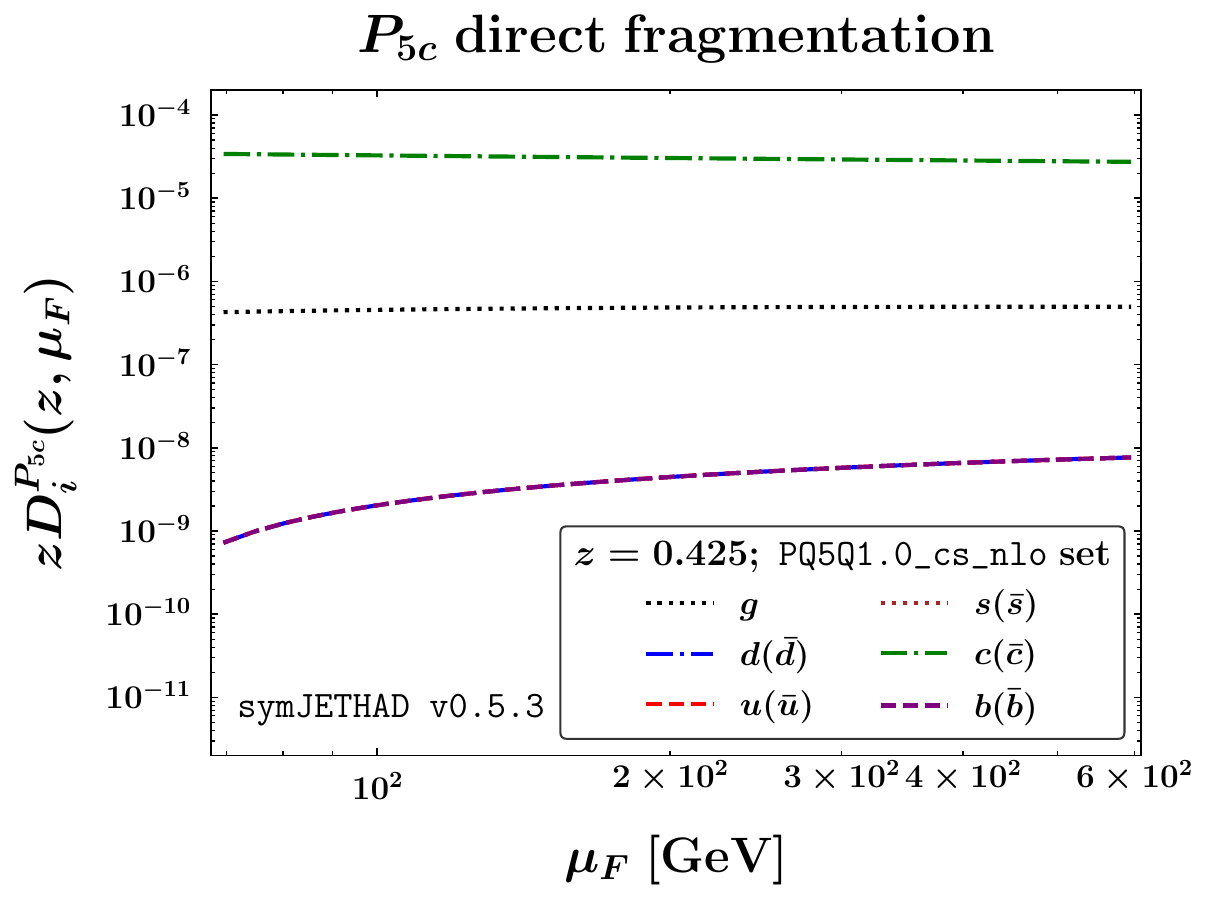}
   \includegraphics[scale=0.400,clip]{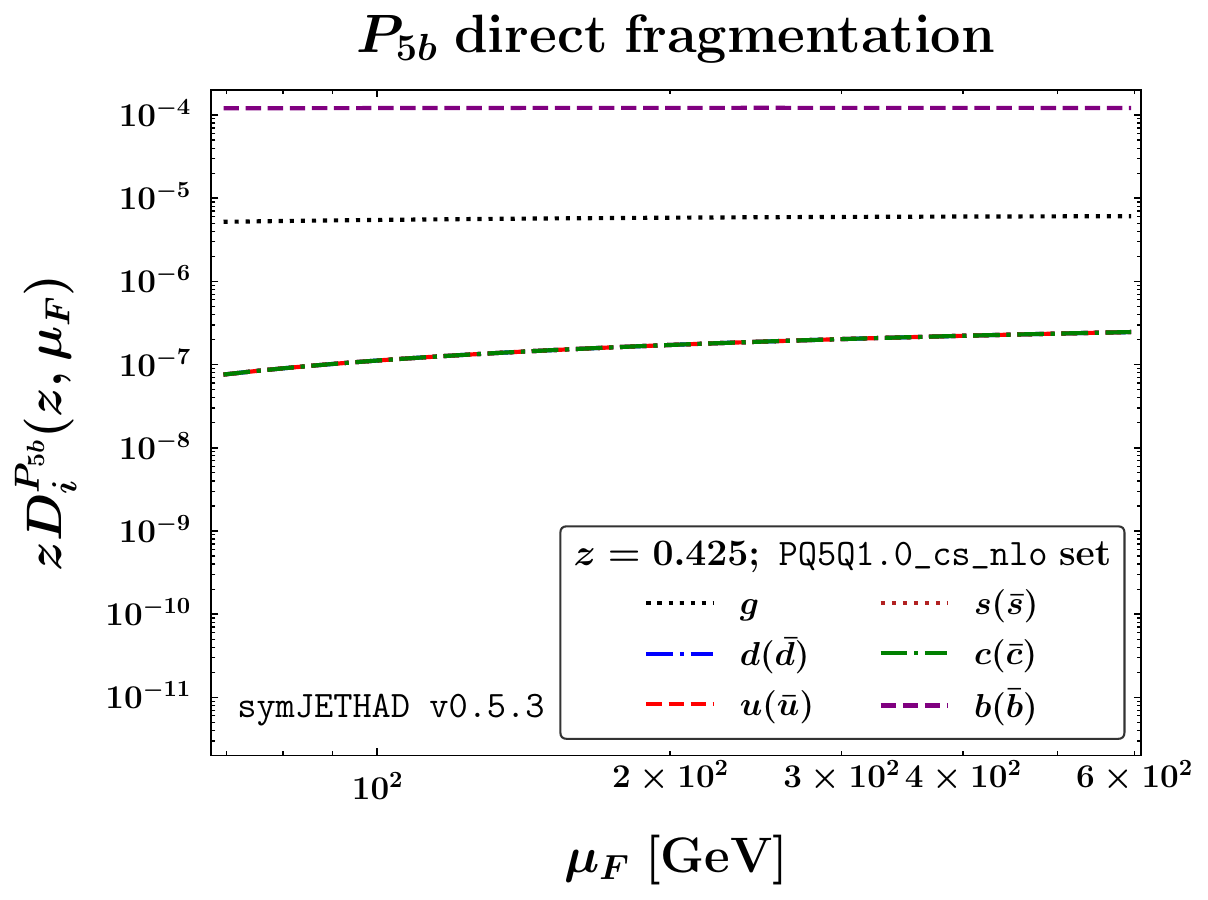}

   \vspace{0.45cm} 

   \includegraphics[scale=0.400,clip]{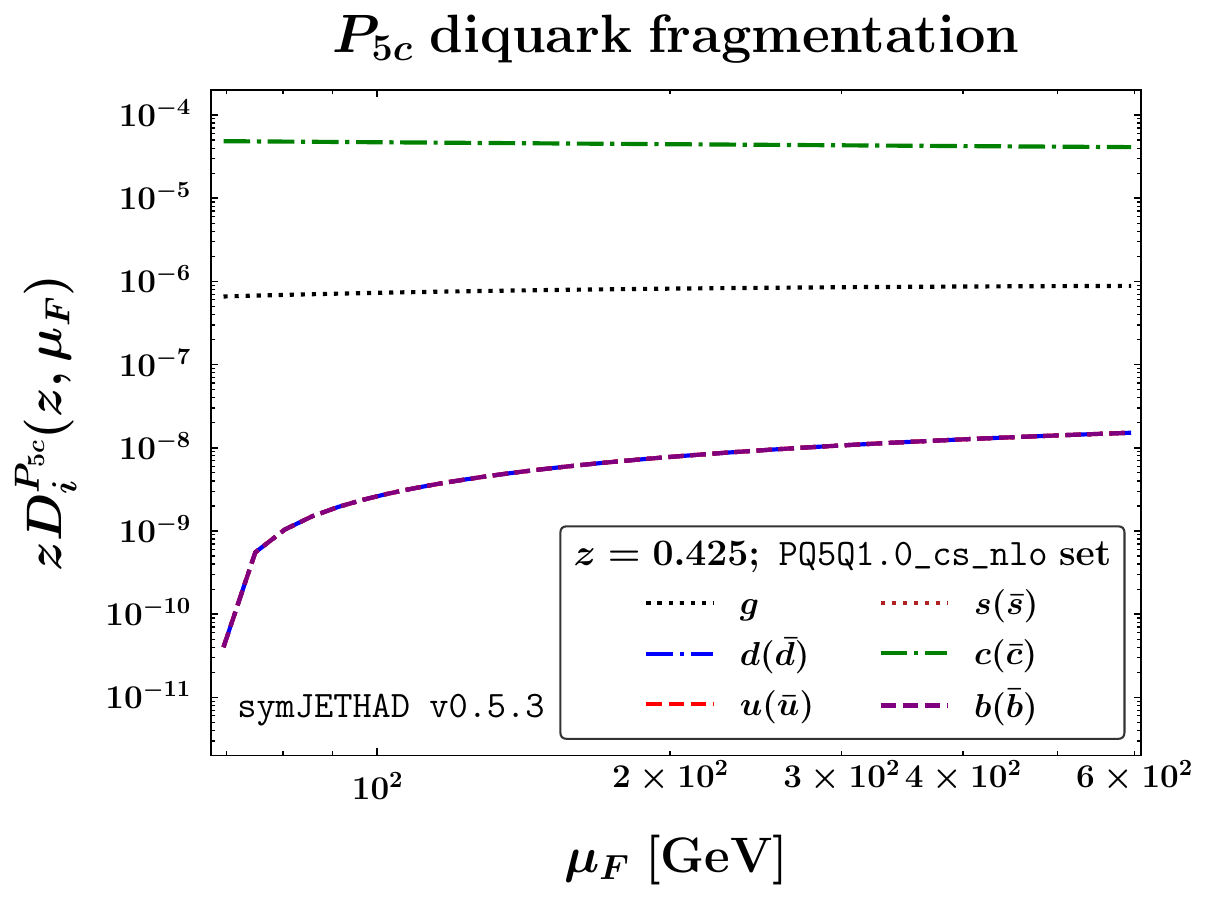}
   \includegraphics[scale=0.400,clip]{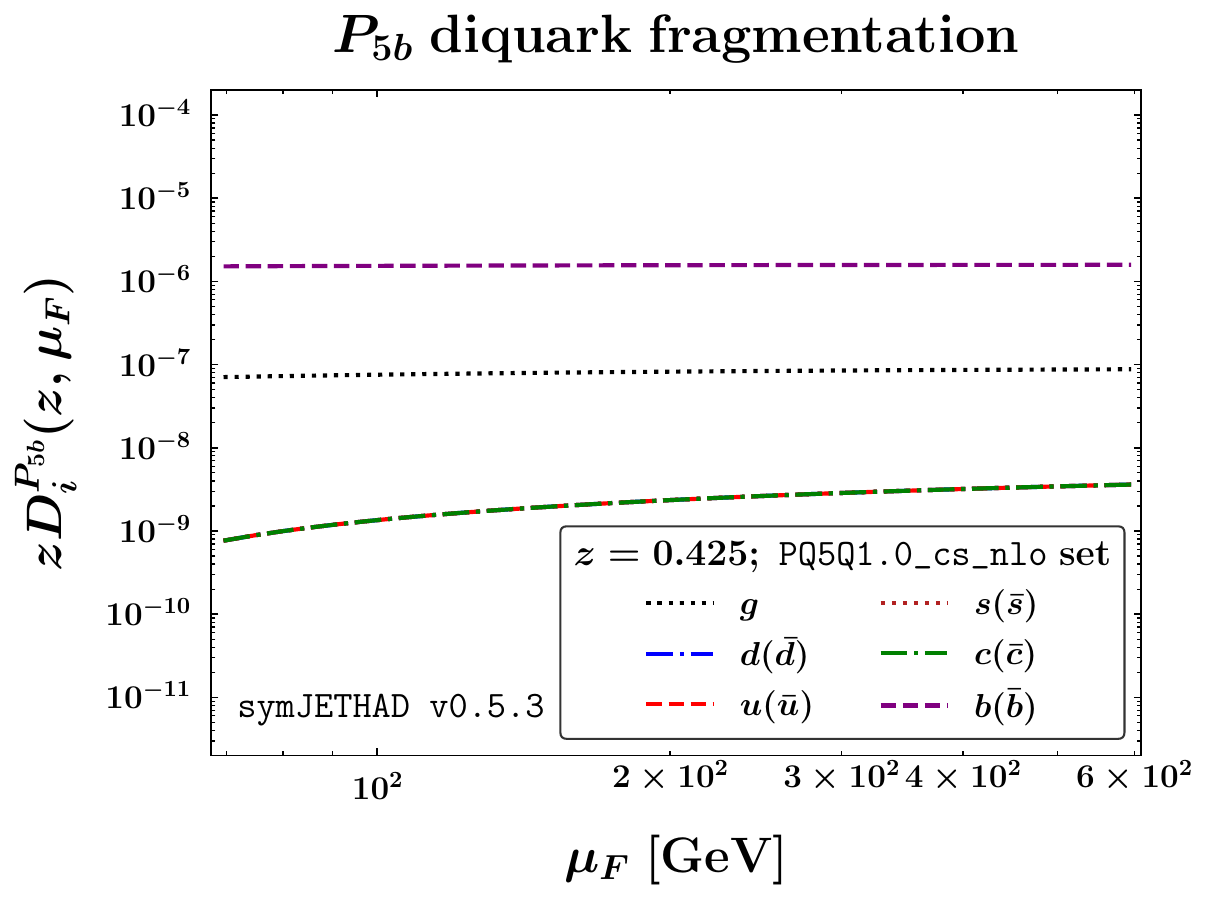}

\caption{Factorization-scale dependence of the {\tt PQ5Q1.0} NLO FFs portraying the ZM-VFNS fragmentation of $\PQc$ (left) and $\PQb$ (right) pentaquarks within direct (upper) or scalar-diquark (bottom) initial-scale inputs.
The hadron momentum fraction is set to $z = 0.425 \simeq \langle z \rangle$.}
\label{fig:NLO_FFs_P5Q}
\end{figure*}

\begin{figure*}[!t]
\centering

   \includegraphics[scale=0.390,clip]{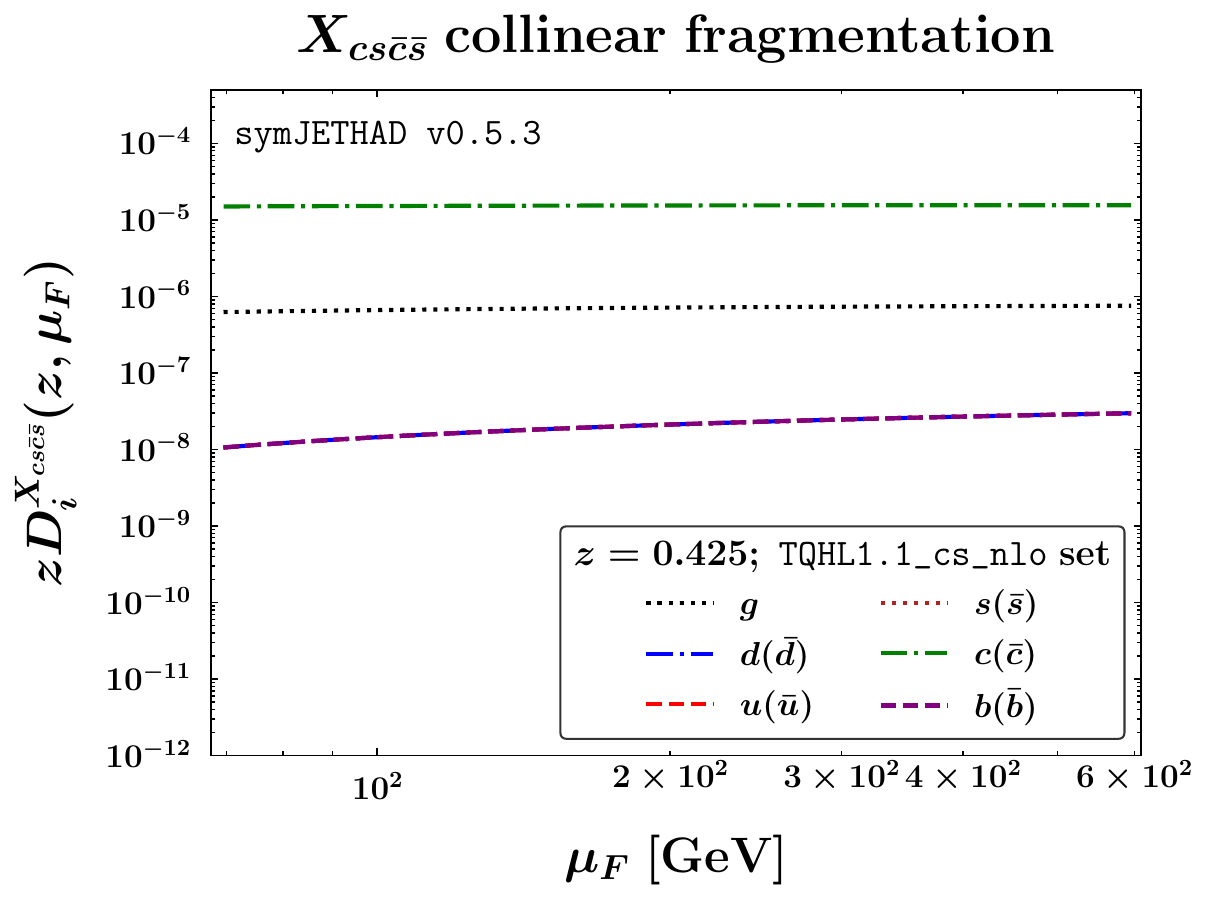}
   \includegraphics[scale=0.390,clip]{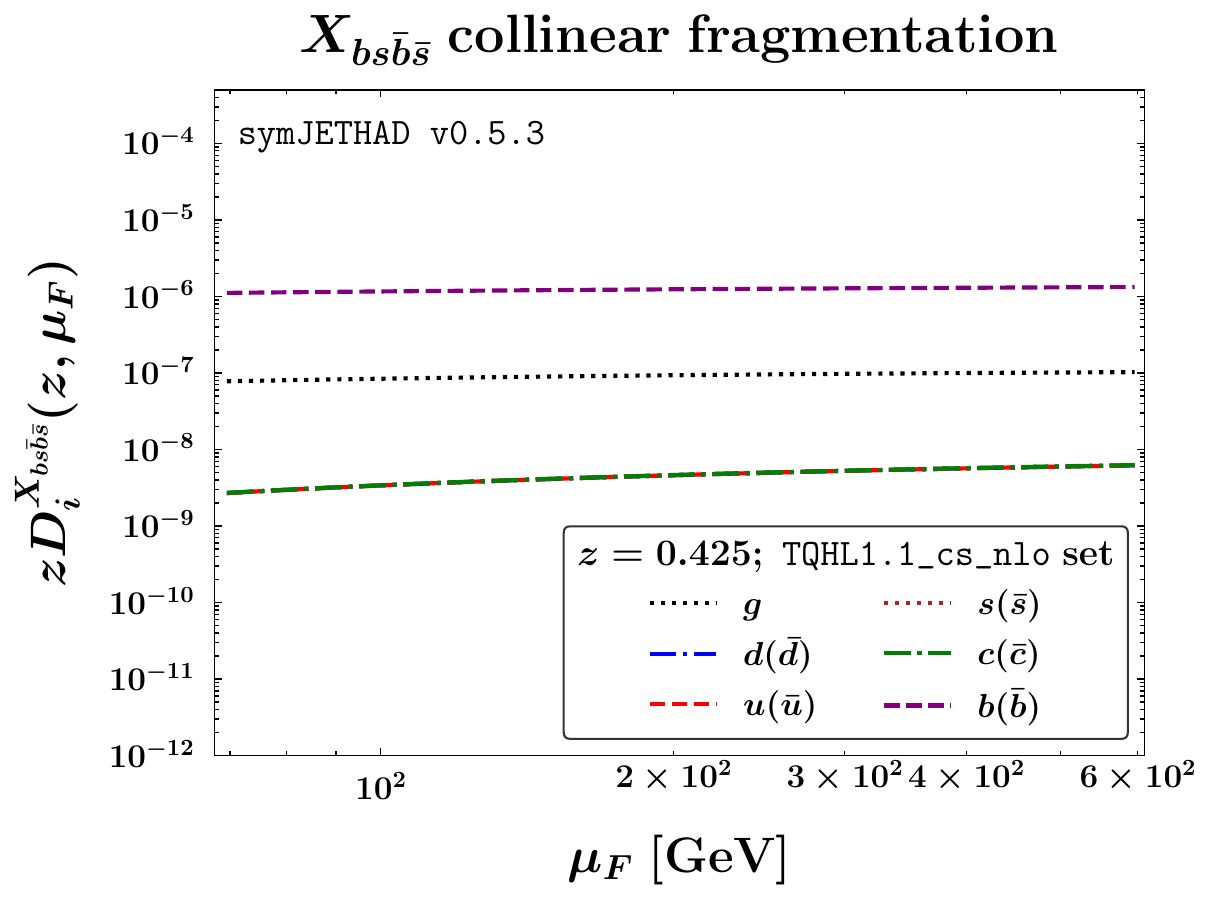}

   \vspace{0.45cm} 

   \includegraphics[scale=0.390,clip]{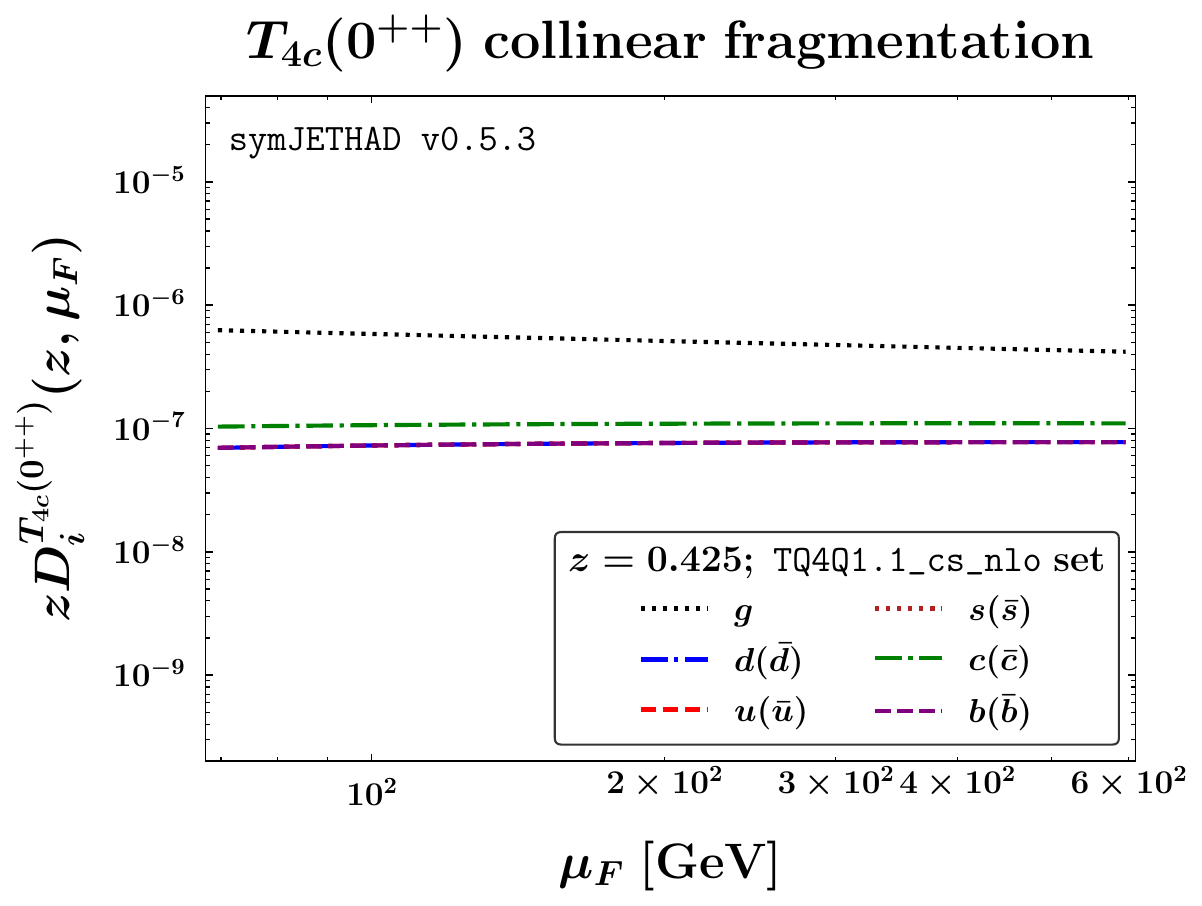}
   \includegraphics[scale=0.390,clip]{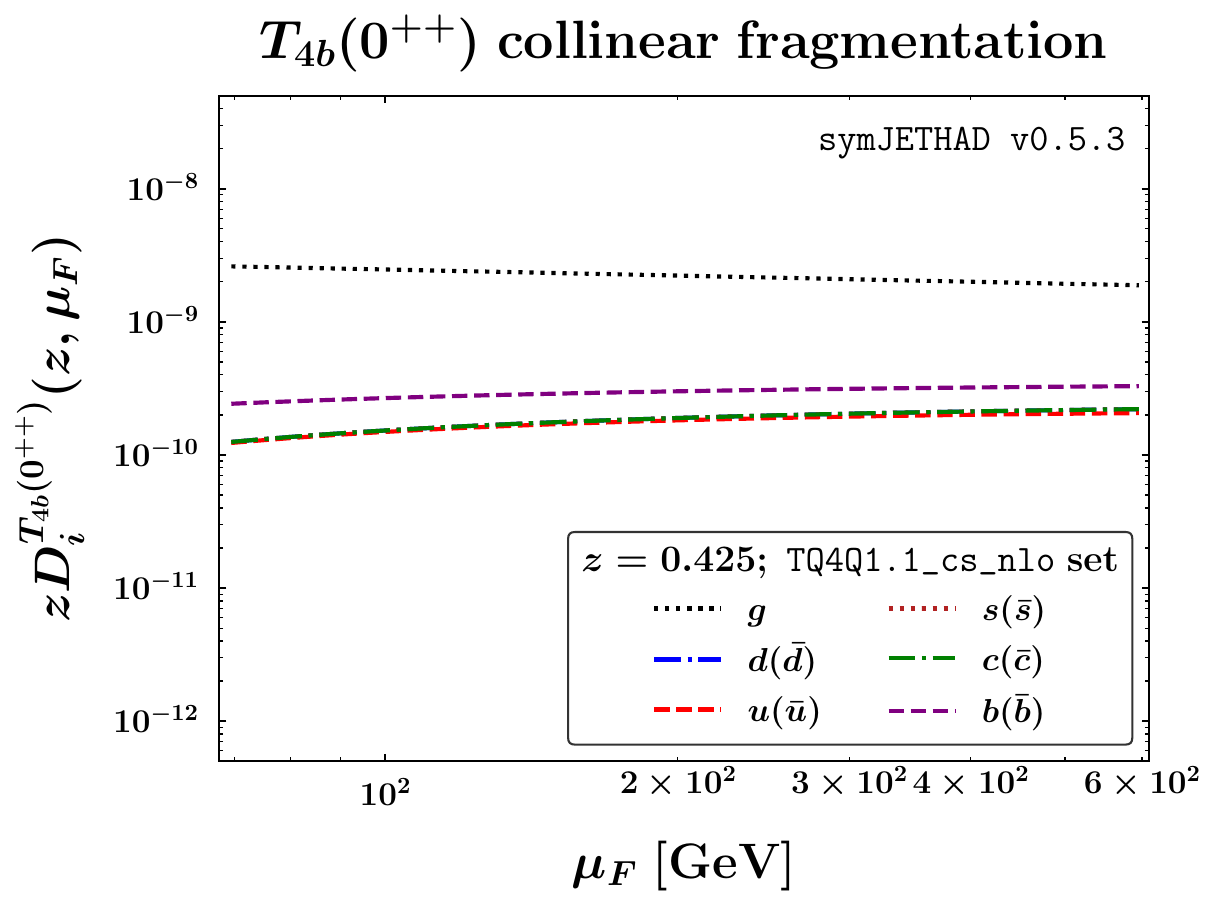}

\caption{Upper plots: factorization-scale dependence of the {\tt TQHL1.1} NLO FFs~\cite{Celiberto:2024beg} portraying the ZM-VFNS fragmentation of $\Xcs$ (left) and $\Xbs$ (right) tetraquarks.
Lower plots: factorization-scale dependence of the {\tt TQ4Q1.1} NLO FFs~\cite{Celiberto:2025dfe,Celiberto:2025ziy} portraying the ZM-VFNS fragmentation of $\TQcZpp$ (left) and $\TQbZpp$ (right) tetraquarks.
The hadron momentum fraction is set to $z = 0.425 \simeq \langle z \rangle$.}
\label{fig:NLO_FFs_XQs_T4Q}
\end{figure*}

The final stage in constructing the {\tt PQ5Q1.0} functions for $\PQQ$ states involves performing a consistent DGLAP evolution of the $[Q,\bar{Q} \to \PQQ]$ initial-scale inputs discussed above.
Based on leading-order kinematics, the minimum invariant mass for the direct-channel splitting $[Q \to (Q \bar{Q} Q Q Q) + \bar{Q} \bar{Q}]$ is $\mu_{F,0}^{\rm [direct]} = 7 m_Q$ (see Fig.~\ref{fig:PQQ_FF_direct}), while the corresponding threshold for the diquark-based splitting $[\bar{Q} \to (Q \bar{Q} Q Q Q) + \bar{Q} \bar{Q} \bar{Q} \bar{Q}]$ is $\mu_{F,0}^{\rm [diquark]} = 9 m_Q$ (see Fig.~\ref{fig:PQQ_FF_diquark}).
These values are adopted as the starting scales for the evolution of the constituent heavy-quark FFs in the two respective scenarios.

As previously emphasized, achieving a fully consistent treatment of heavy-flavor fragmentation requires merging structure-informed input functions with the collinear factorization formalism, thereby establishing well-defined DGLAP evolution thresholds for all relevant parton channels.
This strategy is implemented through the \ac{HF-NRevo} scheme introduced in Refs.~\cite{Celiberto:2024mex,Celiberto:2024bxu,Celiberto:2024rxa,Celiberto:2025xvy}.

Originally developed to evolve FFs from nonrelativistic inputs, {\HFNRevo} has been extended to accommodate more general input models~\cite{Celiberto:2023rzw,Celiberto:2024mab,Celiberto:2024beg,Celiberto:2025dfe,Celiberto:2025ziy,Celiberto:2025ogy}.
It is structured around three essential components: the physical \emph{interpretation} of the input, the \emph{evolution} procedure itself, and the associated \emph{uncertainty} quantification.

Focusing on the evolution step, the DGLAP procedure implemented in {\HFNRevo} follows a two-phase approach.
First, an extended semi-analytical \emph{decoupled} evolution, referred to as {\tt EDevo}, ensures precise handling of threshold effects across all parton species.
This is followed by the numerical \emph{all-order} evolution phase, denoted {\tt AOevo}.

The {\HFNRevo} framework was initially applied to the collinear fragmentation of fully heavy tetraquarks, incorporating both gluon and heavy-quark channels (see Ref.~\cite{Celiberto:2024mab} for the {\tt TQ4Q1.0} set and Section~2.3.4 of Ref.~\cite{Celiberto:2024beg} for the {\tt 1.1} update).

When only a single initial FF channel is available, a reduced version of {\HFNRevo} is employed.
This applies, for example, to the {\tt TQHL1.0} and {\tt TQHL1.1} sets for the fragmentation of doubly heavy $\XQq$ tetraquarks, constructed using only the $[Q,\bar{Q} \to \XQq]$ channel (see Ref.~\cite{Celiberto:2023rzw} and Section~2.2.2 of Ref.~\cite{Celiberto:2024beg}).
In such cases, the {\tt EDevo} step is omitted, and the procedure proceeds directly to the numerical {\tt AOevo} phase.
This second step is currently implemented using the {\APFELpp} evolution library~\cite{Bertone:2013vaa,Carrazza:2014gfa,Bertone:2017gds}, while future developments will enable interfacing {\symJethad} with the {\EKO} framework~\cite{Candido:2022tld,Hekhorn:2023gul}.

Turning back to the fragmentation of fully heavy pentaquarks, we note that, to the best of our knowledge, only the constituent-quark channel has been modeled at the initial scale so far.
As done for the $\XQq$ tetraquarks~\cite{Celiberto:2023rzw,Celiberto:2024beg}, our {\tt PQ5Q1.0} sets are constructed by directly applying the {\tt AOevo} evolution to the $[Q,\bar{Q} \to \PQQ]$ input functions.
All other partonic channels are generated dynamically through DGLAP evolution.

In our formulation, the heavy-quark mass $m_Q$ explicitly enters the initial condition for the constituent channel via the mass-dependent dual-mode scenario introduced in previous sections.
This reflects the physical expectation that fragmentation into a $\PQQ$ state occurs near threshold, where finite-mass effects are non-negligible.

Accordingly, the evolution is carried out within a fixed-flavor scheme with $n_f = 5$, where no crossing of heavy-quark thresholds takes place.
The evolution itself is performed in the ZM-VFNS, where all partons---including charm and bottom---are treated as massless.
This choice is standard in studies of heavy-hadron fragmentation (see, \emph{e.g.}, Ref.~\cite{Cacciari:2024kaa}), based on the assumption that mass effects become negligible at large scales.

One might question the absence of initial-scale inputs for light partons and nonconstituent heavy quarks, which in our setup emerge only radiatively at higher energies.
However, previous analyses, such as Ref.~\cite{Nejad:2021mmp,Celiberto:2023rzw,Celiberto:2024omj} for doubly heavy tetraquarks and Refs.~\cite{Celiberto:2022dyf,Celiberto:2023fzz} for vector quarkonia, indicate that these contributions are subdominant at low scales.
A more complete treatment including all relevant initial channels is left for future developments.

In Fig.~\ref{fig:NLO_FFs_P5Q}, we present the factorization-scale dependence of our {\tt PQ5Q1.0} FFs at NLO within the ZM-VFNS scheme.
The plots show $z D_i^{\PQQ}(z, \mu_F)$ as a function of $\mu_F$ for both $\PQc$ (left column) and $\PQb$ (right column), distinguishing between the direct (top) and scalar-diquark (bottom) initial-scale configurations.
The hadron momentum fraction is fixed at $z = 0.425 \simeq \langle z \rangle$, consistent with typical values probed in semi-inclusive observables (see Refs.~\cite{Celiberto:2016hae,Celiberto:2017ptm,Celiberto:2020wpk,Celiberto:2021dzy,Celiberto:2021fdp,Celiberto:2022dyf,Celiberto:2022keu,Celiberto:2022kxx,Celiberto:2024omj,Celiberto:2025euy}).

In all four panels, we observe that the FFs originating from constituent heavy quarks, namely $c, \bar{c}$ for $\PQc$ and $b, \bar{b}$ for $\PQb$, dominate over all other partonic channels across the full $\mu_F$ range shown.
These FFs are typically larger by one to two orders of magnitude compared to the gluon contribution, and even more so with respect to those from light quarks or nonconstituent heavy flavors.
This behavior reflects the strong localization of fragmentation near threshold, where constituent dynamics play a central role.
The overall pattern is preserved when switching from the direct to the diquark mode, although the absolute normalization and $\mu_F$-slopes may vary slightly depending on the input configuration.

Then, consistent with previous findings for exotic hadrons, the gluon FFs for pentaquark production exhibit a smooth and monotonic growth with increasing $\mu_F$.
This pattern closely resembles what was previously observed in the case of doubly heavy tetraquarks, as illustrated in the upper panels of Fig.~\ref{fig:NLO_FFs_XQs_T4Q}, where the {\tt TQHL1.1} NLO sets~\cite{Celiberto:2024beg} are shown.
A similar trend is also visible for the gluon FFs in the lower panels, corresponding to scalar $\TQQ(0^{++})$ tetraquarks, described by the {\tt TQ4Q1.1} functions~\cite{Celiberto:2025dfe,Celiberto:2025ziy}.
Such stability and gradual enhancement with energy make the gluon channel an important secondary source in semi-inclusive production mechanisms---even though it remains subleading with respect to the constituent heavy-quark channel at the initial scale.

Finally, recent studies have shown that gluon FFs, characterized by a smooth dependence on the factorization scale $\mu_F$, act as effective \emph{stabilizers} for high-energy observables sensitive to the semi-inclusive production of singly~\cite{Celiberto:2021dzy,Celiberto:2021fdp} and multiply~\cite{Celiberto:2022dyf,Celiberto:2022keu,Celiberto:2023rzw,Celiberto:2025ziy,Celiberto:2025ipt,Celiberto:2025ogy} heavy-flavored hadrons.
This behavior underpins the so-called \emph{natural stability} of high-energy resummation~\cite{Celiberto:2022grc} (see Section~\ref{sec:HE_resummation}), a key feature that will play a prominent role in the phenomenological analyses presented in Section~\ref{sec:phenomenology}.

\section{NLL/NLO$^+$ hybrid factorization for pentaquark~$+$~jet systems}
\label{sec:HE_resummation}

The first part of this section contains a brief recap of recent phenomenological progresses of high-energy resummation in QCD (see Section~\ref{ssec:HE_QCD}).
The second part provides us with a formal description of the our pentaquark-sensitive observables described by the hands of the hybrid high-energy and collinear factorization at $\NLLp$ (see Section~\ref{ssec:hybrid_factorization}).

\subsection{High-energy QCD phenomenology: An incomplete summary}
\label{ssec:HE_QCD}

Accurate predictions for high-energy observables rely on the ability to separate long-distance dynamics from short-distance effects in hadronic scatterings.
This separation enables the factorization of nonperturbative contributions from perturbative calculations through the well-established collinear framework~\cite{Collins:1989gx,Sterman:1995fz}.

However, specific regions of phase space present challenges due to the emergence of large logarithmic corrections.
These logarithms grow with the perturbative order, eventually compensating the smallness of the QCD running coupling and undermining the convergence of the perturbative series.
In such regimes, the standard collinear factorization must be supplemented by all-order resummations.

In the semi-hard regime of QCD~\cite{Gribov:1983ivg}, characterized by a strong hierarchy among energy scales, $\sqrt{s} \gg Q \gg \LQCD$, logarithms of the type $\ln(s/Q^2)$ enter perturbative expansions with powers increasing at each order, thus requiring resummation~\cite{Celiberto:2017ius,Bolognino:2021bjd,Mohammed:2022gbk,Gatto:2025kfl}.

The \ac{BFKL} approach~\cite{Fadin:1975cb,Kuraev:1976ge,Kuraev:1977fs,Balitsky:1978ic} provides the most suitable framework for performing resummation at high energies.
Specifically, BFKL resums all terms of the form $(\alpha_s \ln s)^n$ (the so-called LL approximation) as well as terms proportional to $\alpha_s(\alpha_s \ln s)^n$, corresponding to the \ac{NLL} accuracy.

Within the BFKL formalism, a given scattering amplitude can be factorized into a convolution of a universal Green's function with two singly off-shell, transverse-momentum-dependent emission vertices---also known as impact factors---describing the production of a forward-state particle from the remnants of each incoming hadron.
In BFKL terminology, these are referred to as forward-production impact factors.

The Green's function evolves according to an integral equation, whose kernel has been determined at NLO accuracy~\cite{Fadin:1998py,Ciafaloni:1998gs,Fadin:1998jv,Fadin:2000kx,Fadin:2000hu,Fadin:2004zq,Fadin:2005zj} (see also ongoing efforts in higher-order corrections~\cite{Caola:2021izf,Falcioni:2021dgr,DelDuca:2021vjq,Byrne:2022wzk,Fadin:2023roz,Byrne:2023nqx}).

The predictive power of high-energy resummation within the BFKL framework at NLL accuracy is currently limited by the availability of off-shell emission functions computed at NLO.
These include:
a) impact factors for incoming partons (quarks and gluons)\cite{Fadin:1999de,Fadin:1999df}, which are essential for calculating
b) forward-jet~\cite{Bartels:2001ge,Bartels:2002yj,Caporale:2011cc,Ivanov:2012ms,Colferai:2015zfa} and
c) forward light-hadron~\cite{Ivanov:2012iv} emission functions.

Additional NLO-calculated impact factors include:
d) virtual photon to light vector meson transitions~\cite{Ivanov:2004pp},
e) light-by-light scattering contributions~\cite{Bartels:2000gt,Bartels:2001mv,Bartels:2002uz,Bartels:2004bi,Fadin:2001ap,Balitsky:2012bs}, and
f) forward Higgs boson production in the infinite top-mass limit~\cite{Hentschinski:2020tbi,Celiberto:2022fgx,Hentschinski:2022sko} and at finite top mass~\cite{Celiberto:2024bbv,Celiberto:2025ece} (see Ref.~\cite{DelDuca:2025vux} for pioneering extensions to the next-to-NLO accuracy).

At LO accuracy, several additional channels have been studied, such as:
Drell--Yan pair production~\cite{Hentschinski:2012poz,Motyka:2014lya},
heavy-quark pair production~\cite{Celiberto:2017nyx,Bolognino:2019ccd,Bolognino:2019yls}, and
forward $J/\psi$ production at low transverse momentum~\cite{Boussarie:2017oae} (see also Refs.~\cite{Boussarie:2015jar,Boussarie:2016gaq,Boussarie:2017xdy}).

Among the most promising observables for testing high-energy QCD at hadron colliders are those associated with so-called gold-plated channels.
These include a variety of semi-inclusive processes that are particularly sensitive to BFKL dynamics, such as:
Mueller--Navelet dijet production at NLO~\cite{Mueller:1986ey,Colferai:2010wu,Ducloue:2013hia,Caporale:2013uva,Colferai:2015zfa,Caporale:2015uva,Ducloue:2015jba,Celiberto:2015yba,Celiberto:2015mpa,Celiberto:2016ygs,Celiberto:2016vva,Caporale:2018qnm,deLeon:2020myv,deLeon:2021ecb,Celiberto:2022gji,Baldenegro:2024ndr},
correlated dihadron emissions~\cite{Celiberto:2016hae,Celiberto:2016zgb,Celiberto:2017ptm,Celiberto:2017uae,Celiberto:2017ydk},
multi-jet configurations tagged at large rapidity intervals~\cite{Caporale:2015vya,Caporale:2015int,Caporale:2016soq,Caporale:2016vxt,Caporale:2016xku,Celiberto:2016vhn,Caporale:2016djm,Caporale:2016pqe,Chachamis:2016qct,Chachamis:2016lyi,Caporale:2016lnh,Caporale:2016zkc,Caporale:2017jqj,Chachamis:2017vfa},
as well as more exclusive configurations such as hadron-plus-jet~\cite{Bolognino:2018oth,Bolognino:2019cac,Bolognino:2019yqj,Celiberto:2020wpk,Celiberto:2020rxb,Celiberto:2022kxx},
Higgs-plus-jet final states~\cite{Celiberto:2020tmb,Celiberto:2021fjf,Celiberto:2021tky,Celiberto:2021txb,Celiberto:2021xpm},
heavy-light dijets~\cite{Bolognino:2021mrc,Bolognino:2021hxx},
and heavy-flavored hadron production~\cite{Boussarie:2017oae,Celiberto:2017nyx,Bolognino:2019ouc,Bolognino:2019yls,Bolognino:2019ccd,Celiberto:2021dzy,Celiberto:2021fdp,Bolognino:2022wgl,Celiberto:2022dyf,Celiberto:2022grc,Bolognino:2022paj,Celiberto:2022qbh,Celiberto:2022keu,Celiberto:2022zdg,Celiberto:2022kza,Celiberto:2024omj,Celiberto:2025euy}.

A particularly clean and sensitive probe of BFKL dynamics is offered by processes involving the detection of single forward particles.
These are directly linked to the behavior of the proton's \ac{UGD} at small-$x$, whose energy evolution is governed by the BFKL Green's function.
Examples include:
light vector meson leptoproduction, extensively studied at HERA~\cite{Anikin:2009bf,Anikin:2011sa,Besse:2013muy,Bolognino:2018rhb,Bolognino:2018mlw,Bolognino:2019bko,Bolognino:2019pba,Celiberto:2019slj,Luszczak:2022fkf} and projected for future EIC measurements~\cite{Bolognino:2021niq,Bolognino:2021gjm,Bolognino:2022uty,Celiberto:2022fam,Bolognino:2022ndh},
exclusive quarkonium photoproduction~\cite{Bautista:2016xnp,Garcia:2019tne,Hentschinski:2020yfm,Peredo:2023oym,Hentschinski:2025ovo},
inclusive Drell--Yan dilepton detection~\cite{Motyka:2014lya,Brzeminski:2016lwh,Motyka:2016lta,Celiberto:2018muu},
and the emission of bottom-tagged jets~\cite{Chachamis:2015ona,Chachamis:2013bwa,Chachamis:2009ks}.

The insights provided by the BFKL UGD at small-$x$ have proven essential for improving the theoretical description of collinear \ac{PDFs} supplemented by small-$x$ resummation~\cite{Ball:2017otu,Abdolmaleki:2018jln,Bonvini:2019wxf,Silvetti:2022hyc,Silvetti:2023suu,Rinaudo:2024hdb,Celiberto:2025nnq}.
In parallel, they establish a bridge with model-based studies of twist-two gluon TMD distributions enhanced at low-$x$~\cite{Bacchetta:2020vty,Celiberto:2021zww,Bacchetta:2021oht,Bacchetta:2021lvw,Bacchetta:2021twk,Bacchetta:2022esb,Bacchetta:2022crh,Bacchetta:2022nyv,Celiberto:2022omz,Bacchetta:2023zir,Bacchetta:2024fci,Bacchetta:2024uxb}.

Further developments in Refs.~\cite{Hentschinski:2021lsh,Mukherjee:2023snp} offer a more precise characterization of the relationship between low-$x$ dynamics and the TMD formalism,
while recent work~\cite{Boroun:2023goy,Boroun:2023ldq} explores how the UGD connects to dipole cross sections in color-glass-like frameworks.

Analyses of small-$x$ resummed inclusive and differential distributions for Higgs and heavy-flavor production have also been pursued via the {\Hell} method~\cite{Bonvini:2018ixe,Silvetti:2022hyc}, based on the \ac{ABF} formalism~\cite{Ball:1995vc,Ball:1997vf,Altarelli:2001ji,Altarelli:2003hk,Altarelli:2005ni,Altarelli:2008aj,White:2006yh}. This approach merges collinear factorization with small-$x$ resummation, incorporating high-energy factorization theorems~\cite{Catani:1990xk,Catani:1990eg,Collins:1991ty,Catani:1993ww,Catani:1993rn,Catani:1994sq,Ball:2007ra,Caola:2010kv}, and provides a complementary view on high-energy QCD dynamics.

A major challenge in the BFKL description of Mueller--Navelet jet production arises from the destabilizing role of NLL corrections.
Although these next-to-leading logarithmic terms are formally of the same order as the LL contributions, they come with opposite sign.
This results in significant instabilities within the resummed series, especially when analyzing \ac{MHOUs} by varying the energy scales around their natural values.

As a consequence, Mueller--Navelet observables can yield unphysical predictions when the rapidity separation between the jets becomes large.
Similarly, azimuthal-angle correlations exhibit abnormal behaviors both at low and high rapidities.
Several strategies have been developed to mitigate these effects.
Among them, the \ac{BLM} scale-setting procedure~\cite{Brodsky:1996sg,Brodsky:1997sd,Brodsky:1998kn,Brodsky:2002ka}, specifically adapted to semi-hard processes~\cite{Caporale:2015uva}, has shown partial success in restoring stability---particularly in azimuthal distributions---leading to modest improvements in agreement with experimental data.

However, the effectiveness of the BLM method proves limited---especially in observables involving light dihadron or hadron-plus-jet emissions in semi-hard regimes.
This limitation stems from the fact that the optimal renormalization scales suggested by BLM tend to lie significantly above the natural physical scales of the processes under consideration~\cite{Celiberto:2017ius,Bolognino:2018oth,Celiberto:2020wpk}.
As a result, total cross sections in these channels suffer from a marked suppression, leading to poor event statistics.

Encouraging signs of improved stability in high-energy resummation---under both higher-order corrections and scale variations---have recently emerged in processes involving final states that are sensitive to Higgs boson production~\cite{Celiberto:2020tmb,Celiberto:2023rtu,Celiberto:2023uuk,Celiberto:2023eba,Celiberto:2023nym,Celiberto:2023dkr,Celiberto:2023rqp,Celiberto:2024mdt,Celiberto:2024bfu,Celiberto:2025edg}.
This stabilizing trend was first clearly observed in studies of semi-inclusive emissions of $\Lambda_c$ hyperons~\cite{Celiberto:2021dzy} and singly bottom-flavored hadrons~\cite{Celiberto:2021fdp} at the LHC.
Notably, it was found that such stability is intimately linked to the distinctive behavior of ZM-VFNS collinear FFs, which govern the production of these singly heavy-flavored hadrons at high transverse momentum.

Subsequent analyses on vector quarkonia~\cite{Celiberto:2022dyf,Celiberto:2023fzz}, charmed $B$ mesons~\cite{Celiberto:2022keu,Celiberto:2024omj}, heavy tetraquarks~\cite{Celiberto:2023rzw,Celiberto:2024mab,Celiberto:2024mrq,Celiberto:2024beg,Celiberto:2025dfe,Celiberto:2025ziy}, and rare $\Omega$ baryons~\cite{Celiberto:2025ogy}, clarified that this remarkable property, known as \emph{natural stability} of the high-energy resummation in QCD~\cite{Celiberto:2022grc}, emerges as an intrinsic feature inherently associated with final states sensitive to heavy flavor.

\subsection{Hybrid-factorization studies at $\NLL$ and beyond}
\label{ssec:hybrid_factorization}

\begin{figure*}[!t]
\centering

\includegraphics[width=0.575\textwidth]{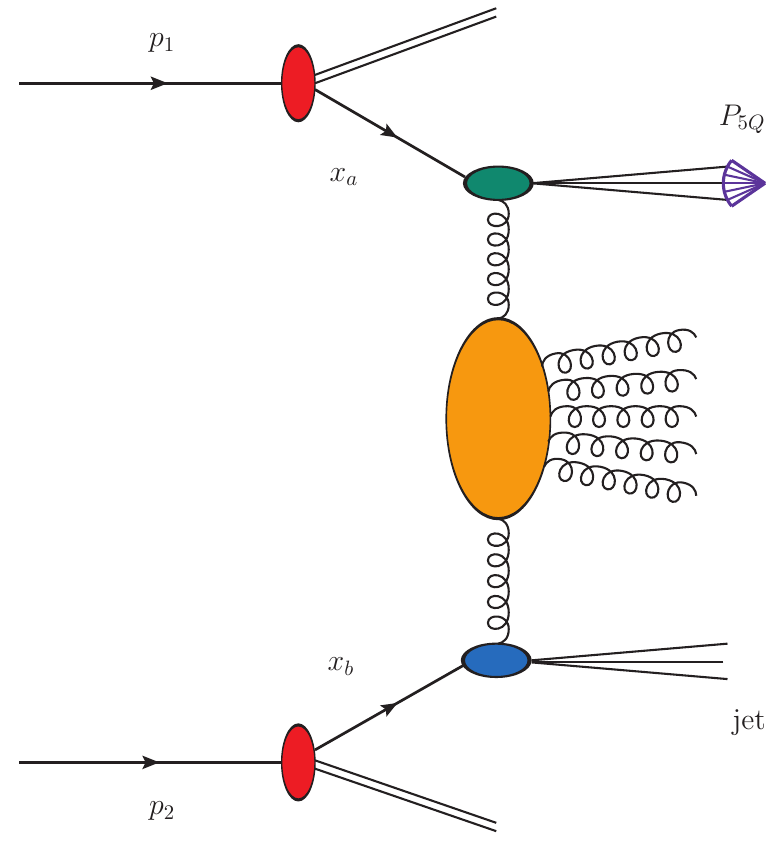}

\caption{Sketch of the pentaquark~$+$~jet semi-inclusive hadroproduction within $\NLLp$ factorization (diagram made with {\tt JaxoDraw 2.0}~\cite{Binosi:2008ig}). 
Red ovals portray collinear PDFs. 
The off-shell vertex, part of the hadron (jet) emission function, is depicted by green (blue) blobs. 
Pentaquarks emissions are depicted by indigo arrows. 
The big orange oval at the center of the diagram stands for the BFKL Green's function.}
\label{fig:process}
\end{figure*}

We investigate the reaction represented in Fig.~\ref{fig:process}
\begin{equation}
\label{process}
\setlength{\jot}{10pt} 
\begin{split}
    {\rm p}(p_1) \;+\; {\rm p}(p_2) &\;\rightarrow\; \PQb(\vec \kappa_1, y_1) \;+\; {\cal X} \;+\; {\rm jet}(\vec \kappa_2, y_2) \; ,
\end{split}
\end{equation}
where a fully bottomed pentaquark $\PQb$ is emitted in association with a light-flavored jet.
The final-state particles possess large transverse momenta, $|\vec \kappa_{1,2}| \gg \Lambda_{\rm QCD}$, and their rapidity separation is $ \DY = y_1 - y_2$. 

Large observed transverse momenta and large rapidity distances are needed for dealing with semi-hard final-state configurations.
Furthermore, transverse-momentum ranges have to be large enough to make the validity of the ZM-VFNS collinear fragmentation be the dominant mechanism for the production of heavy hadrons.

Incoming protons' four-momenta can decomposed as Sudakov vectors satisfying $p_1^2= p_b^2=0$ and $2 (p_1\cdot p_b) = s$.
In this way, $\kappa_1$ and $\kappa_2$ can be cast as
\begin{equation}\label{sudakov}
\kappa_{1,2} = x_{1,2} p_{1,2} - \frac{\kappa_{1,2\perp}^{\,2}}{x_{1,2} s}p_{2,1} + \kappa_{1,2\perp} \ , \qquad
\vec \kappa_{1,2}^{\,2} \equiv -\kappa_{1,2\perp}^2\;.
\end{equation}
Final-state object's longitudinal fractions, $x_{1,2}$, depend on rapidities as
$y_{1,2}=\pm\frac{1}{2}\ln\frac{x_{1,2}^2 s}
{\vec \kappa_{1,2}^2}$.
Thus one have $\drv y_{1,2} = \pm \frac{\drv x_{1,2}}{x_{1,2}}$, and $\DY \equiv y_1 - y_2 = \ln\frac{x_1 x_2 s}{|\vec \kappa_1||\vec \kappa_2|}$.

The LO cross section of our reaction (Eq.~\eqref{process}) in pure collinear QCD would take the form of a one-dimensional convolution among protons' PDFs, hadrons' FFs, and partonic hard factors.
One has
\begin{equation}
\label{sigma_collinear}
\begin{split}
 \frac{\drv \sigma_{[p \;+\; p \;\;\to\;\; {\PQb} \;+\; {\rm jet}]}^{\rm LO}}{\drv x_1 \drv x_2 \drv^2 \vec \kappa_1 \drv^2 \vec \kappa_2}
 &=\sum_{a,b} \int_0^1 \drv x_a \int_0^1 \drv x_b\ 
 f_a(x_a, \mu_F) f_b(x_b, \mu_F)
\int_{x_1}^1 \frac{\drv z}{z}D^{\PQb}_{a}\left(\frac{x_1}{z}\right) 
\frac{\drv {\hat\sigma}_{a,b}}
{\drv x_1\drv x_2\drv z\,\drv ^2\vec \kappa_1\drv ^2\vec \kappa_2}\;,
\end{split}
\end{equation}
where the ($a,b$) indices run over quarks, antiquarks, and the gluon, $f_{a,b}$ are proton PDFs, $D^{\PQb}_{a,b}$ denote pentaquark FFs, $x_{a,b}$ stand for the longitudinal fractions of the struck partons, $z$ is the longitudinal fraction of the outgoing parton fragmenting into $\PQb$, and $\drv \hat\sigma_{a,b}$ are partonic-subprocess cross sections.

In contrast, deriving the high-energy resummed cross section within our hybrid factorization framework involves a two-step procedure.
The first step consists in applying the high-energy factorization scheme dictated by BFKL resummation.
This is then supplemented by a collinear refinement, through the inclusion of PDFs and FFs.
The resulting differential cross section is naturally written as a Fourier series in the azimuthal-angle difference, with coefficients that encode the dynamics of the process.
We write
\begin{equation}
 \label{dsigma_Fourier}
 \frac{\drv \sigma^\NLLp}{\drv y_1 \drv y_2 \drv \vec \kappa_1 \drv \vec \kappa_2 \drv \varphi_1 \drv \varphi_2} =
 \frac{1}{(2\pi)^2} \left[{\cal C}_0^\NLLp + 2 \sum_{n=1}^\infty \cos \left(n (\phi - \pi)\right) \,
 {\cal C}_n^\NLLp \right]\, ,
\end{equation}
with $\varphi_{1,2}$ being the final-state azimuthal angles and $\phi = \varphi_1 - \varphi_2$.
The azimuthal coefficients are evaluated within the BFKL framework and encapsulate the effects of LL and NLL resummation of high-energy logarithms.
Our calculation is carried out in the $\MSb$ renormalization scheme~\cite{PhysRevD.18.3998}, which provides a consistent setup for organizing perturbative corrections and handling ultraviolet divergences.
One has
\begin{equation}
\label{Cn_NLLp_MSb}
\begin{split}
 \CnNLLp &= \int_0^{2\pi} \drv \varphi_1 \int_0^{2\pi} \drv \varphi_2\,
 \cos \left(n (\phi - \pi)\right) \,
 \frac{\drv \sigma^\NLLp}{\drv y_1 \drv y_2\, \drv |\vec \kappa_1| \, \drv |\vec \kappa_2| \drv \varphi_1 \drv \varphi_2}\;
\\
 &= \; \frac{e^{\DY}}{s} 
 \int_{-\infty}^{+\infty} \drv \nu \, e^{{\DY} \bar \alpha_s(\mu_R)\chi^\NLO(n,\nu)}
\\
 &\times \; \alpha_s^2(\mu_R) \, 
 \biggl\{
 \F_1^\NLO(n,\nu,|\vec \kappa_1|, x_1)[\F_2^\NLO(n,\nu,|\vec \kappa_2|,x_2)]^*\,
\\ 
 &+ \,
 \left.
 \bar \alpha_s^2(\mu_R)
 \, \DY
 \frac{\beta_0}{4 N_c}\chi(n,\nu)f(\nu)
 \right\} \;.
\end{split}
\end{equation}
Here, $\bar{\alpha}_s(\mu_R) \equiv \alpha_s(\mu_R) N_c / \pi$, with $N_c$ the number of colors, and $\beta_0 = (11 N_c - 2 n_f)/3$ is the leading coefficient of the QCD $\beta$-function, with $n_f$ denoting the number of active flavors.
We employ a two-loop running coupling, initialized at $\alpha_s(M_Z) = 0.118$, and account for a dynamic flavor number $n_f$ throughout the evolution.
The resummation kernel appearing in the exponent of Eq.~\eqref{Cn_NLLp_MSb} is given by
\begin{eqnarray}
 \label{chi}
 \chi^\NLO(n,\nu) = \chi(n,\nu) + \bar\alpha_s \hat \chi(n,\nu) \;,
\end{eqnarray}
where
\begin{eqnarray}
 \label{kernel_LO}
 \chi\left(n,\nu\right) = -2\gamma_{\rm E} - 2 \, {\rm Re} \left\{ \psi\left(\frac{1+n}{2} + i \nu \right) \right\} \, 
\end{eqnarray}
represent the LO kernel eigenvalues, $\gamma_{\rm E}$ the Euler-Mascheroni constant, and $\psi(z) \equiv \Gamma^\prime
(z)/\Gamma(z)$ the logarithmic derivative of the Gamma function. 
The $\hat\chi(n,\nu)$ function in Eq.~\eqref{chi} stands for the NLO kernel correction
\begin{equation}
\begin{split}
\label{chi_NLO}
\hat \chi\left(n,\nu\right) &= \bar\chi(n,\nu)+\frac{\beta_0}{8 N_c}\chi(n,\nu)
\left(-\chi(n,\nu)+10/3+2\ln\frac{\mu_R^2}{\mu_C^2}\right) \;,
\end{split}
\end{equation}
The mass of the pentaquark is fixed at $m_{\PQQ} = 5 m_Q$, where $m_Q$ denotes the mass of the constituent heavy quark, $Q$.
Since the jet on the light-flavor side does not involve a heavy mass scale, its transverse mass simply reduces to its transverse momentum, $|\vec \kappa_2|$.
The characteristic function $\bar\chi(n, \nu)$ entering the BFKL exponent was derived in Ref.~\cite{Kotikov:2000pm,Kotikov:2002ab} and it reads
\begin{equation}
 \label{kernel_NLO}
 \bar \chi(n,\nu)\,=\, - \frac{1}{4}\left\{\frac{\pi^2 - 4}{3}\chi(n,\nu) - 6\zeta(3) - \frac{\drv^2 \chi}{\drv\nu^2} + \,2\,\phi(n,\nu) + \,2\,\phi(n,-\nu)
 \right.
\end{equation}
\[
 \left.
 +\; \frac{\pi^2\sinh(\pi\nu)}{2\,\nu\, \cosh^2(\pi\nu)}
 \left[
 \left(3+\left(1+\frac{n_f}{N_c^3}\right)\frac{11+12\nu^2}{16(1+\nu^2)}\right)
 \delta_{n0}
 -\left(1+\frac{n_f}{N_c^3}\right)\frac{1+4\nu^2}{32(1+\nu^2)}\delta_{n2}
\right]\right\} \, ,
\]
where
\begin{equation}
\label{kernel_NLO_phi}
 \phi(n,\nu)\,=\,-\int_0^1 \drv x\,\frac{x^{-1/2+i\nu+n/2}}{1+x}\left\{\frac{1}{2}\left(\psi^\prime\left(\frac{n+1}{2}\right)-\zeta(2)\right)+\mbox{Li}_2(x)+\mbox{Li}_2(-x)\right.
\end{equation}
\[
\left.
 +\; \ln x\left[\psi(n+1)-\psi(1)+\ln(1+x)+\sum_{k=1}^\infty\frac{(-x)^k}{k+n}\right]+\sum_{k=1}^\infty\frac{x^k}{(k+n)^2}\left[1-(-1)^k\right]\right\}
\]
\[
 =\; \sum_{k=0}^\infty\frac{(-1)^{k+1}}{k+(n+1)/2+i\nu}\left\{\psi^\prime(k+n+1)-\psi^\prime(k+1)\right.
\]
\[
 \left.
 +\; (-1)^{k+1}\left[\beta_{\psi}(k+n+1)+\beta_{\psi}(k+1)\right]-\frac{\psi(k+n+1)-\psi(k+1)}{k+(n+1)/2+i\nu}\right\} \; ,
\]
with
\begin{equation}
\label{kernel_NLO_phi_beta_psi}
 \beta_{\psi}(z)=\frac{1}{4}\left[\psi^\prime\left(\frac{z+1}{2}\right)
 -\psi^\prime\left(\frac{z}{2}\right)\right] \; ,
\end{equation}
and
\begin{equation}
\label{dilog}
\mbox{Li}_2(x) = \int^x_0 \drv \omega \,\frac{\ln(1-\omega)}{\omega} \; .
\end{equation}
The expressions for the singly off-shell emissions functions are
\begin{equation}
\label{IFs}
\F_{1,2}^\NLO(n,\nu,|\vec \kappa_{1,2}|,x_{1,2}) =
\F_{1,2}(n,\nu,|\vec \kappa_{1,2}|,x_{1,2}) +
\alpha_s(\mu_R) \, \hat \F_{1,2}(n,\nu,|\vec \kappa_{1,2}|,x_{1,2}) \;.
\end{equation}
Here, the LO functions depicting the production of a forward hadron and a forward jet are
\begin{equation}
\label{LOHEF}
\begin{split}
\F_h(n,\nu,|\vec \kappa_h|,x_h) 
&= 2 \, \sqrt{\frac{C_F}{C_A}}
|\vec \kappa_h|^{2i\nu-1}\,\int_{x_h}^1\frac{\drv z}{z}
\left(\frac{z}{x_h} \right)
^{2 i\nu-1} 
 \left[\frac{C_A}{C_F}f_g(z)D_g^h\left(\frac{x_h}{z}\right)
 +\sum_{a=q,\bar q}f_a(z)D_a^h\left(\frac{x_h}{z}\right)\right] 
\end{split}
\end{equation}
and
\begin{equation}
 \label{LOJEF}
 \F_J(n,\nu,|\vec \kappa_J|,x_J) =  2 \sqrt{\frac{C_F}{C_A}}
 |\vec \kappa_J|^{2i\nu-1}\,\left[\frac{C_A}{C_F}f_g(x_J)
 +\sum_{b=q,\bar q}f_b(x_J)\right] \;,
\end{equation}
with $C_F \equiv (N_c^2-1)/(2N_c)$ and $C_A \equiv N_c$ the Casimir QCD factors.
The $f(\nu)$ function encodes the logarithmic derivative of LO functions
\begin{equation}
 f(\nu) = \frac{i}{2} \, \frac{\drv}{\drv \nu} \ln\left(\frac{\F_1}{\F_2^*}\right) + \ln\left(|\vec \kappa_1| |\vec \kappa_2|\right) \;.
\label{fnu}
\end{equation}

In Eq.~\eqref{Cn_NLLp_MSb}, the remaining ingredients are the NLO corrections to the emission functions, denoted as $\hat \F_{1,2}$.
The NLO term associated with the forward hadron was derived in Ref.~\cite{Ivanov:2012iv}, and its explicit expression is reported in Appendix~\hyperlink{app:NLOHEF}{A} of this review.
For the forward jet, we adopt the prescription of Refs.~\cite{Ivanov:2012iv,Ivanov:2012ms}.
To facilitate numerical implementation, we rely on a jet selection function\footnote{Jet algorithms are generally classified into two main categories: \emph{cone-type} and \emph{sequential-clustering} algorithms. For a detailed discussion, see Refs.~\cite{Chekanov:2002rq,Salam:2010nqg} and references therein. A prominent example of the latter is the (anti-)$\kappa_\perp$ algorithm~\cite{Catani:1993hr,Cacciari:2008gp}.} evaluated within the \ac{SCA}~\cite{Furman:1981kf,Aversa:1988vb}, using its cone-type version\cite{Colferai:2015zfa} with jet radius ${\cal R} = 0.5$.
The analytical form of the NLO jet emission function is provided in Appendix~\hyperlink{app:NLOJEF}{B}.

A rigorous phenomenological comparison between our hybrid factorization framework and fixed-order predictions would require a dedicated numerical setup capable of evaluating NLO distributions for two-particle production in hadronic collisions.
To the best of our knowledge, such a tool is currently not available.
As a pragmatic alternative, we benchmark our results against fixed-order estimates by truncating the azimuthal-coefficient expansion in Eq.~\eqref{Cn_NLLp_MSb} at ${\cal O}(\alpha_s^3)$.
This yields an effective high-energy fixed-order approximation ($\HENLOp$), which retains the leading-power asymptotic behavior of full NLO calculations while discarding all terms suppressed by inverse powers of the partonic center-of-mass energy.
We report here the $\MSb$ formula for the azimuthal coefficients at $\HENLOp$:
\begin{equation}
\label{Cn_HENLO_MSb}
 \CnHENLOp =
 \frac{e^{\DY}}{s}
 \int_{-\infty}^{+\infty} \drv \nu \,
 \alpha_s^2(\mu_R) \,
 \left[ 1 + \bar \alpha_s(\mu_R) \DY \chi(n,\nu) \right] \,
 \F_1^\NLO(n,\nu,|\vec \kappa_1|, x_1) \,[\F_2^\NLO(n,\nu,|\vec \kappa_2|,x_2)]^*
 \;,
\end{equation}
where the resummation kernel has been expanded and truncated at ${\cal O}(\alpha_s)$.
For comparison, we will also present results at a pure LL accuracy:
\begin{equation}
\label{Cn_LL_MSb}
  \CnLL = \frac{e^{\DY}}{s} 
 \int_{-\infty}^{+\infty} \drv \nu \, e^{{\DY} \bar \alpha_s(\mu_R)\chi(n,\nu)} \, \alpha_s^2(\mu_R) \, \F_1(n,\nu,|\vec \kappa_1|, x_1)[\F_2(n,\nu,|\vec \kappa_2|,x_2)]^* \,.
\end{equation}

Equations~\eqref{Cn_NLLp_MSb} to~\eqref{Cn_LL_MSb} outline the construction of our hybrid factorization framework.
In accordance with the BFKL approach, the hadronic cross section is expressed as a transverse-momentum convolution of the Green's function with two off-shell emission functions (impact factors).
These impact factors incorporate collinear PDFs and FFs, effectively linking high-energy and collinear dynamics.

The $\NLLp$ label signals that energy logarithms are fully resummed at next-to-leading logarithmic accuracy using NLO-calculated perturbative ingredients.
The `$+$' superscript in Eq.~\eqref{Cn_NLLp_MSb} indicates that certain next-to-NLL contributions---specifically, those arising from the cross-product of the two NLO impact-factor corrections---are also included.

Renormalization and factorization scales are chosen according to the natural kinematic prescription,
\begin{equation}
\label{natural scales}
 \mu_R = \mu_F = \mu_N = m_{1 \perp} + m_{2 \perp} \;,
\end{equation}
where $m_{i \perp}$ denotes the transverse mass of final-state particles.

For the collinear PDFs, we employ the {\tt NNPDF4.0} NLO set~\cite{NNPDF:2021uiq,NNPDF:2021njg}, accessed through the {\tt LHAPDFv6.5.4} interface~\cite{Buckley:2014ana}.
These distributions are obtained via global fits using the replica method~\cite{Forte:2002fg}, a technique originally introduced in neural-network training and now widely used in multi-dimensional proton-structure analyses~\cite{Bacchetta:2017gcc,Scimemi:2019cmh,Bacchetta:2019sam,Bacchetta:2022awv,Bury:2022czx,Moos:2023yfa}.
For a critical assessment of inter-set correlations and associated uncertainties, we refer to Ref.~\cite{Ball:2021dab}.

All results in this review are derived within the $\MSb$ renormalization scheme~\cite{PhysRevD.18.3998}.

\section{Fully heavy pentaquarks from HL-LHC to FCC with {\Jethad}}
\label{sec:phenomenology}

All predictions presented in this review were obtained using {\Jethad}, a hybrid, multimodular interface that integrates both \textsc{Python}- and \textsc{Fortran}-based components.  
Designed for the computation, management, and processing of physical distributions across different theoretical frameworks~\cite{Celiberto:2020wpk,Celiberto:2022rfj,Celiberto:2023fzz,Celiberto:2024mrq,Celiberto:2024swu,Celiberto:2025csa}, {\Jethad} enabled precise numerical evaluations of our pentaquark-sensitive observables.  
Specifically, differential distributions were computed using selected \textsc{Fortran 2008} modular routines within {\Jethad}, while the built-in \textsc{Python 3.0} analyzer was employed for final data processing and elaboration.

Section~\ref{ssec:jethad} summarizes the key features of the current {\Jethad} framework ({\tt v0.5.3}). 
Its (super)modules have been validated against collider data in related processes; see Sec.~\ref{sssec:data_validation} for a brief overview.
A comprehensive discussion of the uncertainty quantification strategy is presented in Section~\ref{ssec:uncertainty}, while Section~\ref{ssec:final_state} details the kinematic cuts applied to final-state particles.
Numerical predictions for the $\PQb$ plus jet system, including rapidity-interval and transverse-momentum distributions, are provided in Sections~\ref{ssec:I_rates} and~\ref{ssec:pT_rates}, respectively.
The present phenomenological study expands and completes the analysis in Section~4.2 of Ref.~\cite{Celiberto:2025ipt}, which focused exclusively on $\PQc$ plus jet final states.

\subsection{The {\Jethad} {\tt v0.5.3} technology}
\label{ssec:jethad}

The {\Jethad} project originated in late 2017, driven by the need for accurate predictions of semi-hard hadron~\cite{Celiberto:2016hae,Celiberto:2017ptm} and jet~\cite{Celiberto:2015yba,Celiberto:2016ygs,Bolognino:2018oth} final states at the LHC. These processes, proposed as key channels to explore high-energy resummation in QCD, called for the development of a dedicated numerical framework to compute and analyze high-energy observables.  

The first named release, {\Jethad} {\tt v0.2.7}, facilitated an initial quantitative study comparing BFKL and DGLAP dynamics in the context of semi-inclusive hadron-plus-jet emissions at the LHC~\cite{Celiberto:2020wpk}. Subsequent updates introduced new functionalities, such as the selection of forward heavy-quark pair observables ({\tt v0.3.0}~\cite{Bolognino:2019yls}), the study of Higgs emissions and transverse-momentum spectra ({\tt v0.4.2}~\cite{Celiberto:2020tmb}), and the integration of a \textsc{Python} analyzer with the core \textsc{Fortran} module ({\tt v0.4.3}~\cite{Bolognino:2021mrc}).  

Progress continued with the inclusion of heavy-flavored hadrons within the ZM-VFNS framework at NLO ({\tt v0.4.4} \cite{Celiberto:2021dzy}). The \deffont{\rlap{D}Unamis} (\textsc{DYnamis}) package, dedicated to forward Drell--Yan dilepton production~\cite{Celiberto:2018muu}, was incorporated in {\tt v0.4.5}. 
Additionally, integration with the \textsc{LExA} modular code enabled investigations into proton structure at low-$x$ using small-$x$ TMD densities in {\tt v0.4.6}~\cite{Bolognino:2021niq}.  

The release of {\tt v0.4.7}~\cite{Celiberto:2022dyf} introduced quarkonium-sensitive observables derived from NRQCD leading-power fragmentation. Further developments in {\tt v0.5.0}~\cite{Celiberto:2023fzz} and {\tt v0.5.1}~\cite{Celiberto:2024omj} featured an improved system for MHOU studies, an expanded suite of observables (including singly and doublydifferential transverse-momentum distributions~\cite{Celiberto:2022gji,Celiberto:2022kxx,Celiberto:2024omj}) and capabilities for \emph{matching} procedures with collinear factorization~\cite{Celiberto:2023rtu,Celiberto:2023uuk,Celiberto:2023eba,Celiberto:2023nym,Celiberto:2023rqp}. 
The most notable addition in~{\tt v0.5.2} is {\symJethad}, a \textsc{Mathematica}~\cite{Mathematica_V14-2} plugin designed for symbolic computations in high-energy QCD and proton structure studies.
Version~{\tt v0.5.3} added major support to exotic matter studies~\cite{Celiberto:2024beg}.

From its foundational core to specialized modules and routines, {\Jethad} is engineered for high computational efficiency. The multidimensional integrators are optimized for parallel computing, allowing the software to dynamically select the most effective integration algorithm based on the specific characteristics of the integrand.  

Processes implemented within {\Jethad} can be flexibly managed through an intuitive, \emph{structure-based} smart interface. 
Final-state particles are represented as \emph{object} prototypes, encapsulating physical attributes such as mass, charge, kinematic limits, and rapidity labels. These objects are initially retrieved from a master database through a \emph{particle generation} routine, with support for user-defined particles. They are then \emph{cloned} into a final-state vector and \emph{injected} into the corresponding process module via a dedicated \emph{controller}.  

The flexibility in constructing final states extends to the selection of initial-state conditions. A unique \emph{particle-ascendancy} attribute enables {\Jethad} to automatically determine whether an object originates from hadroproduction, electroproduction, photoproduction, or another mechanism. This adaptive feature ensures that only relevant computational modules are initialized, optimizing performance.  

Designed as an \emph{object-based} system, {\Jethad} operates independently of the specific process being studied. While initially developed for high-energy QCD and TMD factorization, its modular architecture facilitates the straightforward incorporation of alternative theoretical approaches through additional (super)modules. These new components can be seamlessly integrated into the framework via the built-in \emph{point-to-routine} system, making {\Jethad} a versatile computational tool for particle physics.  

With the goal of providing the scientific community with a standardized computational environment capable of handling diverse processes across different theoretical formalisms, we anticipate releasing the first public version of {\Jethad} in the near future.

\subsubsection{Validation against collider data}
\label{sssec:data_validation}

Our framework has already undergone data driven checks in channels that are sensitive to the same high-energy dynamics relevant here. 
A first validation was presented in Ref.~\cite{Celiberto:2020wpk}, where $\NLL$ predictions for Mueller--Navelet dijet azimuthal correlations at $\sqrt{s}=7$ TeV were confronted with CMS measurements~\cite{CMS:2016qng}. 
This was later strengthened in Ref.~\cite{Celiberto:2022gji}, upgrading the accuracy to $\NLLp$, benchmarking against the $\HENLOp$ expansion, and validating both full and truncated azimuthal moments.

A second validation used Drell--Yan lepton pair data from LHCb~\cite{LHCb:2015okr} and ATLAS~\cite{ATLAS:2014ape}, providing a direct test of the \deffont{\rlap{D}Unamis} (\textsc{DYnamis}) supermodule of {\Jethad}~\cite{Celiberto:2018muu} at partial $\NLL$ accuracy. 
Since these are semi inclusive processes, the agreement offers an almost direct check of the ingredients that also enter the observables studied in this review.

Within BFKL, semi-hard cross sections have a universal convolution structure---two singly off-shell emission functions linked by the Green's function. 
This universality underpins the above validations and supports the reliability of our numerical implementation. 
While current hadron initiated datasets do not yet include identified light or heavy hadrons plus a companion jet at large rapidity separations, forthcoming measurements will enable a dedicated validation in hadron tagged channels, to which our setup is readily applicable.

An additional, more indirect validation comes from exclusive processes. 
Within {\Jethad}, the \textsc{LExA} supermodule has been benchmarked via helicity amplitudes for $\rho$ leptoproduction computed in our approach~\cite{Bolognino:2018rhb}, which describe HERA data~\cite{H1:2009cml,ZEUS:2007iet}; the same machinery has been extended to polarized observables and to EIC projections~\cite{Bolognino:2021niq}.

\subsection{Uncertainty estimation}
\label{ssec:uncertainty}

A standard approach to estimating the impact of MHOUs is to analyze how our observables respond to variations in the renormalization and factorization scales around their characteristic values.  
MHOUs are recognized as a major contributor to theoretical uncertainties~\cite{Celiberto:2022rfj}. To evaluate their influence, we simultaneously vary $\mu_R$ and $\mu_F$ in the range $\mu_N/2$ to $2 \mu_N$. The parameter $C_{\mu}$, displayed in the figures of Sections~\ref{ssec:I_rates} and~\ref{ssec:pT_rates}, is defined as $C_\mu \equiv \mu_{F}/\mu_N = \mu_{R}/\mu_N$.  

Another source of uncertainty is related to proton PDFs. 
Studies on high-energy production processes indicate that different choices of PDF parametrizations or members within the same set have only a minor impact on predictions~\cite{Bolognino:2018oth,Celiberto:2020wpk,Celiberto:2021fdp,Celiberto:2022rfj}. Consequently, we base our calculations on the central member of the {\tt NNPDF4.0} set.  

Further uncertainties may arise from the \emph{collinear improvement} of the NLO kernel, which introduces renormalization-group (RG) terms to ensure a smooth transition between the BFKL and DGLAP equations in the collinear limit, as well as from changes in the renormalization scheme~\cite{Salam:1998tj,Ciafaloni:2003rd,Ciafaloni:2003ek,Ciafaloni:2000cb,Ciafaloni:1999yw,Ciafaloni:1998iv,SabioVera:2005tiv}. However, the impact of these refinements on semi-hard rapidity-dependent distributions remains within the uncertainty bands associated with MHOUs~\cite{Celiberto:2022rfj}.  

The transition from the $\MSb$~\cite{PhysRevD.18.3998} to the MOM~\cite{Barbieri:1979be,PhysRevLett.42.1435} renormalization scheme was explored in Ref.~\cite{Celiberto:2022rfj}, showing that MOM predictions for rapidity distributions tend to be systematically larger. However, these deviations still fall within the uncertainty bands induced by MHOUs. A fully self-consistent MOM study would require the availability of PDFs and FFs evolved in the MOM scheme, which are not yet available.  

Finally, uncertainty bands in our distributions are obtained by combining MHOUs with the numerical errors inherent to multidimensional integration (see Section~\ref{ssec:final_state}). The latter remains consistently below $1\%$, thanks to the efficiency of the {\Jethad} integrators.

\subsection{Final-state kinematic ranges}
\label{ssec:final_state}

We consider two main observables: the rapidity-interval distribution, which represents the cross section differential in the rapidity separation, $\Delta Y = y_1 - y_2$, between the two produced particles, and the transverse-momentum distribution differential in $|\vec{\kappa}_1|$.

The first observable directly relates to the ${\cal C}_0$ angular coefficient, defined in Section~\ref{ssec:hybrid_factorization}, integrated over transverse momenta and rapidities of the final-state particles and evaluated at fixed rapidity separation $\Delta Y$ between the pentaquark and the light jet. It is given by  
\begin{equation}
 \label{obs:I}
 \frac{\drv \sigma(\DY, s)}{\drv \DY} =
 \int_{|\vec \kappa_2|^{\rm min}}^{|\vec \kappa_2|^{\rm max}} 
 \!\!\drv |\vec \kappa_1|
 \int_{|\vec \kappa_2|^{\rm min}}^{|\vec \kappa_2|^{\rm max}} 
 \!\!\drv |\vec \kappa_2|
 \int_{\max (y_1^{\rm min}, \, \DY + y_2^{\rm min})}^{\min (y_1^{\rm max}, \, \DY + y_2^{\rm max})} \drv y_1
 \, \,
 {\cal C}_0^{\rm [res]}
\Bigm \lvert_{y_2 = y_1 - \DY}
 \;,
\end{equation}
where the label `${\rm [res]}$' denotes different resummation schemes: $\NLLp$, $\HENLOp$, or $\LL$.  
To eliminate one of the rapidity integrations, specifically over $y_2$, we employed the constraint $\delta(\Delta Y - (y_1 - y_2))$.  

The transverse momenta of the forward $\PQb$ state are constrained within $60 < |\vec \kappa_1|/{\rm GeV} < 120$, ensuring compatibility with the ZM-VFNS-based fragmentation approach, where energy scales must be sufficiently above the thresholds for DGLAP evolution of heavy quarks.  
The jet is selected in a transverse-momentum range slightly different yet aligned with current LHC and future HL-LHC studies~\cite{Khachatryan:2016udy}, namely $50 < |\vec \kappa_2|/{\rm GeV} < 60$.  

This \emph{asymmetric} transverse-momentum selection helps distinguish high-energy dynamics from fixed-order effects~\cite{Celiberto:2015yba,Celiberto:2015mpa,Celiberto:2020wpk}.  
Additionally, it suppresses large Sudakov logarithms associated with nearly back-to-back configurations, which would otherwise require further resummation~\cite{Mueller:2013wwa,Marzani:2015oyb,Mueller:2015ael,Xiao:2018esv,Hatta:2020bgy,Hatta:2021jcd}.  
Furthermore, this choice mitigates instabilities from radiative corrections~\cite{Andersen:2001kta,Fontannaz:2001nq} and prevents violations of energy-momentum conservation~\cite{Ducloue:2014koa}.

For rapidity coverage, we follow constraints used in current LHC studies.  
Pentacharm states are exclusively detected in the barrel calorimeter region~\cite{Chatrchyan:2012xg}, spanning $-2.5$ to $2.5$, while light jets can also be identified in the endcap regions~\cite{Khachatryan:2016udy}, extending from $-4.7$ to $4.7$.

The second observable is the transverse-momentum spectrum  
\begin{equation}
\label{obs:I-k1b}
 \frac{\drv \sigma(|\vec \kappa_1|, s)}{\drv |\vec \kappa_1|} =
 \int_{|\vec \kappa_2|^{\rm min}}^{|\vec \kappa_2|^{\rm max}} 
 \!\!\drv |\vec \kappa_2|
 \, \,
 \int_{\DY^{\rm min}}^{\DY^{\rm max}} \drv \DY
 \int_{\max (y_1^{\rm min}, \, \DY + y_2^{\rm min})}^{\min (y_1^{\rm max}, \, \DY + y_2^{\rm max})} \drv y_1
 \, \,
 {\cal C}_{l=0}^{\rm [res]}
\Bigm \lvert_{y_2 = y_1 - \DY}
 \;.
\end{equation}
This distribution is differential in the $\PQQ$ transverse momentum $|\vec \kappa_1|$ but integrated over the jet transverse-momentum range $40~\text{GeV} < |\vec \kappa_2| < 120~\text{GeV}$ within a given $\DY$ bin.
The same rapidity cuts for pentacharm and jet detection apply.

This $|\vec \kappa_1|$ distribution provides a useful testbed for investigating potential connections between the $\NLLp$ factorization and alternative formalisms.  
Allowing $|\vec \kappa_1|$ to vary from $10$ to $100~\text{GeV}$ enables exploration of a broad kinematic range, where additional resummation effects may become relevant.  
Recent studies on high-energy Higgs~\cite{Celiberto:2020tmb} and heavy-jet~\cite{Bolognino:2021mrc} tagging confirm the reliability of our hybrid approach in the moderate transverse-momentum domain, specifically when $|\vec \kappa_1|$ and $|\vec \kappa_2|$ are of similar magnitude.  

Conversely, large $|\vec \kappa_1|$ values are expected to enhance threshold logarithms in the hard regime ($|\vec \kappa_1| > |\vec \kappa_2|^{\rm max}$), while soft logarithms become dominant in the low-momentum region ($|\vec \kappa_1| \ll |\vec \kappa_2|^{\rm min}$).  
Thus, analyzing $|\vec \kappa_1|$ distributions serves both as a consistency check for our framework and as a step toward incorporating additional resummation techniques.

\subsection{Rapidity-interval rates}
\label{ssec:I_rates}

\begin{figure*}[!t]
\centering

   \hspace{0.00cm}
   \includegraphics[scale=0.375,clip]{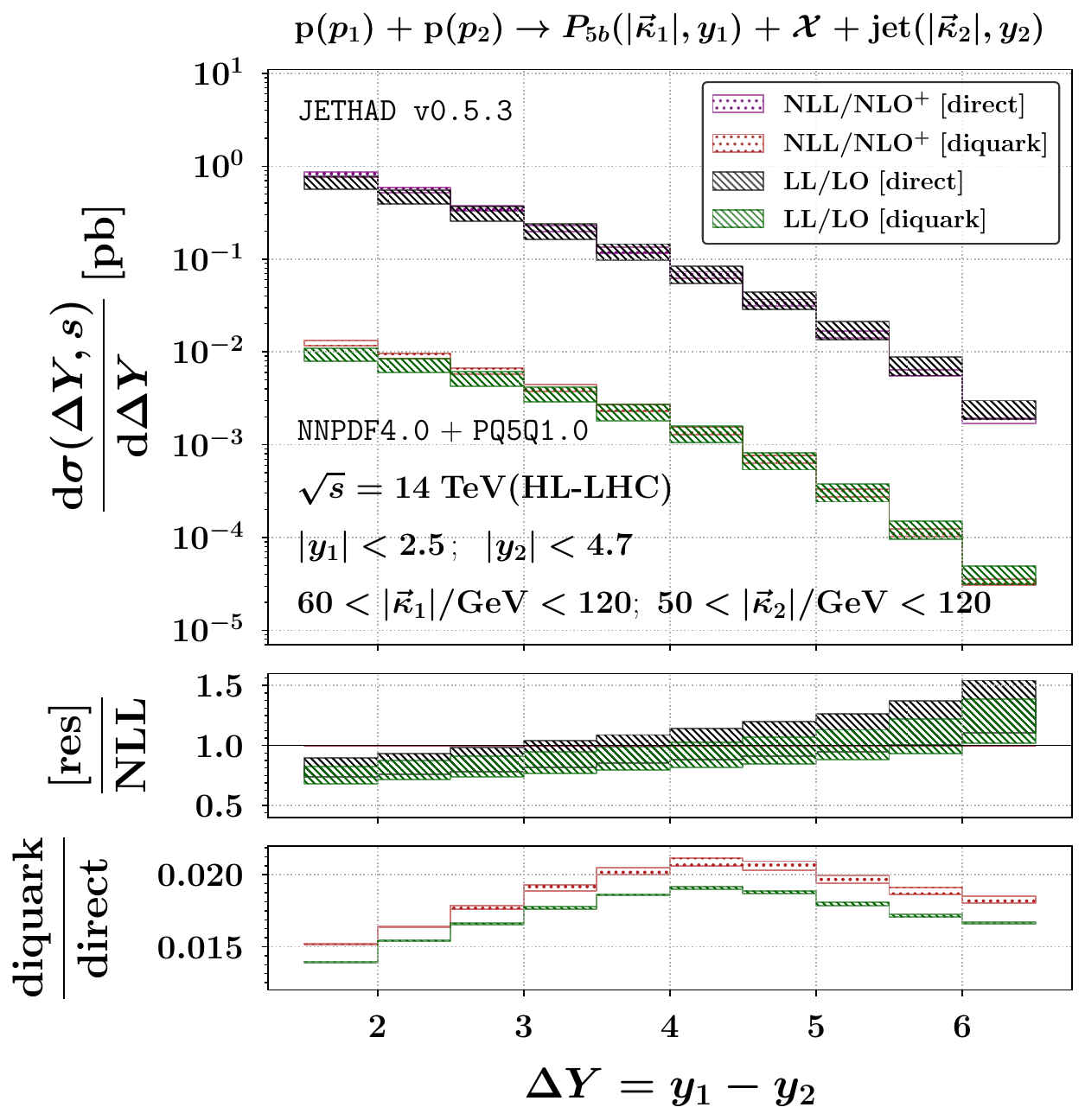}
   \hspace{0.30cm}
   \includegraphics[scale=0.375,clip]{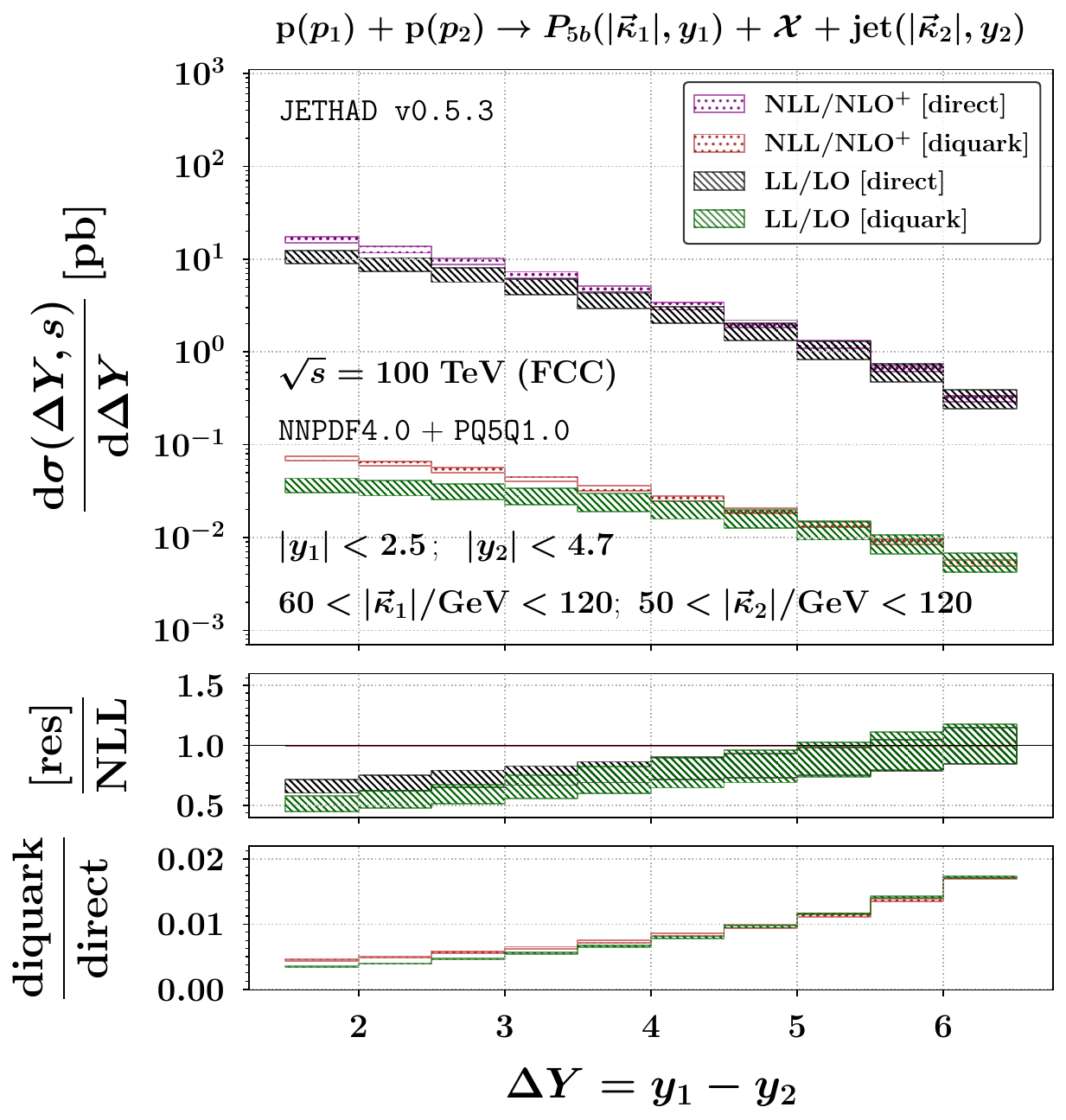}

\caption{Rapidity-interval rates for the semi-inclusive production of $\PQb$ plus jet systems at $\sqrt{s} = 14$ TeV (HL-LHC, left) and $100$ TeV (nominal FCC, right).
The first set of ancillary panels beneath the main plots displays the ratio between $\LL$ and $\NLLp$ predictions.
The second set illustrates the relative weight of the diquark-induced and direct fragmentation channels at the initial scale.
The uncertainty bands represent the combined effect of MHOUs and multidimensional phase-space integration.}
\label{fig:I}
\end{figure*}

Figure~\ref{fig:I} presents our resummed predictions for $\PQb$ plus jet rapidity-interval ($\DY$) distributions at the $14$~TeV HL-LHC (left panel) and the $100$~TeV FCC (right panel).
To support meaningful comparisons with forthcoming experimental measurements, we adopt uniform $\DY$ bins of fixed width $\Delta Y = 0.5$ throughout the analysis.

The first set of sub-panels beneath the main distributions shows the ratio of pure $\LL$ results to those obtained at $\NLLp$ accuracy.
The second set displays the relative contribution of the diquark fragmentation mechanism compared to the direct channel.

Cross-section values span a promising range, from about $3 \times 10^{-5}$~pb up to $20$~pb.
Notably, the overall event yield increases by roughly an order of magnitude when moving from HL-LHC to FCC energies (we emphasize that the vertical scales differ between the two panels.)

All $\DY$ distributions display a consistent pattern: the cross section decreases as $\Delta Y$ increases.
This behavior emerges from the interplay between two competing mechanisms.
On one side, the partonic hard factor grows with energy (and thus with $\Delta Y$), as expected from high-energy resummation.
On the other, this growth is counteracted by suppression effects originating from collinear convolutions with PDFs and FFs within the emission functions (see Eqs.~\eqref{LOHEF} and~\eqref{LOJEF}).

From a resummation standpoint, two important observations can be made.
First, the predictions demonstrate a remarkable degree of stability against MHOUs.
The relative size of the uncertainty bands remains below a factor of 1.5 for the HL-LHC setup and tightens further to around 1.25 at FCC energies (see first ancillary panels).
While these bands may seem broad when compared to fixed-order benchmarks, they are notably narrow within the context of high-energy resummation.
Indeed, in standard BFKL analyses of light-particle production, such as dijet or hadron-jet final states~\cite{Celiberto:2016hae,Celiberto:2017ptm,Bolognino:2018oth,Celiberto:2020wpk}, NLL or NLO$^{(+)}$ corrections often induce variations spanning one or even two orders of magnitude.
In contrast, our results for exotic $\PQb$
baryons remain well-behaved across the full $\Delta Y$ range.

Second, we observe clear signs of convergence when moving from $\LL$ to $\NLLp$ accuracy: the $\NLLp$ bands are systematically narrower than the $\LL$ ones and gradually approach them at large rapidity intervals, eventually exhibiting partial overlap in that region.
This behavior echoes previous findings in the production of doubly and fully charmed tetraquarks~\cite{Celiberto:2023rzw,Celiberto:2024mab}, and further supports the idea that forward semi-inclusive production of exotic states via single-parton fragmentation constitutes a highly stable channel for exploring high-energy QCD dynamics.

Turning to the fragmentation mechanisms, we find that the diquark-induced channel consistently dominates over the direct channel, particularly in the moderate and low $\Delta Y$ region.
This enhancement points to a potentially strong interplay between heavy-flavor fragmentation dynamics and high-energy resummation, which deserves further dedicated investigation.

\subsection{Transverse-momentum rates}
\label{ssec:pT_rates}

\begin{figure*}[!t]
\centering

   \hspace{0.00cm}
   \includegraphics[scale=0.375,clip]{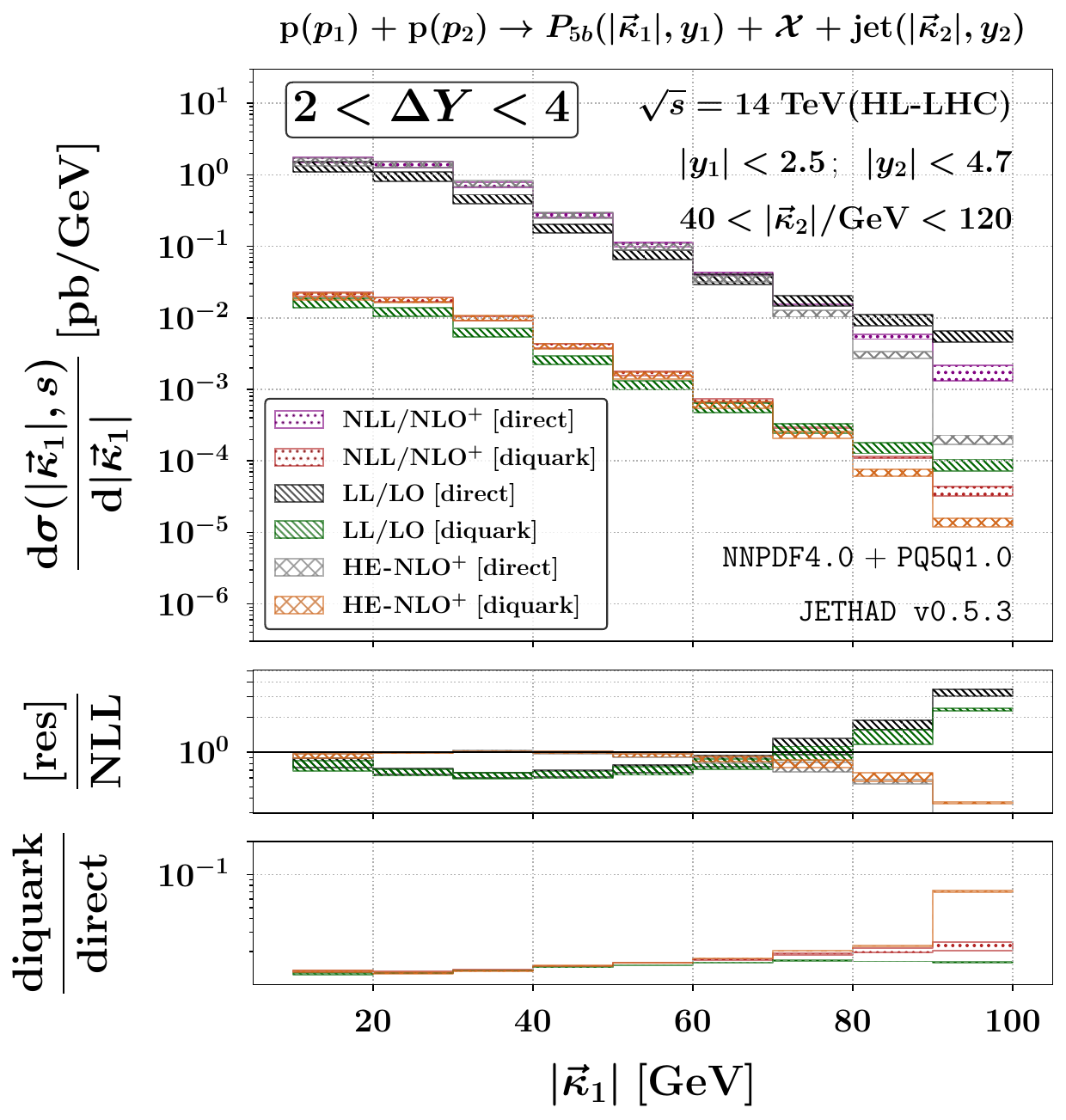}
   \hspace{0.30cm}
   \includegraphics[scale=0.375,clip]{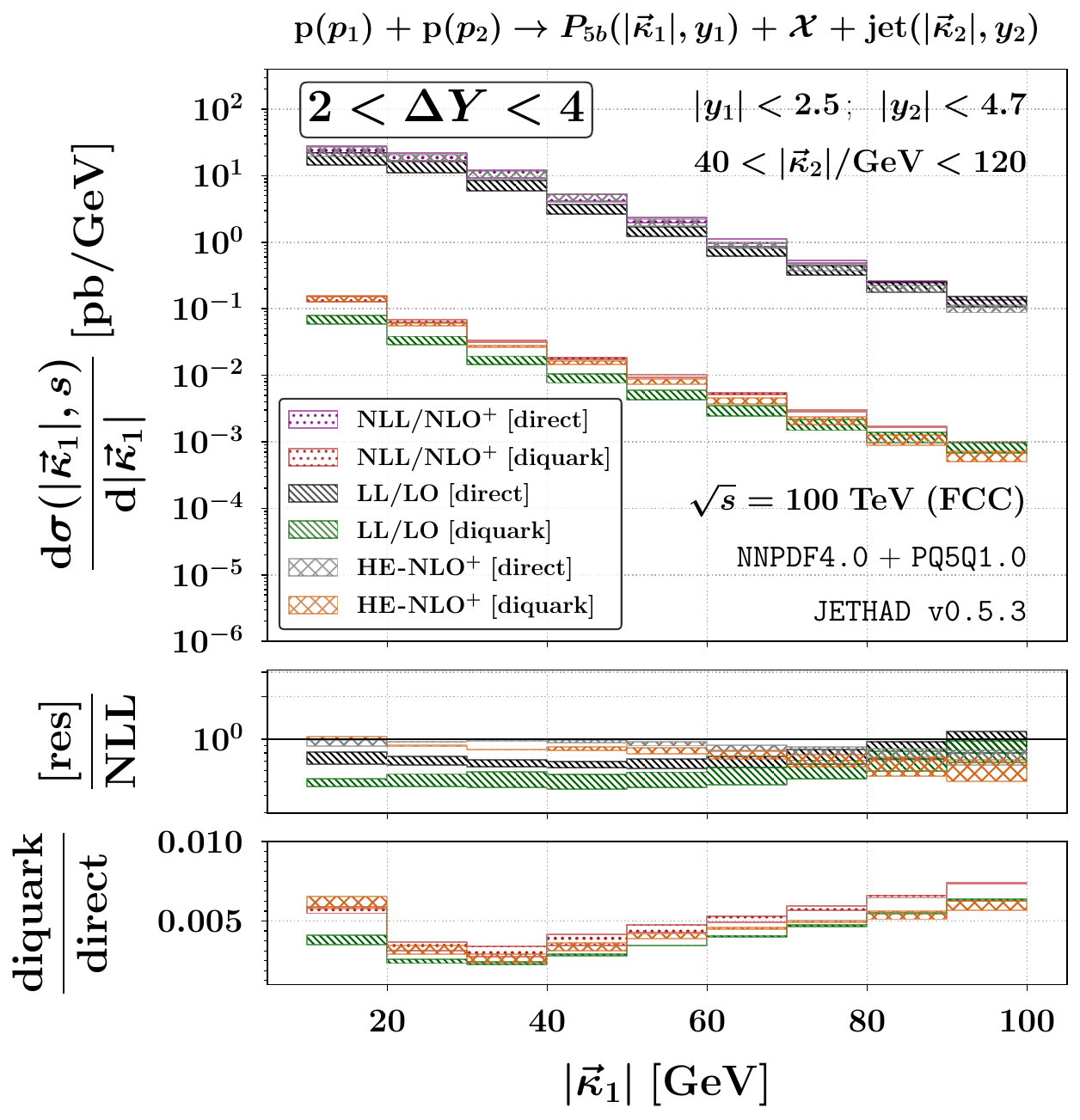}

   \vspace{0.45cm}  

   \hspace{0.00cm}
   \includegraphics[scale=0.375,clip]{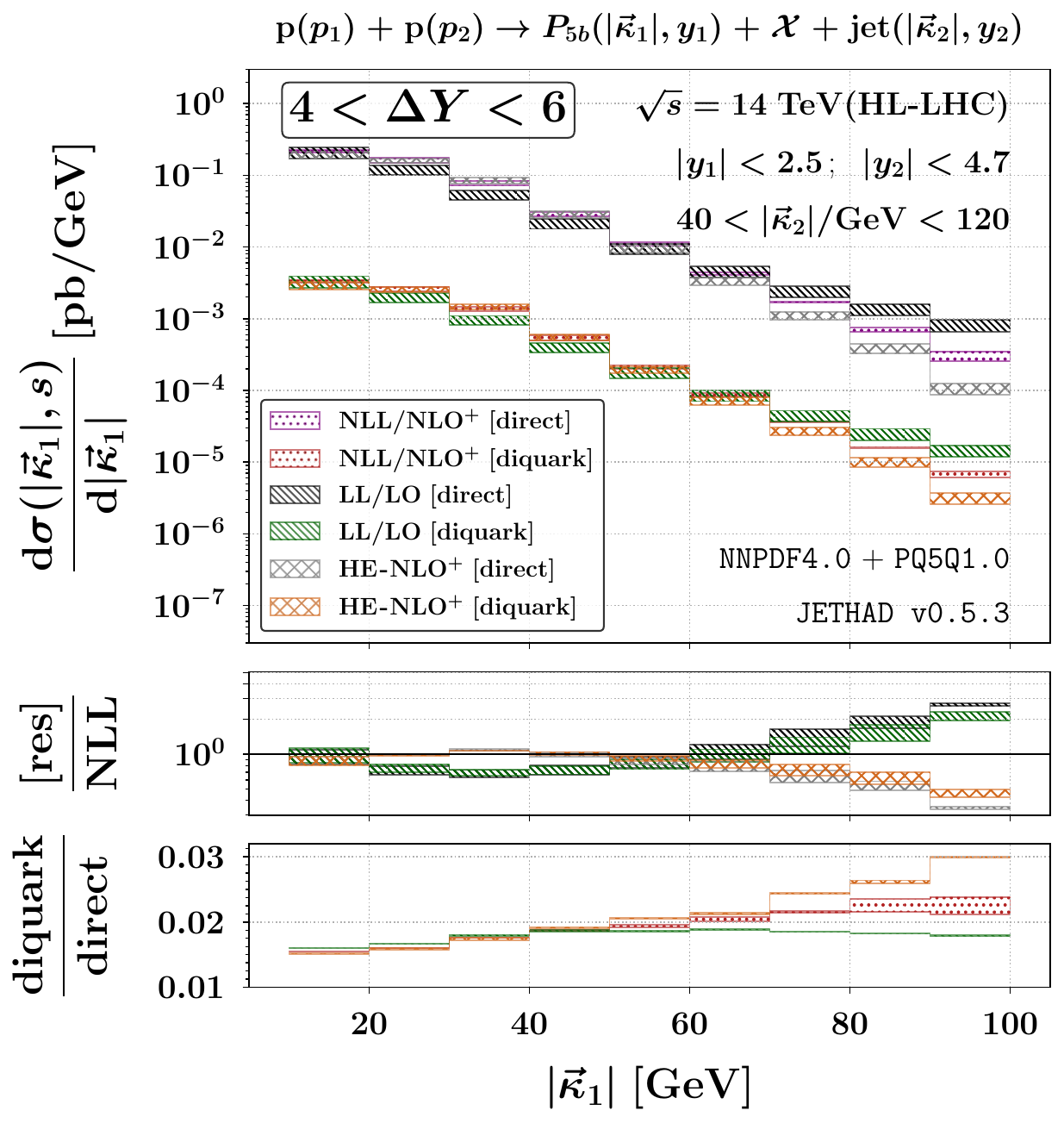}
   \hspace{0.30cm}
   \includegraphics[scale=0.375,clip]{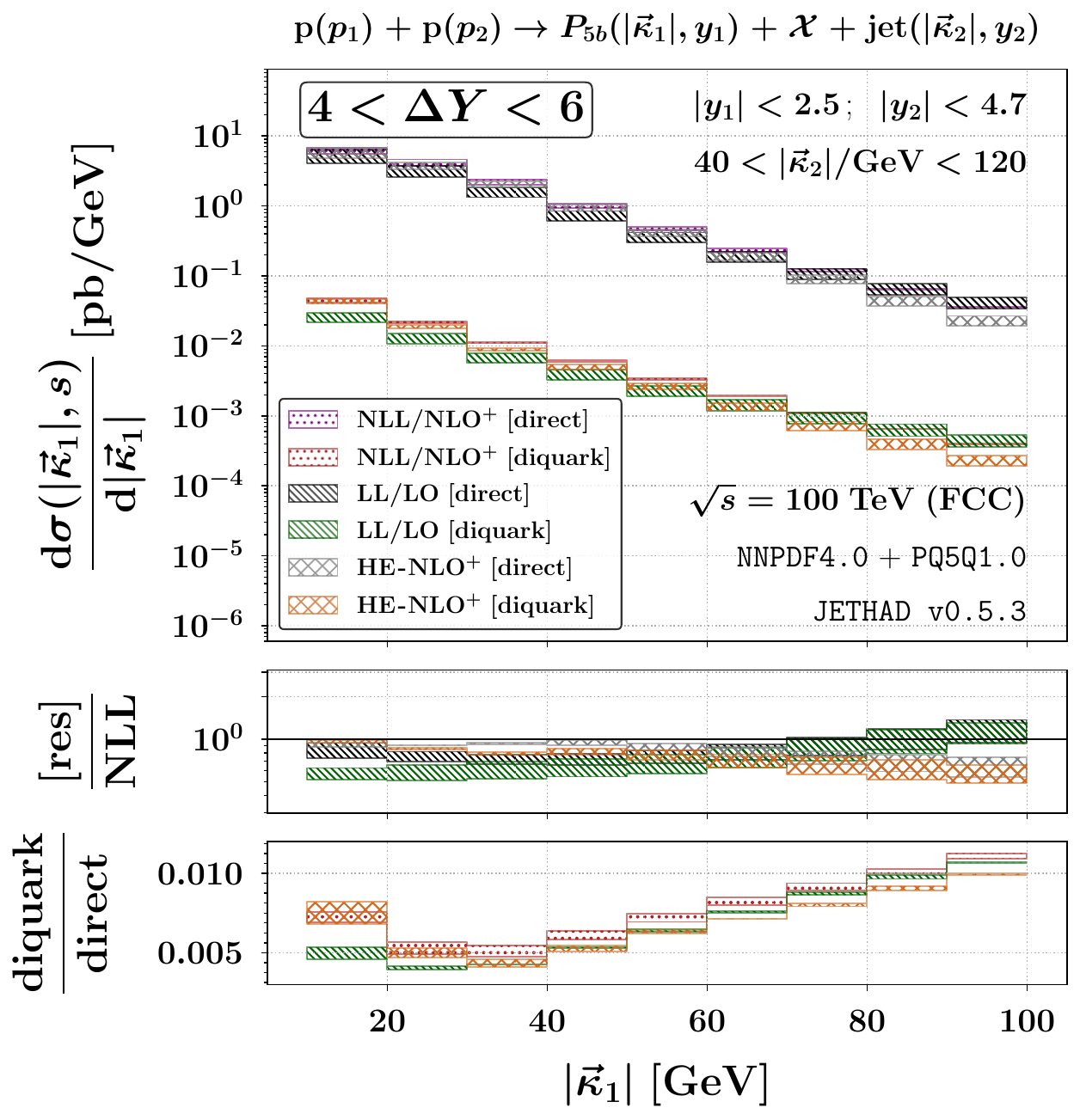}

\caption{Transverse-momentum rates for the semi-inclusive production of $\PQb$ plus jet systems at $\sqrt{s} = 14$ TeV (HL-LHC, left) and $100$ TeV (nominal FCC, right), and for $2 < \DY < 4$ (lower) or $4 < \DY < 6$ (lower).
The first set of ancillary panels beneath the main plots displays the ratio between $\LL$ or $\HENLOp$ and $\NLLp$ predictions.
The second set illustrates the relative weight of the diquark-induced and direct fragmentation channels at the initial scale.
The uncertainty bands represent the combined effect of MHOUs and multidimensional phase-space integration.}
\label{fig:I-k1b}
\end{figure*}

The upper panels of Fig.~\ref{fig:I-k1b} display the transverse-momentum ($|\vec{\kappa}_1|$) rates for the production of $\PQb$ plus jet systems at the 14~TeV HL-LHC (left) and the 100~TeV FCC (right).
These predictions are integrated over a ``lower'' rapidity-interval bin, $2 < \Delta Y < 4$, and use fixed-width transverse-momentum bins of 10~GeV to facilitate future comparisons with experimental data.

The lower panels of Fig.~\ref{fig:I-k1b} show the same rates computed in the adjacent ``upper'' $\Delta Y$ bin, $4 < \Delta Y < 6$.
In all panels, the first set of ancillary subplots shows the ratio of resummed results---either $\LL$ or $\HENLOp$, collectively labeled as ${\rm [res]}$---to the $\NLLp$ baseline.
The second ancillary panels show the relative contribution from the diquark versus direct initial-scale fragmentation mechanisms.

Across both collider energies and rapidity bins, the transverse-momentum spectra exhibit the expected steep falloff with increasing $|\vec{\kappa}_1|$.
The predictions also display notable stability under energy-scale variations, with the uncertainty bands in the first ancillary panels remaining below a relative width of 20\%.

We observe that the $\HENLOp$ to $\NLLp$ ratio typically stays below unity, and decreases further as $|\vec{\kappa}_1|$ increases.
In contrast, the $\LL$ to $\NLLp$ ratio follows an opposite trend: starting below one at low transverse momentum and rising progressively with increasing $|\vec{\kappa}_1|$.
This contrasting behavior stems from a complex interplay of dynamical effects.

On the one hand, previous studies have shown that the relative behavior between NLL-resummed signals and NLO high-energy backgrounds can vary significantly depending on the observable.
For instance, in cascade-baryon plus jet production, the $\HENLOp$ to $\NLLp$ ratio remains above one across all $|\vec{\kappa}_1|$ values (see Fig.7 of Ref.~\cite{Celiberto:2022kxx}).
By contrast, preliminary results on Higgs-jet observables, evaluated within a partially matched NLL/NLO framework, suggest a more nuanced picture~\cite{Celiberto:2023dkr}.

The fact that the $\HENLOp$ to $\NLLp$ ratio remains below one for $\PQb$ plus jet systems appears to be a distinctive signature of this specific process.
Interestingly, similar suppression patterns are found in transverse-momentum spectra for tetraquark plus jet final states~\cite{Celiberto:2023rzw,Celiberto:2024mab,Celiberto:2024beg}, pointing to unique dynamics in exotic-hadron production that amplify the interplay between heavy-flavor fragmentation and high-energy resummation.

On the other hand, the trend of the $\LL$ to $\NLLp$ ratio is closely tied to the structure of NLO corrections to the underlying emission functions.
For jet production, the NLO corrections to the jet function are known to be predominantly negative~\cite{Ivanov:2012ms,Colferai:2015zfa}.
In contrast, the hadron-side contributions receive mixed NLO corrections: the $C_{gg}$ coefficient contributes positively, while the $C_{gq}$, $C_{qg}$, and $C_{qq}$ coefficients yield negative terms~\cite{Ivanov:2012iv}.
This compensation can lead to partial cancellations in certain kinematic regions, modifying the shape and magnitude of the $\LL$ to $\NLLp$ ratio.

For example, in the case of $\Xi^-/\bar\Xi^+$ plus jet final states, the $\LL$ to $\NLLp$ ratio exceeds one across most of the $|\vec{\kappa}_1|$ spectrum~\cite{Celiberto:2022kxx}.
However, this enhancement is much milder in channels such as doubly charmed tetraquark plus jet production~\cite{Celiberto:2023rzw}, underscoring the process-dependent nature of the balance between leading and subleading logarithmic contributions.

These differences become especially relevant at large $|\vec{\kappa}_1|$, where the system moves away from the quasi-symmetric regime ($|\vec{\kappa}_1| \simeq |\vec{\kappa}_2|$).
Such asymmetric configurations lie at the edge of the formal applicability of high-energy resummation.
They are more sensitive to subleading effects and to the interplay between resummed and fixed-order contributions, often resulting in wider uncertainty bands and more pronounced deviations between $\LL$ and $\NLLp$ results.
This behavior is consistent with earlier observations in forward observables involving heavy-flavored final states~\cite{Bolognino:2021mrc,Celiberto:2024beg,Celiberto:2025ziy}, and reflects an intrinsic limitation of high-energy resummation when applied far from its optimal domain.

Finally, an inspection of the second ancillary panels in Fig.~\ref{fig:I-k1b} reveals that the diquark-like initial-scale fragmentation of the $\PQb$ state yields systematically higher rates than the direct channel.
In most cases, the diquark contribution increases with $|\vec{\kappa}_1|$, and this enhancement becomes more pronounced when either truncated or fully resummed NLL corrections are included.

\section{Final remarks}
\label{sec:conclusions}

In this work, we studied the leading-power fragmentation of fully heavy $S$-wave pentaquark states $\PQQ$ at future high-energy hadron colliders.
To this end, we introduced a new set of hadron-structure-oriented, \emph{multimodal} collinear FFs, labeled {\tt PQ5Q1.0}, which incorporate an improved modeling of the initial-scale input for the constituent (anti)bottom fragmentation channel.
These functions are designed to capture short-distance emissions of both compact multiquark states and diquark-like configurations.

To assess the phenomenological impact of these functions, we employed the data-validated {\Jethad} numerical framework and its symbolic plugin, {\symJethad}~\cite{Celiberto:2020wpk,Celiberto:2022rfj,Celiberto:2023fzz,Celiberto:2024mrq,Celiberto:2024swu,Celiberto:2025csa}, analyzing semi-inclusive production rates for $\PQb$ systems within a hybrid factorization formalism that combines NLL high-energy and NLO$^+$ collinear resummation.
Our predictions span a wide energy range, from 14~TeV (HL-LHC) up to 100~TeV (FCC).

We found that applying ZM-VFNS collinear fragmentation to describe $\PQb$ production at high transverse momentum significantly stabilizes the high-energy resummation framework.
In particular, it suppresses instabilities driven by NLL corrections and unresummed threshold logarithms.
This emergent \emph{natural stability} enhances the predictive power and convergence of our approach across the full collider energy spectrum considered.

To reach a higher level of precision, we plan to upgrade our $\NLLp$ factorization framework through a multilateral strategy that incorporates additional resummation techniques.
An initial step will involve establishing connections with soft-gluon~\cite{Hatta:2020bgy,Hatta:2021jcd,Caucal:2022ulg,Taels:2022tza} and jet-radius resummations~\cite{Dasgupta:2014yra,Dasgupta:2016bnd,Banfi:2012jm,Banfi:2015pju,Liu:2017pbb}.
In parallel, exploring potential synergies with ongoing developments in the study of jet angularities~\cite{Luisoni:2015xha,Caletti:2021oor,Reichelt:2021svh} represents an exciting direction for future investigation.

The leading-power fragmentation of heavy-flavored hadrons offers a unique and valuable intersection between hadronic structure and precision QCD.
The presence of one or more heavy quarks in the lowest Fock state has a twofold implication.
On the one hand, it complicates the modeling of the initial-scale fragmentation, compared to light hadrons, requiring a hadron-structure-oriented treatment of the nonperturbative component, potentially encoding momentum and spin correlations among constituent partons.
On the other hand, the large mass of the heavy quarks relative to $\LQCD$ allows for a controlled application of perturbative methods in computing the short-distance component.
As such, a precise description of heavy-flavor fragmentation demands a hybrid approach that draws from both hadron-structure modeling and high-precision QCD techniques.

Further progress will also depend on establishing connections between our framework and NLO studies of single-forward or nearly back-to-back semi-inclusive emissions within the gluon saturation formalism (see, for instance, Refs.~\cite{Gelis:2010nm,Kovchegov:2012mbw,Chirilli:2012jd,Boussarie:2014lxa,Benic:2016uku,Benic:2018hvb,Roy:2019hwr,Roy:2019cux,Beuf:2020dxl,Iancu:2021rup,Iancu:2023lel,vanHameren:2023oiq,Wallon:2023asa,Agostini:2024xqs,Altinoluk:2024zom,Altinoluk:2025dwd,Altinoluk:2025tms} and references therein).
These analyses have addressed the role of soft-gluon radiation in generating angular asymmetries in dijet and dihadron production (see Refs.~\cite{Hatta:2020bgy,Hatta:2021jcd,Caucal:2021ent,Caucal:2022ulg,Taels:2022tza}).

The NLO saturation framework opens access to the (un)polarized gluon content of protons and nucleons in the small-$x$ regime~\cite{Kotko:2015ura,vanHameren:2016ftb,Altinoluk:2020qet,Altinoluk:2021ygv,Boussarie:2021ybe,Caucal:2023nci,Cheung:2024qvw,Caucal:2025mth}.
In this context, several works have explored heavy-hadron production in proton-proton and proton-nucleus collisions while incorporating small-$x$ effects~\cite{Kang:2013hta,Ma:2014mri,Ma:2015sia,Ma:2018qvc,Stebel:2021bbn}.
A natural extension of our current program is to investigate the interplay between tetraquark production via NLLp hybrid factorization and NLO saturation-based calculations for exclusive emissions of heavy particles~\cite{Mantysaari:2021ryb,Mantysaari:2022kdm}.

A key milestone in refining our understanding of tetraquark fragmentation will involve benchmarking our {\tt PQ5Q1.0} FFs against those extracted from global data.
In this direction, machine-learning-based techniques---already employed for the collinear fragmentation of lighter hadron species~\cite{Nocera:2017qgb,Bertone:2017xsf,Bertone:2017tyb,Bertone:2018ecm,Khalek:2021gxf,Khalek:2022vgy,Soleymaninia:2022qjf,Soleymaninia:2022alt}---will serve as a valuable tool.

A key milestone in advancing our understanding of exotic pentaquark fragmentation will be the benchmarking of the {\tt PQ5Q1.0} FFs against determinations extracted from global data.
In this context, machine-learning-based methodologies, already applied to the collinear fragmentation of lighter hadron species~\cite{Nocera:2017qgb,Bertone:2017xsf,Bertone:2017tyb,Bertone:2018ecm,Khalek:2021gxf,Khalek:2022vgy,Soleymaninia:2022qjf,Soleymaninia:2022alt}, will represent a powerful tool for both fitting and modeling purposes.

Looking ahead, we aim to enhance our framework by extending the {\HFNRevo} scheme~\cite{Celiberto:2024mex,Celiberto:2024bxu,Celiberto:2024rxa,Celiberto:2025xvy} with systematic uncertainty quantification, potentially incorporating MHOU-related effects~\cite{Kassabov:2022orn,Harland-Lang:2018bxd,Ball:2021icz,McGowan:2022nag,NNPDF:2024dpb,Pasquini:2023aaf}.
In parallel, the inclusion of additional fragmentation channels---such as those initiated by light and bottom quarks---will allow for a more complete and threshold-aware implementation of DGLAP evolution.
This, in turn, will provide deeper insight into the distinct production mechanisms of fully heavy and heavy-light pentaquark states.

A deeper understanding of hadron structure will emerge as we further unravel the fundamental dynamics governing quarkonium formation and the genesis of exotic matter.
This progress will be fueled by data from the FCC~\cite{FCC:2018byv,FCC:2018evy,FCC:2018vvp,FCC:2018bvk,FCC:2025lpp,FCC:2025uan,FCC:2025jtd} and other next-generation collider facilities~\cite{Chapon:2020heu,Anchordoqui:2021ghd,Feng:2022inv,AlexanderAryshev:2022pkx,Arbuzov:2020cqg,Accettura:2023ked,InternationalMuonCollider:2024jyv,MuCoL:2024oxj,MuCoL:2025quu,Black:2022cth,Accardi:2023chb}.
A particularly promising channel for probing the proton's intrinsic charm content was recently proposed: unresolved photoproduction of a $\Jpsi$ accompanied by a charmed jet~\cite{Flore:2020jau} at the future EIC~\cite{AbdulKhalek:2021gbh,Khalek:2022bzd,Hentschinski:2022xnd,Amoroso:2022eow,Abir:2023fpo,Allaire:2023fgp}.
Such measurements would enable a direct probe of valence-like intrinsic-charm PDFs~\cite{Brodsky:1980pb,Brodsky:2015fna,Jimenez-Delgado:2014zga,Ball:2016neh,Hou:2017khm,Ball:2022qks,Guzzi:2022rca,NNPDF:2023tyk}, potentially bridging intrinsic-charm dynamics and the phenomenology of exotic hadrons~\cite{Vogt:2024fky}.

A connection between intrinsic charm and the formation of doubly charmed pentaquarks was explored in Ref.~\cite{Mikhasenko:2012km}, while the role of intrinsic charm in $\Lambda_b$ decays to pentaquark final states was examined in Ref.~\cite{Hsiao:2015nna}.
Furthermore, multi-$\Jpsi$ production events observed by the NA3 experiment in the 1980s, and later revisited at the LHC and Tevatron, may offer supporting evidence for mechanisms involving pion-induced double intrinsic charm and intermediate tetraquark resonances~\cite{NA3:1982qlq,NA3:1985rmd}.

Another compelling avenue involves exploiting tetraquark and tetraquark-in-jet observables to probe the QCD ``dead-cone effect'', a distinctive suppression of collinear gluon radiation from heavy quarks first predicted in the early 1990s~\cite{Dokshitzer:1991fd} and recently observed at the ALICE experiment~\cite{ALICE:2021aqk}.
Programs focused on the high-energy production of exotic states will provide fertile ground for investigating this and other emergent QCD phenomena.

Studies on pentaquark structure and formation mechanisms offer a novel testbed for QCD in its nonperturbative domain, where confinement and hadronization shape the spectrum of observable states~\cite{Jaffe:2004ph,Jaffe:2005md,Shifman:2005zn,Shifman:2007xn,Esposito:2016noz}.
In addition, their study contributes to a broader effort to map the internal quark-gluon structure of hadrons, complementing the physics of conventional baryons and mesons~\cite{Klempt:2009pi,Ketzer:2019wmd}.

Although further research is needed, both on theoretical and phenomenological grounds, we believe that the present study paves the way for new discovery opportunities.
It offers a compelling avenue to deepen our understanding of the underlying structure of exotic matter, potentially accessible at HL-LHC and next-generation colliders~\cite{FCC:2018byv,FCC:2018evy,FCC:2018vvp,FCC:2018bvk,FCC:2025lpp,FCC:2025uan,FCC:2025jtd,Chapon:2020heu,LHCspin:2025lvj,Anchordoqui:2021ghd,Feng:2022inv,Hentschinski:2022xnd,Accardi:2012qut,AbdulKhalek:2021gbh,Khalek:2022bzd,Acosta:2022ejc,AlexanderAryshev:2022pkx,LinearCollider:2025lya,LinearColliderVision:2025hlt,Brunner:2022usy,Arbuzov:2020cqg,Abazov:2021hku,Bernardi:2022hny,Amoroso:2022eow,Celiberto:2018hdy,Klein:2020nvu,2064676,MuonCollider:2022xlm,Aime:2022flm,MuonCollider:2022ded,Black:2022cth,Accettura:2023ked,InternationalMuonCollider:2024jyv,MuCoL:2024oxj,InternationalMuonCollider:2025sys,Vignaroli:2023rxr,Dawson:2022zbb,Bose:2022obr,Begel:2022kwp,Abir:2023fpo,Accardi:2023chb,Gessner:2025acq,Altmann:2025feg}.

\section*{Data availability}
\label{sec:data_availability}
\addcontentsline{toc}{section}{\nameref{sec:data_availability}}

The two grids for the {\tt PQ5Q1.0} FF family
\begin{itemize}
    \item NLO, $\PQc$\,: \,{\tt PQ5Q10\_cs\_P5c\_nlo};
    \item NLO, $\PQb$\,: \,{\tt PQ5Q10\_cs\_P5b\_nlo},
\end{itemize}
can be publicly accessed from the following url: \url{https://github.com/FGCeliberto/Collinear_FFs/}.
The central value (replica 0) of each set is for the direct fragmentation channel, whereas replica 1 portrays the diquark channel.

Data underlying figures presented in this review can be made available upon a reasonable request.

\section*{Acknowledgments}
\label{sec:acknowledgments}
\addcontentsline{toc}{section}{\nameref{sec:acknowledgments}}

The author thanks colleagues of \textbf{Quarkonia As Tools} and \textbf{EXOTICO} conferences for fruitful conversations.
The author is grateful to Alessandro~Papa, Seyed~Mohammad~Moosavi~Nejad, and Alessandro~Pilloni for insightful discussions.
This work was supported by the Atracci\'on de Talento Grant n. 2022-T1/TIC-24176 of the Comunidad Aut\'onoma de Madrid, Spain.

\begin{appendices}

\printacronyms

\setcounter{appcnt}{0}
\hypertarget{app:NLOHEF}{
\section*{Appendix~A: NLO correction to the heavy-hadron impact factor}}
\label{app:NLOHEF}

The analytical expression for the NLO correction to the forward heavy-hadron singly off-shell emission function reads~\cite{Ivanov:2012iv}

\begin{equation}
  \label{NLOHEF}
  \hat \F_h(n,\nu,|\vec \kappa_h|,x_h)=
  \frac{1}{\pi}\sqrt{\frac{C_F}{C_A}}
  \left(|\vec \kappa_h|^2\right)^{i\nu-\frac{1}{2}}
  \int_{x_h}^1\frac{\drv x}{x}
  \int_{\frac{x_h}{x}}^1\frac{\drv \eta}{\eta}
  \left(\frac{x\eta}{x_h}\right)^{2i\nu-1}
\end{equation}
  \[ \times \,
  \left[
  \frac{C_A}{C_F}f_g(x)D_g^h\left(\frac{x_h}{x\eta}\right){\cal C}_{gg}
  \left(x,\eta\right)+\sum_{i=q\bar q}f_i(x)D_i^h
  \left(\frac{x_h}{x\eta}
  \right){\cal C}_{qq}\left(x,\eta\right)
  \right.
  \]
  \[ + \,
  \left.D_g^h\left(\frac{x_h}{x\eta}\right)
  \sum_{i=q\bar q}f_i(x){\cal C}_{qg}
  \left(x,\eta\right)+\frac{C_A}{C_F}f_g(x)\sum_{i=q\bar q}D_i^h
  \left(\frac{x_h}{x\eta}\right){\cal C}_{gq}\left(x,\eta\right)
  \right]\, ,
  \]
with the ${\cal C}_{ij}$ partonic coefficients being
\begin{equation}
\stepcounter{appcnt}
\label{Cgg_hadron}
 {\cal C}_{gg}\left(x,\eta\right) =  P_{gg}(\eta)\left(1+\eta^{-2\gamma}\right)
 \ln \left( \frac {|\vec \kappa_h|^2 x^2 \eta^2 }{\mu_F^2 x_h^2}\right)
 -\frac{\beta_0}{2}\ln \left( \frac {|\vec \kappa_h|^2 x^2 \eta^2 }
 {\mu^2_R x_h^2}\right)
\end{equation}
\[
 + \, \delta(1-\eta)\left[C_A \ln\left(\frac{s_0 \, x^2}{|\vec \kappa_h|^2 \,
 x_h^2 }\right) \chi(n,\gamma)
 - C_A\left(\frac{67}{18}-\frac{\pi^2}{2}\right)+\frac{5}{9}n_f
 \right.
\]
\[
 \left.
 +\frac{C_A}{2}\left(\psi^\prime\left(1+\gamma+\frac{n}{2}\right)
 -\psi^\prime\left(\frac{n}{2}-\gamma\right)
 -\chi^2(n,\gamma)\right) \right]
 + \, C_A \left(\frac{1}{\eta}+\frac{1}{(1-\eta)_+}-2+\eta\bar\eta\right)
\]
\[
 \times \, \left(\chi(n,\gamma)(1+\eta^{-2\gamma})-2(1+2\eta^{-2\gamma})\ln\eta
 +\frac{\bar \eta^2}{\eta^2}{\cal I}_2\right)
\]
\[
 + \, 2 \, C_A (1+\eta^{-2\gamma})
 \left(\left(\frac{1}{\eta}-2+\eta\bar\eta\right) \ln\bar\eta
 +\left(\frac{\ln(1-\eta)}{1-\eta}\right)_+\right) \ ,
\]

\begin{equation}
\stepcounter{appcnt}
\label{Cgq_hadron}
 {\cal C}_{gq}\left(x,\eta\right)=P_{qg}(\eta)\left(\frac{C_F}{C_A}+\eta^{-2\gamma}\right)\ln \left( \frac {|\vec \kappa_h|^2 x^2 \eta^2 }{\mu_F^2 x_h^2}\right)
\end{equation}
\[
 + \, 2 \, \eta \bar\eta \, T_R \, \left(\frac{C_F}{C_A}+\eta^{-2\gamma}\right)+\, P_{qg}(\eta)\, \left(\frac{C_F}{C_A}\, \chi(n,\gamma)+2 \eta^{-2\gamma}\,\ln\frac{\bar\eta}{\eta} + \frac{\bar \eta}{\eta}{\cal I}_3\right) \ ,
\]

\begin{equation}
\stepcounter{appcnt}
\label{qg}
 {\cal C}_{qg}\left(x,\eta\right) =  P_{gq}(\eta)\left(\frac{C_A}{C_F}+\eta^{-2\gamma}\right)\ln \left( \frac {|\vec \kappa_h|^2 x^2 \eta^2 }{\mu_F^2 x_h^2}\right)
\end{equation}
\[
 + \eta\left(C_F\eta^{-2\gamma}+C_A\right) + \, \frac{1+\bar \eta^2}{\eta}\left[C_F\eta^{-2\gamma}(\chi(n,\gamma)-2\ln\eta)+2C_A\ln\frac{\bar \eta}{\eta} + \frac{\bar \eta}{\eta}{\cal I}_1\right] \ ,
\]
and
\begin{equation}
\stepcounter{appcnt}
\label{Cqq_hadron}
 {\cal C}_{qq}\left(x,\eta\right)=P_{qq}(\eta)\left(1+\eta^{-2\gamma}\right)\ln \left( \frac {|\vec \kappa_h|^2 x^2 \eta^2 }{\mu_F^2 x_h^2}\right)-\frac{\beta_0}{2}\ln \left( \frac {|\vec \kappa_h|^2 x^2 \eta^2 }{\mu^2_R x_h^2}\right)
\end{equation}
\[
 + \, \delta(1-\eta)\left[C_A \ln\left(\frac{s_0 \, x_h^2}{|\vec \kappa_h|^2 \, x^2 }\right) \chi(n,\gamma)+ C_A\left(\frac{85}{18}+\frac{\pi^2}{2}\right)-\frac{5}{9}n_f - 8\, C_F \right.
\]
\[
 \left. +\frac{C_A}{2}\left(\psi^\prime\left(1+\gamma+\frac{n}{2}\right)-\psi^\prime\left(\frac{n}{2}-\gamma\right)-\chi^2(n,\gamma)\right) \right] + \, C_F \,\bar \eta\,(1+\eta^{-2\gamma})
\]
\[
 +\left(1+\eta^2\right)\left[C_A (1+\eta^{-2\gamma})\frac{\chi(n,\gamma)}{2(1-\eta )_+}+\left(C_A-2\, C_F(1+\eta^{-2\gamma})\right)\frac{\ln \eta}{1-\eta}\right]
\]
\[
 +\, \left(C_F-\frac{C_A}{2}\right)\left(1+\eta^2\right)\left[2(1+\eta^{-2\gamma})\left(\frac{\ln (1-\eta)}{1-\eta}\right)_+ + \frac{\bar \eta}{\eta^2}{\cal I}_2\right] \; ,
\]

The $s_0$ scale is a BFKL-typical energy-normalization parameter, usually set to $s_0 = \mu_C$.
Furthermore, one has $\bar \eta \equiv 1 - \eta$ and $\gamma \equiv - \frac{1}{2} + i \nu$. The LO DGLAP $P_{i j}(\eta)$ splitting functions read
\begin{eqnarray}
\stepcounter{appcnt}
\label{DGLAP_kernels}
 P_{gq}(z)&=&C_F\frac{1+(1-z)^2}{z} \; , \\ \nonumber
 P_{qg}(z)&=&T_R\left[z^2+(1-z)^2\right]\; , \\ \nonumber
 P_{qq}(z)&=&C_F\left( \frac{1+z^2}{1-z} \right)_+= C_F\left[ \frac{1+z^2}{(1-z)_+} +{3\over 2}\delta(1-z)\right]\; , \\ \nonumber
 P_{gg}(z)&=&2C_A\left[\frac{1}{(1-z)_+} +\frac{1}{z} -2+z(1-z)\right]+\left({11\over 6}C_A-\frac{n_f}{3}\right)\delta(1-z) \; ,
\end{eqnarray}
while the ${\cal I}_{2,1,3}$ functions are
\begin{equation}
\stepcounter{appcnt}
\label{I2}
{\cal I}_2=
\frac{\eta^2}{\bar \eta^2}\left[
\eta\left(\frac{{}_2F_1(1,1+\gamma-\frac{n}{2},2+\gamma-\frac{n}{2},\eta)}
{\frac{n}{2}-\gamma-1}-
\frac{{}_2F_1(1,1+\gamma+\frac{n}{2},2+\gamma+\frac{n}{2},\eta)}{\frac{n}{2}+
\gamma+1}\right)\right.
\end{equation}
\[
 \stepcounter{appcnt}
 \left.
 +\eta^{-2\gamma}\left(\frac{{}_2F_1(1,-\gamma-\frac{n}{2},1-\gamma-\frac{n}{2},\eta)}{\frac{n}{2}+\gamma}-\frac{{}_2F_1(1,-\gamma+\frac{n}{2},1-\gamma+\frac{n}{2},\eta)}{\frac{n}{2} -\gamma}\right)
\right.
\]
\[
 \left.
 +\left(1+\eta^{-2\gamma}\right)\left(\chi(n,\gamma)-2\ln \bar \eta \right)+2\ln{\eta}\right] \; ,
\]
\begin{equation}
\stepcounter{appcnt}
\label{I1}
 {\cal I}_1=\frac{\bar \eta}{2\eta}{\cal I}_2+\frac{\eta}{\bar \eta}\left[\ln \eta+\frac{1-\eta^{-2\gamma}}{2}\left(\chi(n,\gamma)-2\ln \bar \eta\right)\right] \; ,
\end{equation}
and
\begin{equation}
\stepcounter{appcnt}
\label{I3}
 {\cal I}_3=\frac{\bar \eta}{2\eta}{\cal I}_2-\frac{\eta}{\bar \eta}\left[\ln \eta+\frac{1-\eta^{-2\gamma}}{2}\left(\chi(n,\gamma)-2\ln \bar \eta\right)\right] \; .
\end{equation}
Moreover, ${}_2F_1$ stands for the Gauss hypergeometric function.

The \emph{plus~prescription} in Eqs.~\eqref{Cgg_hadron} and~\eqref{Cqq_hadron} is given by
\begin{equation}
\label{plus-prescription}
\stepcounter{appcnt}
\int^1_\zeta \drv x \frac{f(x)}{(1-x)_+}
=\int^1_\zeta \drv x \frac{f(x)-f(1)}{(1-x)}
-\int^\zeta_0 \drv x \frac{f(1)}{(1-x)}\; ,
\end{equation}
where $f(x)$ represents a regular-behaved generic function at $x=1$.

\setcounter{appcnt}{0}
\hypertarget{app:NLOJEF}{
\section*{Appendix~B: NLO correction to the light-jet impact factor}}
\label{app:NLOJEF}

The analytical expression for the NLO correction to the forward light-jet singly off-shell emission function within the small-cone algorithm reads~\cite{Colferai:2015zfa}
\begin{equation}
\stepcounter{appcnt}
\label{NLOJEF}
 \hat \F_{J}(n,\nu,|\vec \kappa_J|,x_J)=
 \frac{1}{\pi}\sqrt{\frac{C_F}{C_A}}
 \left(|\vec \kappa_J|^2 \right)^{i\nu-1/2}
 \int^1_{x_J}\frac{\drv \eta}{\eta}
 \eta^{-\bar\alpha_s(\mu_R)\chi(n,\nu)}
\end{equation}
\[
\times\;
\left\{\sum_{i=q,\bar q} f_i \left(\frac{x_J}{ \eta}\right)\left[\left(P_{qq}(\eta)+\frac{C_A}{C_F}P_{gq}(\eta)\right)
\ln\frac{|\vec \kappa_J|^2}{\mu_F^2}\right.\right.
\]
\[
-\;2\eta^{-2\gamma} \ln \frac{{\cal R}}{\max(\eta, \bar \eta)} \,
\left\{P_{qq}(\eta)+P_{gq}(\eta)\right\}-\frac{\beta_0}{2}
\ln\frac{|\vec \kappa_J|^2}{\mu_R^2}\delta(1-\eta)
\]
\[
+\;C_A\delta(1-\eta)\left[\chi(n,\gamma)\ln\frac{s_0}{|\vec \kappa_J|^2}
+\frac{85}{18}
\right.
\]
\[
\left.
+\;\frac{\pi^2}{2}+\frac{1}{2}\left(\psi^\prime
\left(1+\gamma+\frac{n}{2}\right)
-\psi^\prime\left(\frac{n}{2}-\gamma\right)-\chi^2(n,\gamma)\right)
\right]
\]
\[
+\;(1+\eta^2)\left\{C_A\left[\frac{(1+\eta^{-2\gamma})\,\chi(n,\gamma)}
{2(1-\eta)_+}-\eta^{-2\gamma}\left(\frac{\ln(1-\eta)}{1-\eta}\right)_+
\right]
\right.
\]
\[
\left.
+\;\left(C_F-\frac{C_A}{2}\right)\left[ \frac{\bar \eta}{\eta^2}{\cal I}_2
-\frac{2\ln\eta}{1-\eta}
+2\left(\frac{\ln(1-\eta)}{1-\eta}\right)_+ \right]\right\}
\]
\[
+\;\delta(1-\eta)\left(C_F\left(3\ln 2-\frac{\pi^2}{3}-\frac{9}{2}\right)
-\frac{5n_f}{9}\right)
+C_A\eta+C_F\bar \eta
\]
\[
\left.
+\;\frac{1+\bar \eta^2}{\eta}
\left(C_A\frac{\bar \eta}{\eta}{\cal I}_1+2C_A\ln\frac{\bar\eta}{\eta}
+C_F\eta^{-2\gamma}(\chi(n,\gamma)-2\ln \bar \eta)\right)\right]
\]
\[
+\;f_{g}\left(\frac{x_J}{\eta}\right)\frac{C_A}{C_F}
\left[
\left(P_{gg}(\eta)+2 \,n_f \frac{C_F}{C_A}P_{qg}(\eta)\right)
\ln\frac{|\vec \kappa_J|^2}{\mu_F^2}
\right.
\]
\[
\left.
-\;2\eta^{-2\gamma} \ln \frac{{\cal R}}{\max(\eta, \bar \eta)} \left(P_{gg}(\eta)+2 \,n_f P_{qg}(\eta)\right)
-\frac{\beta_0}{2}\ln\frac{|\vec \kappa_J|^2}{4\mu_R^2}\delta(1-\eta)
\right.
\]
\[
\left.
+\; C_A\delta(1-\eta)
\left(
\chi(n,\gamma)\ln\frac{s_0}{|\vec \kappa_J|^2}+\frac{1}{12}+\frac{\pi^2}{6}
\right.\right.
\]
\[
\left.
+\;\frac{1}{2}\left[\psi^\prime\left(1+\gamma+\frac{n}{2}\right)
-\psi^\prime\left(\frac{n}{2}-\gamma\right)-\chi^2(n,\gamma)\right]
\right)
\]
\[
+\,2C_A (1-\eta^{-2\gamma})\left(\left(\frac{1}{\eta}-2
+\eta\bar\eta\right)\ln \bar \eta + \frac{\ln (1-\eta)}{1-\eta}\right)
\]
\[
+\,C_A\, \left[\frac{1}{\eta}+\frac{1}{(1- \eta)_+}-2+\eta\bar\eta\right]
\left((1+\eta^{-2\gamma})\chi(n,\gamma)-2\ln\eta+\frac{\bar \eta^2}
{\eta^2}{\cal I}_2\right)
\]
\[
\left.\left.
+\,n_f\left[\, 2\eta\bar \eta \, \frac{C_F}{C_A} +(\eta^2+\bar \eta^2)
\left(\frac{C_F}{C_A}\chi(n,\gamma)+\frac{\bar \eta}{\eta}{\cal I}_3\right)
-\frac{1}{12}\delta(1-\eta)\right]\right]\right\} \; ,
\]
with ${\cal R}$ denoting the jet-cone radius.

\end{appendices}

\bibliographystyle{elsarticle-num}

\bibliography{references}

\end{document}